\documentclass[twocolumn]{aastex631}

\usepackage{microtype}
\usepackage{bold-extra}
\usepackage{graphicx}
\usepackage{amsmath}
\usepackage{amssymb}
\usepackage{bm}
\usepackage{hyperref}

\definecolor{notes}{HTML}{C70039}
\definecolor{softgreen}{HTML}{468465}
\definecolor{edits}{HTML}{BB8F00}

\newcommand{\mgii}{\hbox{Mg\,{\sc ii}}}     
\newcommand{\oii}{\hbox{\sc [O\,ii]}}     
\newcommand{\hg}{\hbox{\sc H$\gamma$}}      
\newcommand{\hb}{\hbox{\sc H$\beta$}}       
\newcommand{\oiii}{\hbox{\sc [O\,iii]}}     
\newcommand{\oi}{\hbox{\sc [O\,i]}}     
\newcommand{\ha}{\hbox{\sc H$\alpha$}}      
\newcommand{\nii}{\hbox{[N\,{\sc ii}]}}     
\newcommand{\sii}{\hbox{[S\,{\sc ii}]}}     

\newcommand{\neiii}{\hbox{[Ne\,{\sc iii]}}}  

\newcommand{\met}{12 + log(O/H)}

\newcommand{\niiha}{N2\ha}
\newcommand{\niioii}{N2O2}

\newcommand{\siiha}{S2\ha}
\newcommand{\oiha}{O1\ha}

\newcommand{\HST}{{\it HST}}
\newcommand{\jwst}{{\it JWST}}
\newcommand{\JWST}{{\it JWST}}

\newcommand{\waz}{Waz Arc}

\newcommand{\jdrp}{\texttt{jwst}}

\newcommand{\lam}{$\lambda$}
\newcommand{\unit}[1]{\ensuremath{\mathrm{\,#1}}\xspace}

\newcommand{\ang}{$\mbox{\AA}$}

\newcommand{\kms}{km \,s$^{-1}$}

\newcommand{\Msolar}{M${_\odot}$}

\newcommand{\e}{\unit{e^{-}}}

\newcommand{\logU}{log$_{10}$U}

\mathchardef\mhyphen="2D

\newlength{\dhatheight}

\shorttitle{highly spatially-resolved physical conditions at $z=5$}
\shortauthors{Hutchison et al.}

\graphicspath{{./}{}}

\usepackage{fontspec}

\usepackage[english]{babel} 
\babelprovide{punjabi} 
\babelfont[punjabi]{rm}[
  Renderer=Harfbuzz,
  Script=Gurmukhi
]{FreeSerif.otf} 

\newcommand{\Punj}[1]{\selectlanguage{punjabi}{#1}\selectlanguage{english}}

\begin{document}

\title{JWST \& the Waz Arc I: Spatially Resolving the Physical Conditions within \\ a Post-Starburst Galaxy at Redshift 5 with NIRSpec IFS}

\author[0000-0001-6251-4988]{Taylor A. Hutchison}
\altaffiliation{NASA Postdoctoral Fellow}
\affiliation{Astrophysics Science Division, Code 660, NASA Goddard Space Flight Center, 8800 Greenbelt Rd., Greenbelt, MD 20771, USA}
\affiliation{Department of Astronomy, University of Maryland, Baltimore Country, MD 21250, USA}
\affiliation{Center for Research and Exploration in Space Science and Technology, NASA/GSFC, Greenbelt, MD 20771 USA}

\author[0000-0002-3475-7648]{Gourav Khullar (\Punj{ਗੌਰਵ ਖੁੱਲਰ})}
\altaffiliation{Baum Postdoctoral Fellow for Innovative Astronomy}
\affiliation{Department of Astronomy, University of Washington, Physics-Astronomy Building, Box 351580, Seattle, WA 98195-1700, USA}
\affiliation{eScience Institute, University of Washington, Physics-Astronomy Building, Box 351580, Seattle, WA 98195-1700, USA}

\author[0000-0002-7627-6551]{Jane R. Rigby}
\affiliation{Astrophysics Science Division, Code 660, NASA Goddard Space Flight Center, 8800 Greenbelt Rd., Greenbelt, MD 20771, USA}

\author[0000-0003-1815-0114]{Brian Welch}
\affiliation{International Space Science Institute, Hallerstrasse 6, 3012 Bern, Switzerland}

\author[0000-0001-5097-6755]{Michael K. Florian}
\affiliation{Observational Cosmology Lab Code 665, NASA Goddard Space Flight Center, Greenbelt, MD 20771, USA}

\author[0000-0002-7559-0864]{Keren Sharon}
\affiliation{Department of Astronomy, University of Michigan, 1085 S. University Ave, Ann Arbor, MI 48109, USA}


\author[0000-0002-2323-303X]{Isaac Sierra}
\affiliation{Department of Astronomy and Astrophysics, University of Chicago, 5640 South Ellis Avenue, Chicago, IL 60637, USA}

\author[0000-0002-5293-3975]{Julissa Sarmiento}
\affiliation{Department of Physics and Astronomy, University of Pittsburgh, Pittsburgh, PA 15260, USA}

\author[0000-0003-3266-2001]{Guillaume Mahler}
\affiliation{STAR Institute, Quartier Agora - All\'ee du six Ao\^ut, 19c B-4000 Li\`ege, Belgium}

\author[0000-0001-7151-009X]{Nikko J.\ Cleri}
\affiliation{Department of Astronomy and Astrophysics, The Pennsylvania State University, University Park, PA 16802, USA}
\affiliation{Institute for Computational \& Data Sciences, The Pennsylvania State University, University Park, PA 16802, USA}
\affiliation{Institute for Gravitation and the Cosmos, The Pennsylvania State University, University Park, PA 16802, USA}

\author[0000-0001-5063-8254]{Rachel Bezanson}
\affiliation{Department of Physics and Astronomy and PITT PACC, University of Pittsburgh, Pittsburgh, PA 15260, USA}

\author[0000-0003-1370-5010]{Michael D. Gladders}
\affiliation{Department of Astronomy and Astrophysics, University of Chicago, 5640 South Ellis Avenue, Chicago, IL 60637, USA}
\affiliation{Kavli Institute for Cosmological Physics, University of Chicago, 5640 South Ellis Avenue, Chicago, IL 60637, USA}

\author[0000-0003-1074-4807]{Matthew B. Bayliss}
\affiliation{Department of Physics, University of Cincinnati, Cincinnati, OH 45221, USA}

\author[0000-0002-1728-8042]{Juliana S. M. Karp}
\affiliation{Department of Astronomy, University of Washington, Physics-Astronomy Building, Box 351580, Seattle, WA 98195-1700, USA}

\author[0009-0000-5333-9970]{Dylan Berry}
\affiliation{Department of Astronomy, University of Washington, Physics-Astronomy Building, Box 351580, Seattle, WA 98195-1700, USA}

\author[0009-0005-8103-5823]{Alex Ross}
\affiliation{Department of Astronomy, University of Washington, Physics-Astronomy Building, Box 351580, Seattle, WA 98195-1700, USA}

\author[0000-0002-9204-3256]{T.\ Emil Rivera-Thorsen}
\affiliation{The Oskar Klein Centre, Department of Astronomy, Stockholm University, AlbaNova, 10691 Stockholm, Sweden}

\author[0000-0003-1343-197X]{Suhyeon C. Choe}
\affiliation{The Oskar Klein Centre, Department of Astronomy, Stockholm University, AlbaNova, 10691 Stockholm, Sweden}


\author[0000-0003-2200-5606]{H{\aa}kon Dahle}
\affiliation{Institute of Theoretical Astrophysics, University of Oslo, P.O. Box 1029, Blindern, NO-0315 Oslo, Norway}

\author[0000-0002-0302-2577]{John Chisholm}
\affiliation{Department of Astronomy, University of Texas at Austin, 2515 Speedway, Austin, Texas 78712, USA}

\author[0000-0003-3216-7190]{Erini L.\ Lambrides}
\altaffiliation{NASA Postdoctoral Fellow}
\affiliation{Astrophysics Science Division, Code 660, NASA Goddard Space Flight Center, 8800 Greenbelt Rd., Greenbelt, MD 20771, USA}
\affiliation{Department of Astronomy, University of Maryland, College Park, MD 20742, USA}
\affiliation{Center for Research and Exploration in Space Science and Technology, NASA/GSFC, Greenbelt, MD 20771 USA}

\author[0000-0003-2366-8858]{Rebecca L.\ Larson}
\altaffiliation{Giacconi Postdoctoral Fellow}
\affil{Space Telescope Science Institute, 3700 San Martin Drive, Baltimore, MD 21218, USA}

\author[0000-0002-4606-4240]{Grace M.\ Olivier}
\affiliation{The Observatories of the Carnegie Institution for Science, 813 Santa Barbara Street, Pasadena, CA 91101, USA}

\author[0000-0002-2862-307X]{Riley Owens}
\affiliation{University of California, Berkeley, Department of Astronomy, 501 Campbell Hall, Berkeley, CA, 94720 USA}
\affiliation{Department of Physics, University of Cincinnati, Cincinnati, OH 45221, USA}

\author[0000-0003-3302-0369]{Erik Solhaug}
\affiliation{Department of Astronomy and Astrophysics, University of Chicago, 5640 South Ellis Avenue, Chicago, IL 60637, USA}
\affiliation{Kavli Institute for Cosmological Physics, University of Chicago, 5640 South Ellis Avenue, Chicago, IL 60637, USA}


\begin{abstract}

We present NIRSpec/IFS observations of a restframe UV-bright, massive ($M_* \sim 10^{10}$ \Msolar, $z_{AB}=20.5$) galaxy highly magnified by gravitational-lensing observed just after the end of the epoch of reionization ($z=5.04$, $\bar{\mu}\sim90$). With \jwst\ accessing the restframe UV and optical spectrum of this galaxy with high fidelity, we classify this UV-bright galaxy as post-starburst in nature -- due to weak/absent emission lines and strong absorption features -- making this an example of a new class of UV-bright but significantly quenched galaxies being discovered in this epoch. With a median $E(B-V)=0.44\pm0.14$, we identify the presence of stellar absorption across the arc both in Balmer lines and the \mgii\ doublet, indicative of older stellar populations dominated by A stars (and potentially B stars). Using spatially-resolved maps of rest-optical strong emission lines, we find a heterogeneous distribution of nebular metallicities across the arc, potentially hinting at different enrichment processes.  With a low median lensing-corrected \ha\ star formation rate of SFR$_{H\alpha} = 0.024 \pm 0.001$ \Msolar\ yr$^{-1}$,  we find in the most ``star-forming'' clumps indications of lower ionization (log$_{10}$U $\sim -3.2$), lower nebular metallicities (12+log$_{10}$O/H $\lesssim$ 8.3), and hints of higher densities that suggest a possible recent infall of more pristine (low metallicity) gas onto the galaxy. Investigating the regions with no detectable \hb\ emission, we find (for the first time at $z>5$) signatures of diffuse ionized gas (DIG). Separating DIG from HII regions within a galaxy has predominantly been demonstrated at lower redshifts, where such spatial resolution allows clear separation of such regions -- highlighting the immense power of gravitational lensing to enable studies at the smallest spatial scales at cosmic dawn. 

\keywords{post-starburst galaxies, gravitational lensing, \jwst\, integral field spectroscopy, nebular line diagnostics, star formation histories, SED fitting}

\end{abstract}


\section{Introduction} \label{sec:intro}


The advent of the \jwst\ has enabled studies of stellar mass assembly in the high-redshift Universe, finding massive quiescent galaxies earlier in the Universe than previously expected \citep{carnall2024,Setton.2024,Weibel.2024, long2024,nanayakkara2025,degraaf2025}, as well as systems with bursty star formation \citep{trussler2024,ciesla2024} and rapid quenching \citep{Looser2023b, Strait2023, Looser.2024, covelo2025}. In simulations, early time low-mass galaxies are expected to have stochastic star formation histories (SFHs), as their gravitational potentials are shallow and cannot withstand feedback-based ejection of star-forming gas \citep[e.g.,][]{pallottini_2023}. As galaxies assemble mass, their gravitational potentials deepen and they eventually transition to SFHs driven by secular processes -- with works such as \citet{ciesla2024} reporting a stochastic to secular SFH transition mass of $\log(M_\star/M_{\odot})\lesssim 9$ at $z\sim6-9$.

In this period of stochasticity, we expect to observe some bursty galaxies in ``mini-quenched'' states, i.e., where the galaxy happens to be in a star formation lull (either temporarily slowing or halting its star formation). There are various possible modes of quenching in these systems, both internal (e.g., AGN, supernovae, stellar winds, gas outflows) and external/environmental (e.g., ram pressure stripping). These napping/mini-quenched galaxies show B-type post-starbust galaxy (PSB) spectral features -- strong Balmer absorption lines, bright UV flux and weak nebular emission features -- similar to systems discovered in \cite{Looser2023a} ($z=7.3$), \cite{Strait2023} ($z=5$) and \cite{covelo2025} ($z=5-7$).


At lower redshifts, PSBs have been studied in great detail. 
They were first described as E+A or K+A galaxies \citep[e.g.,][]{dressler1983,goto2004,quintero2004}, with their spectra dominated by A-type stars powering Balmer breaks, strong stellar absorption in the Balmer lines, and a general paucity (but not necessarily absence) of strong nebular emission features. Recently, studies have suggested that some PSB galaxies may have small pockets of star formation still ongoing \citep[e.g.,][]{chen2019,french2021,bezanson2022,cheng2024,zhu2025}.  In this scenario, remaining gas-rich reservoirs have been observed, potentially fueling more star-formation that is too dust-obscured to be measured by UV emission features (such as the \oii\lam\lam3727,3730 doublet) and/or SED fitting methods.

Above $z>1$, previous studies have measured these systems' global properties, since spatially resolved observations are difficult due to cosmological dimming and small angular diameter distances even resolved by space telescopes at these redshifts. Star-forming regions, star clusters, globular clusters, and giant molecular clouds are all well below sub-pixel resolution for \jwst. Because the signatures of quenching mechanisms are hidden at these scales, probing the spatial and temporal scales of star formation and quenching, and unpacking the physics of these regions with nebular diagnostics, requires spectroscopy of the smallest regions. 

The combination of \jwst\ spectroscopy and strong gravitational lensing provides the solution. Integral field spectroscopy (IFS) observations across highly magnified/sheared lensed galaxies are proving to be excellent cosmic telescopes to probe the smallest regions within galaxies at cosmic noon and beyond \citep[e.g.,][Olivier et al., in prep]{Birkin.2023,Rivera-Thorsen.2025,Choe.2025,Florian.2025}. Moreover, serendipitous studies to characterize UV-bright galaxies at $z>3$ have revealed mini-quenched/post-starburst activity in some systems (e.g., \citealt{pascale2025}). Initial investigations show that these are likely globally quenched systems with some star-forming pockets \citep{mowla2022,messa2025}.



Spatially-integrated measurements of high-redshift galaxies with \jwst\ have uncovered a complex mix of physical conditions, helping differentiate between different models of galaxy formation in early galaxies. Prior to \jwst, spatially-resolved measurements of quenched/PSB galaxies were limited to only those relatively nearby \citep{chen2019,cheng2024,Leung2025}. Thanks to \jwst, we are finally in an era of measuring the detailed physical conditions of individual star-forming regions within galaxies in early cosmic time. In particular, recent lensing studies using \jwst\ have revealed significant internal diversity in high-redshift systems and regions within them \citep[e.g.,][]{Perna.2023,Birkin.2023,Marconcini.2024,Rivera-Thorsen.2025,Roy.2025,Ju.2025,Jones.2025,Scholtz.2025,Parlanti.2025}. For example, \cite{2022ApJ...938L..16W} utilized strong lensing and NIRISS slitless spectroscopy to produce one of the first spatially resolved metallicity and nebular excitation maps at $z>3$, discovering a highly inverted metallicity gradient in a lensed dwarf galaxy. Soon after, \citet{Birkin.2023} shared one of the first gas-phase metallicity maps of two $z>3$ lensed dust-obscured, HST-dark star-forming galaxies, including identifying a small region within one source where there may possibly be an additional ionizing source.
\cite{2024Natur.636..332M} show that strongly lensed \jwst\ observations can resolve galaxies at $z\geq8$ into individual star-forming clumps on tens-of-parsec scales, providing evidence that early galaxies may assemble through compact clusters and irregular star formation patterns. However, only a handful of these analyses have targeted PSB systems, as their transient and elusive nature makes them challenging to capture. Thus, their internal physical mechanisms that govern their rapid quenching remain largely unexplored.


COOL J1241+2219 (which we nickname the Waz Arc) is the brightest galaxy at $z>5$ ($z_{AB}=20.47$; \citealt{Khullar.2021}). This source is a classic example of a spatially extended galaxy at $\log\text{M}\sim10~\text{M}_{\odot}$ in the early Universe. The Waz Arc was first discovered by the COOL-LAMPS collaboration \citep{Khullar.2021} in the Dark Energy Camera Legacy Survey (DECaLS), a component of the Dark Energy Spectroscopic Instrument Legacy Imaging Surveys \citep{2019AJ....157..168D}. We observe this galaxy at an opportune moment: at $z=5.043$, we see the Waz Arc at the receding end of the epoch of reionization, with the bulk of its stellar mass having been formed beyond $z>10$ (Khullar et al., in prep). Because this galaxy has an average magnification $\bar{\mu}\sim90$ \citep{Klein.2024}, using \JWST/NIRSpec IFS we are able to study its resolved properties at 10-100pc scales.

In this series of papers, we aim to unpack physical processes at small scales within the \waz, specifically the conditions within nebular regions (those that were forming stars in the past, or still have weak star formation), as well as the mechanisms that are responsible for stellar mass assembly and quenching within the galaxy. Deep imaging and spatially resolved spectroscopy probing $\sim3$ images of this lensed galaxy allow us to robustly create an experimental setup from two perspectives simultaneously: the stars and the nebular regions. 

In this paper (Paper I in this series), we measure and study the spatial variation on sub-kpc scales of diagnostics of nebular gas conditions, namely gas-phase metallicity, ionization parameter, nebular density, and dust attenuation. We also estimate the amount of Balmer absorption from stellar photospheres, as this must be accounted for when determining dust attenuation. We relate the physical quantities of the gas to the stellar populations that influence them, and examine how the interpretation of physical conditions would be different at lower spatial resolution, were this galaxy not lensed.

This paper is structured as follows. In Section \S\ref{sec:extant}, we briefly review the extant studies of the galaxy, placing it in context with this work.    Section \S\ref{sec:data} lays out the details of the \JWST\ observations for \waz, and analysis methods, various tools and frameworks used in this study. Section \S\ref{sec:Results} showcases the results of the various measurements, while \S\ref{sec:discuss} analyzes the inferred properties in the context of the broader field. Finally, we summarize our work in Section \S\ref{sec:conclude}. All reported magnitudes are calibrated to the AB system. The fiducial cosmology model used is a standard flat cold dark matter universe with a cosmological constant $(\Lambda$CDM), corresponding to reported numbers from the nine-year Wilkinson Microwave Anisotropy Probe (WMAP9, \citealt{2013ApJS..208...19H}).
In this work, equivalent width is reported as $W_{line}<0$ for absorption features.

\section{Extant Studies of the Waz Arc}\label{sec:extant}

\subsection{Ground-Based Observations}

The \waz\ (n\'e COOL J1241+2219) represents the brightest strongly-lensed galaxy in the $\text{z}>5$ universe, with an AB magnitude of 20.5 in the $z$ and $H$ filters \citep{Khullar.2021}. The lensed galaxy is multiply imaged ($\sim3\times$, forming the primary arc) by the foreground lensing cluster located at $z$=1.002. It has an average magnification of $\bar{\mu}=92^{+37}_{-31}$, estimated from extant shallow HST observations (ACS/WFC F814W, WFC3/IR F110W and F160W; HST-GO 16484), a source plane reconstruction of the lensed galaxy, and estimating the flux ratio between the counter-image and the lensed arc \citep[for more details, see][]{Klein.2024}. 

Combined with ground-based characterization (described in \citealt{Khullar.2021}), the HST observations use global restframe UV observations (Magellan/PISCO photometry + red spectrum from LDSS3) to measure a UV-based stellar mass (now outdated) of $\log(M_\star/M_{\odot}) = $ \textbf{$9.7\pm0.3$}\ and star formation rate ${\rm SFR_{UV}} =$ \textbf{$10.3^{+7.0}_{-4.4}$} $ M_{\odot}$ yr$^{-1}$, albeit heavily biased towards mass assembly from young stars. 

\begin{figure*}
    \centering
    \hspace*{-7mm}\includegraphics[width=\linewidth]{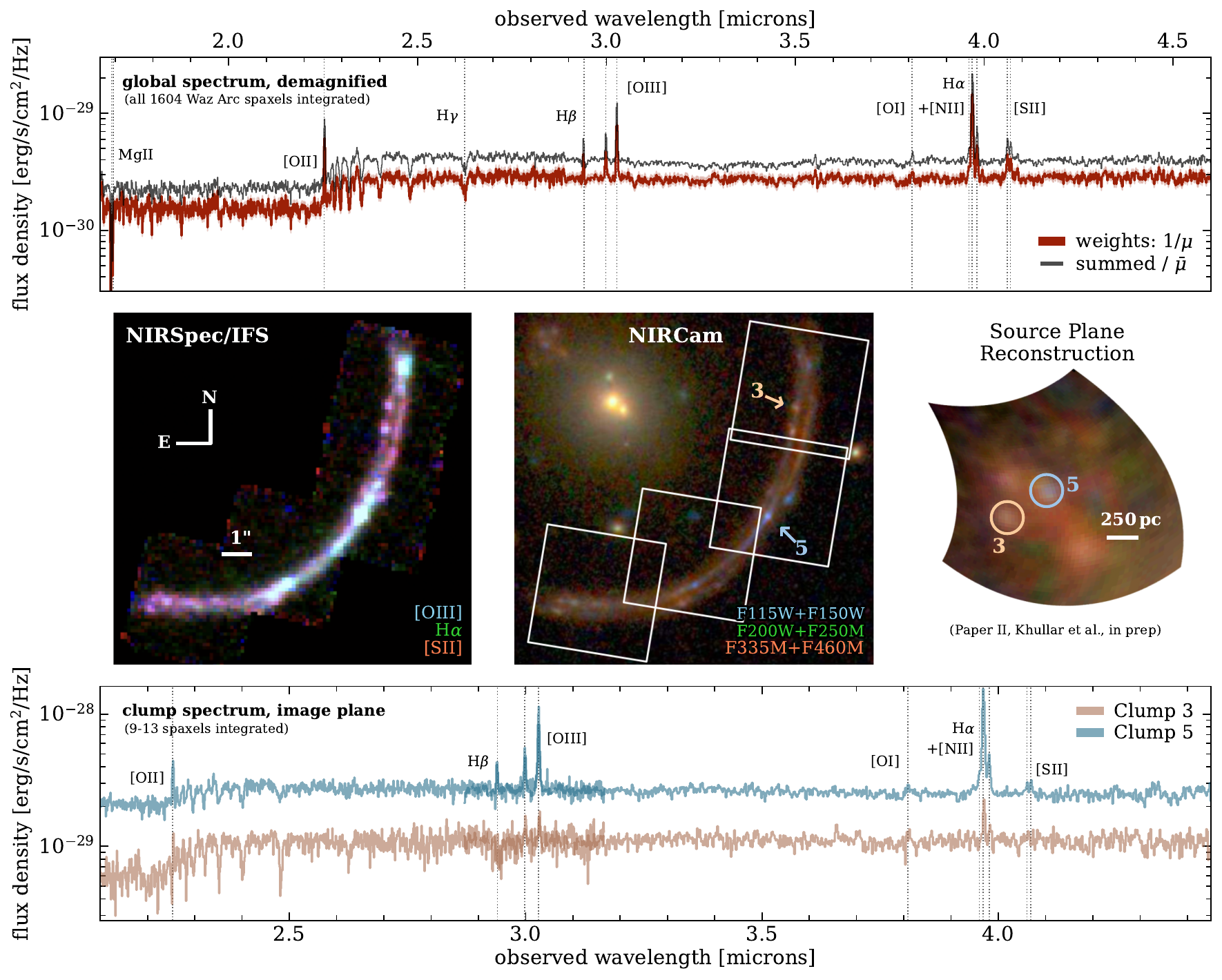}
    \caption{\textbf{Global and spatially-resolved observations from JWST GO-2566 of COOL J1241+2219 (the ``Waz Arc'', \citealt{Khullar.2021}) characterized as a post-starburst galaxy at $\mathbf{z=5}$.}
    (\textit{top}) Demagnified 1D global spectroscopy of the \waz, where every galaxy spaxel has been integrated, showing the global spectrum weighted by the magnification map (red) and the global spectrum integrated and divided by the average magnification ($\bar{\mu}\sim92$; \citealt{Klein.2024}). 
    (\textit{middle left}) A synthetic RGB reconstruction from NIRSpec IFS data, made from \sii\ 6717,6731; \ha; and \oiii\ 5008 line maps. 
    (\textit{middle center}) NIRCam RGB image with NIRSpec IFS pointings overlaid. (\textit{middle right}) The source-plane image of the \waz\ reconstructed from NIRCam photometry is in the top left corner, with scale markers showing the source-plane sizes of various features.  
    (\textit{bottom}) Image plane 1D integrated spectra for the spaxels comprising Clumps 3 and 5 (see nomenclature from Paper II), corresponding to the south east and central clump in the source plane image, respectively. 
    We highlight the diversity in colors and spectral shapes in the \waz, indicating the spatial heterogeneity of this high-z post-starburst galaxy and the promise of spatially-resolved studies.}
    \label{fig:rgb}
\end{figure*}

\begin{deluxetable*}{lcl}[!ht]
\tablecaption{\label{table:properties}Derived Galaxy Properties for the \waz}
\tablecolumns{3}
\tablehead{ \colhead{Property} & \colhead{Global} & \colhead{Reference}}
\startdata
    RA, Dec (J2000) & 190.3743, 22.3282& \cite{Khullar.2021} \\
    $\rm z_{spec}$ & 5.0453 $\pm$ 0.0003 & \cite{Khullar.2021} \\
    Magnification $\rm \mu_{avg.}$ & $92^{+37}_{-31}$ & \cite{Klein.2024} \\
    $\rm \log_{10}(M_\star/M_{\odot})$ & $10.0\pm0.4$ & Paper II \\
    $\rm log_{10}(sSFR_{Prospector} (yr^{-1}))$ & $-9.5^{+0.1}_{-0.2}$ & Paper II\\
\enddata 
\end{deluxetable*}

\subsection{Stellar Population Studies with JWST}\label{subsec:first-global-ref}

We describe the details of stellar populations within the \waz\ in Paper II (Khullar et al. in prep) using observations from \jwst\ GO-02566 (PI: Khullar), and summarize them here. Paper II uses NIRSpec/IFS spatially resolved spectra and NIRCam photometry (see \S\ref{sec:data} for details) to constrain mass assembly within \waz\ via Bayesian SED fitting, using \texttt{Prospector} (see \S\ref{subsec:Methodsstellarabs} and Paper II for specific details). We find the \waz\ exhibiting a declining SFR, and spectral signatures corresponding to a globally post-starburst galaxy -- with classic Balmer absorption and weak emission lines across the arc. The detailed constraints, long wavelength arm, and robust characterization of the emission and absorption features correct the stellar mass and SFR to $\log(M_\star/M_{\odot}) = $ \textbf{$10.1\pm0.4$}\ and a specific star formation rate $\rm log_{10}(sSFR_{Prospector} (yr^{-1}))$ = $-9.5^{+0.1}_{-0.2}$.
The star formation history (SFH) model fits, constrained via a flexible SFH model (continuity-PSB SFH model within \texttt{Prospector}, \citealt{Suess2022a, Setton2023}) show a galaxy exhibiting a rapid burst of SF \textbf{$20^{+80}_{-10}$} Myr before the epoch of observation, and forming the majority of its stellar mass at a formation redshift $z_{form}\sim6-10$.

We map critical features in the source plane of \waz\ using the combination of HST observations, lens modeling from \cite{Klein.2024} and multi-filter NIRCam imaging (see Figure \ref{fig:rgb}). The counter-image, albeit marginally magnified, offers the most accurate representation of the galaxy in the source plane. Clumps 3 and 5 (nomenclature utilized to identify unique and multiply imaged clumps, Paper II) are visibly red (quenched) and blue (star-forming) clumps within \waz, respectively; their 1D co-added spectra in Figure \ref{fig:rgb} show them as such. Morphologically, this galaxy resembles low-redshift irregular/ring-like PSBs observed from SDSS/MaNGA \citep{chen2019}. The galaxy in the source plane is estimated to be $\sim2$ kpc across, with individual regions ranging from 20-200pc resolved spatially with NIRCam imaging.

It is also critical to note that the global stellar mass in these studies was estimated using a magnification-weighted addition of all 1604 spaxels corresponding to \waz\ (Figure \ref{fig:rgb}, red spectrum in top panel), with the additional correction for multiple images of the galaxy in the primary arc ($\sim2.7\times$) -- this correction is applied to both the IFS spectra and NIRCam photometry. The correction is an approximate calculation, since each spaxel/pixel -- even if nominally multiply imaged across a triple-image lensing configuration \citep{Khullar.2021,Klein.2024} -- has a lensing magnification that is marginally different in each ``multiple image''. As the NIRCam/NIRSpec pixel scale stays the same, our effective spatial resolution for each multiply imaged clump in the source plane changes slightly. We explore the modeling of \waz\ in the source plane of the lensing system -- the most precise version of this analysis -- in future study. 

\section{Observations and Methods} \label{sec:data}

\jwst\ observations for the \waz\ were taken using NIRSpec \citep{Boker.2023} integral field spectroscopy and NIRCam \citep{Rieke.2023} imaging as part of program GO-02566 (PI: Khullar, program ID 2566). Here we describe the data and the data reduction process.  Figure \ref{fig:rgb} shows both the NIRCam and the NIRSpec/IFS views of the \waz.

\subsection{JWST/NIRSpec Integral Field Spectroscopy}
NIRSpec observations for this program were conducted in February 2023, spanning four 3\arcsec$\times$3\arcsec\ IFS pointings to cover the entire arc (Figure \ref{fig:rgb}). As is standard practice when observing faint targets, the NRSIRS2 detector readout pattern \citep{Rauscher.2017} was used to minimize detector readnoise.  The improved reference sampling and subtraction (IRS2) readout mode better accounts for correlated noise performance, generating uniform background measurements across the NIRSpec detector.\footnote{\href{https://jwst-docs.stsci.edu/jwst-near-infrared-spectrograph/nirspec-instrumentation/nirspec-detectors/nirspec-detector-readout-modes-and-patterns/nirspec-irs2-detector-readout-mode}{jwst-docs.stsci.edu/jwst-near-infrared-spectrograph/nirspec-instrumentation/nirspec-detectors/nirspec-detector-readout-modes-and-patterns/nirspec-irs2-detector-readout-mode}} For each of the 4 pointings, we employed a standard 4-point dither pattern, integrating for a total of 1.3 hours per pointing, for each of two grating/filter settings. The grating/filter combinations used were G235M/F170LP and G395M/F290LP, which provide medium spectral resolution (R$\sim$1000). Together, this pair of grating/filter combinations obtained all the strong restframe optical emission lines that are commonly-used diagnostics (e.g., \oii, \neiii, \hb, \oiii, \ha, \nii, \sii). 

We reduce the NIRSpec/IFS data as described in detail in \citet{Rigby.2023}, using a modified version of the publicly-available ERS TEMPLATES NIRSpec/IFS notebook\footnote{\href{https://github.com/JWST-Templates/Notebooks/blob/main/nirspec_ifu_cookbook.ipynb}{github.com/JWST-Templates/Notebooks/blob/main/nirspec\_ifu\_cookbook.ipynb}}. We briefly summarize the process here. The reduction uses the \jwst\ calibration pipeline (version 1.17.0) and calibration reference files from the calibration reference data system (CRDS) context 1321.pmap.  We use the NSClean software \citet{Rauscher.2024} to remove residual detector noise from each exposure.  For each grating, individual exposures from every dither and pointing are drizzled together into a mosaic with the pipeline's built-in cube building step, using the 3D drizzle algorithm of \citet{law2023}. The final pipeline product is an IFS cube where each pixel in the image axis has the scale 0\farcs1/pix, while the pixel scale in the spectral direction is defined by the grating used. We use the JWST backgrounds tool\footnote{\href{https://jwst-docs.stsci.edu/jwst-other-tools/jwst-backgrounds-tool}{jwst-docs.stsci.edu/jwst-other-tools/jwst-backgrounds-tool}} to calculate an expected background, which we subtract from each spaxel of the processed data cube. \cite{Rigby.2023} found this method of background subtraction to be robust for similar NIRSpec IFS data. 
Finally, we post-process the cubes to remove outliers missed by the \jdrp\ pipeline, by using the \texttt{baryon-sweep} algorithm\footnote{\href{https://github.com/aibhleog/baryon-sweep}{github.com/aibhleog/baryon-sweep}} described in \citet{Hutchison.2024}. 

For the grating combinations and the redshift of the target, the observations cover restframe wavelengths of $\sim$2750--6690 \ang, which includes the following bright emission lines: \oii\ \lam\lam3727,3730, \hg\ \lam4342, \hb\ \lam4864, \oiii\ \lam\lam4960,5008, \ha\ \lam6564, \nii\ \lam\lam6550,6585, and \sii\ \lam\lam6718,6733.

\subsection{JWST/NIRCam Imaging}

The reduction of NIRCam imaging and measurement practices for photometry are described in detail in Paper II (Khullar et al. in prep). We describe it briefly here. 

We obtain photometry in the NIRCam filters F115W \& F150W for the sharpest PSF and restframe UV constraints, F200W \& F250M to capture the Balmer Break implicitly, and F335M \& F460M which are strategically placed to sample non emission-line regions (in the optical regime) in a given pixel's SED within the Waz Arc. We download process Level 2A data products from MASt via an STScI-adapted custom Python script and the JWST pipeline v1.12.1. We then apply a custom de-striping algorithm to correct for 1/f noise and jumps between amplifiers \citep{cerny2025}, and propagate to Level3 mosaics. The filters are WCS-matched to Gaia as per the specifications of the default pipeline. The imaging is used in this paper to identify critical regions of scientific interest, for matching the overall shape of the SED/spectra extracted from IFS observations, and to calculate the underlying continuum and stellar absorption from SED fitting (see Paper II, Khullar et al., in prep, for more SED fitting details).

\subsection{Correction for Galactic reddening} \label{subsec:MethodsMWdust}
We correct the NIRSpec/IFS data for Galactic reddening using the Schlafly \& Finkbeiner 2011 dust map on the NASA/IPAC Infrared Science Archive\footnote{\href{https://irsa.ipac.caltech.edu/applications/DUST/}{irsa.ipac.caltech.edu/applications/DUST/}}. For the \waz\ coordinates, the reported Galactic color excess is $E(B-V) = 0.0261 \pm 0.0006$ mag, which we apply to the data assuming the Cardelli+89 dust law.

\begin{figure*}[bt]
    \centering
    \includegraphics[width=1.0\linewidth]{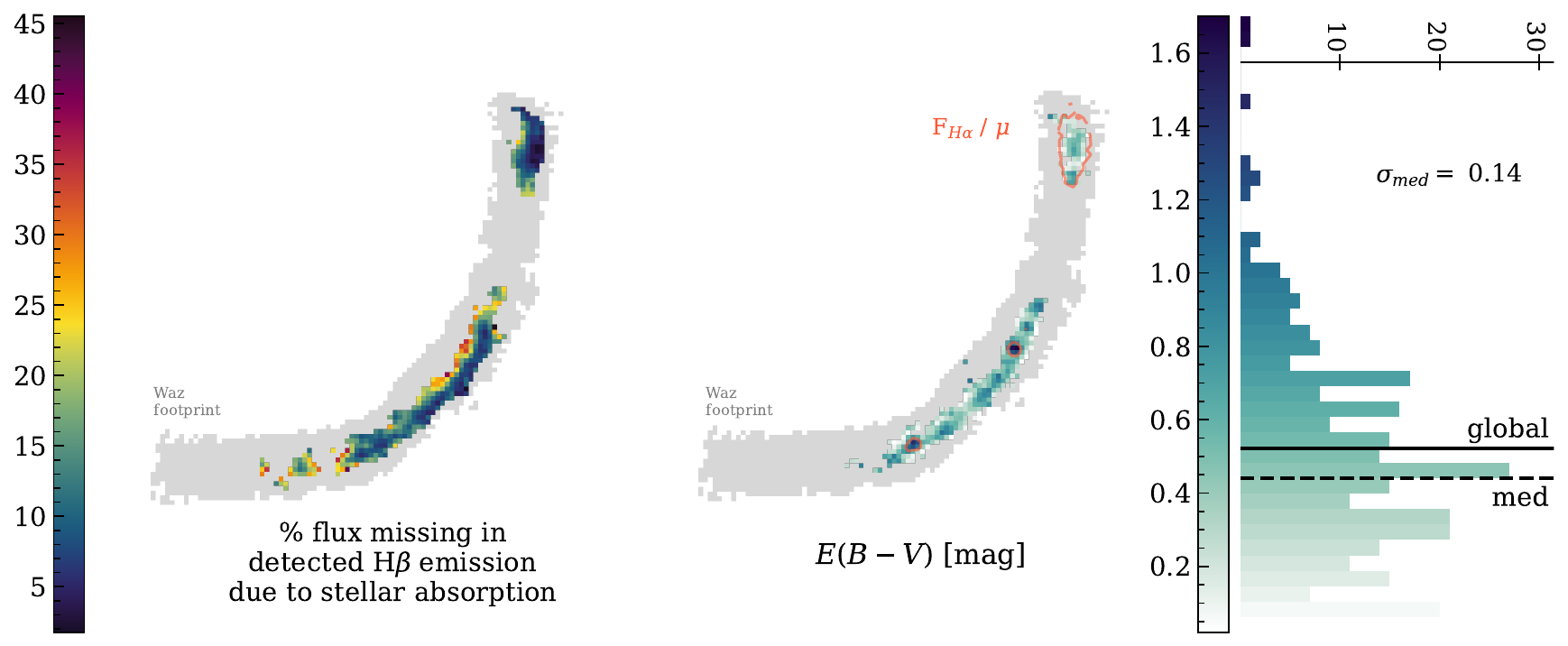}
    \caption{\textbf{Presence of stellar absorption corrected to properly estimate dust attenuation in Waz Arc.}
    (\textit{left}) Percent of \hb\ flux inferred to be missing due to infill from stellar absorption, as derived from photometry and SED modeling. (\textit{right}) Dust attenuation inferred from \ha/\hb, after correction for underlying stellar absorption. Regions where \hb\ emission was not detected are marked in light gray. The histogram attached to the right panel's colorbar shows the dispersion of spatially-resolved color excess values, with the median of the spatial values and the global value shown as dashed and solid lines, respectively. Finally, the orange contours show peaks of lensing-corrected \ha\ flux, highlighting where some of the brightest star-forming regions are located in the arc.
    The dust attenuation varies substantially across the Waz arc, preferentially peaking coincident with star-forming regions, while the global and median $E(B-V)$ are in good agreement.}
    \label{fig:dust}
\end{figure*}

\subsection{Correction for underlying stellar absorption} \label{subsec:Methodsstellarabs}
The observed Balmer spectral features are a combination of the following:  a) stellar continuum that may show Balmer absorption features, depending on the age of the stellar population; b) nebular emission, which contributes Balmer lines in emission due to recombination; and c) dust attenuation, which removes more flux from the bluer Balmer features than the redder features.  In order to use the emission line diagnostics, we must account for the effects of stellar absorption and dust attenuation.

We estimate the varying stellar absorption as follows. We first define 30 circular apertures evenly spaced along the arc.  These apertures will be used both in defining regions for photometry measurements used in SED fitting (and in the 1D spectral extraction used to measure the shape of the observed Balmer features later in this section).  We set the aperture radii to 0\farcs175, such that the apertures do not overlap but cover the arc.
We perform photometry in each aperture using all available NIRCam imaging, and fit stellar population synthesis models using the \texttt{Prospector} framework. We estimate the stellar absorption for each aperture using fixed age bin star formation history models \citep{leja2019,Khullar.2021}, assume a \cite{Chabrier2003} initial mass function, and adopt the \cite{Kriek2013} dust law with $A_v$ and dust index as free parameters. We marginalize over stellar- and gas-phase metallicity (with logarithmic uniform priors across [-2.0,0.2], the range allowed by the MIST isochrone library; \citealt{Choi2016}), and estimate the underlying absorption by removing the nebular line and continuum emission estimated through the in-built Cloudy models within \texttt{Prospector}. We use the \texttt{dynesty} dynamic nested sampling package \citep{Speagle2020} to sample the posterior distributions. 

We sample 1000 of the highest weighted models from the posterior distribution of SEDs, and estimate the stellar absorption in \hb\ and \ha\ for each aperture. We set the median and standard deviation of the models' measured absorption as the final inferred stellar absorption and associated uncertainty, respectively.  


For each spaxel in a given circular region, we assign the inferred \hb\ and \ha\ stellar absorption equivalent widths measured from the associated aperture \texttt{Prospector} result to every spaxel in the region. Each aperture has minimal variation across spaxels, both in terms of spectral flux and magnification -- we take this approximation to be robust. Taking the resulting piece-wise equivalent width correction map, we multiply every spaxel by the local observed continuum around each Balmer line to approximate the missing flux in emission (that had previously been hidden by stellar absorption). The left panel of Figure \ref{fig:dust} shows the percent of \hb\ emission that we infer is missing. The correction ranges from $1\%$ to $45\%$.   We note that a more comprehensive analysis of the stellar populations and generated nebular emission (from the spectro-photometric dataset) on common spatial scales would likely yield consistent corrections, but would improve the precision of this inference (see Paper II, also Sarmiento et al., in prep).


\subsection{Dust attenuation} \label{subsec:methodsdust}
After correcting for Balmer absorption in the underlying stellar continuum, we use the measured Balmer decrement of the nebular \ha\ and \hb\ emission lines to map the dust attenuation within the \waz. We compare the observed Balmer flux measurements to the theoretical \ha/\hb\ ratio of 2.86 (assuming Case B recombination, electron densities of 10$^2$ -- 10$^3$ cm$^{-3}$, and electron temperature of 10,000\,K;  \citealt{Osterbrock.2006}).

We calculate the $E(B-V)$ correction for each spaxel with an \hb\ and \ha\ detection using the Cardelli+89 dust law. We derive the associated uncertainties from bootstrapping the Balmer decrement in each spaxel, using a normal distribution with the observed \ha/\hb\ flux ratio and uncertainty as the mean and standard deviation.

\subsection{Continuum fitting} \label{subsec:continuum-fitting}
To map the emission lines, we must first fit and subtract the stellar continuum. We fit the continuum using the \texttt{continuum} tools provided by the TEMPLATES ERS team\footnote{\href{https://github.com/JWST-Templates/jwst_templates}{github.com/JWST-Templates/jwst\_templates}}, after masking out wavelength slices within 1000~\kms\ of each detected emission line, interpolating over the masked areas, and smoothing the continuum using a boxcar convolution with a boxcar wavelength window of 100\ang\ in the restframe. We subtract the resulting continuum fits for each spaxel to derive a continuum-subtracted cube for each grating.  

\subsection{Spatially-Resolved Emission Line Fitting} \label{subsec:line-fitting}

We create a map for each emission line by fitting the emission feature on a spaxel-by-spaxel basis in the continuum-subtracted cubes, following the methods used in \citet{Birkin.2023} and using the \texttt{curve\_fit} function in the scipy python package \citep{Virtanen.2020.SciPy}.  

We fit isolated emission lines with a single Gaussian.
For line complexes, such as doublets like \oiii\ \lam\lam4960,5008 or blended features like \nii+\ha, we model the lines together as a doublet or triplet of Gaussians with the line widths and line centroids tied to one another. 
When fitting the \oiii\ doublet, we fix the \oiii\ \lam5008 / \oiii\ \lam4960 flux ratio at the theoretical value of 2.98 \citep[e.g.,][]{Osterbrock.2006}.  When fitting the \nii+\ha\ complex, we fix the flux ratio of \nii\ \lam6585 / \lam6550  at the theoretical value of 2.8 \citep[e.g.,][]{Osterbrock.2006}.

For line doublets such as \sii, where the line ratios can vary depending upon gas density and temperature \citep{Berg.2018,Osterbrock.2006}, we allow the flux ratio to vary within the theoretical range (approximately 0.4--1.4 for \sii\ and 0.3--1.5 for \oii; e.g., \citealt{Berg.2018,Sanders.2016}).  
At the spectral resolution of these data, the \oii\ doublet is mostly unresolved.  We fit it with a single Gaussian in the spatially resolved maps, and with two Gaussians to the spatially-integrated (or ``global'') \oii\ line profile.

Table \ref{table:measurements} presents the measured line ratios, relevant to the following sections.

\subsection{Calculating star formation rates}

We calculate the instantaneous star formation rate (SFR) from the \ha\ and \oii\ emission lines, using the luminosity distance calculated from the spectroscopic redshift measured in our NIRSpec observations, and the relationships of Kennicutt+98  and Kewley+2004, respectively.  For the \oii-derived SFR, we use an emission line map smoothed with a 3x3 pixel kernel, such that every spaxel is combined with spaxels $\pm$ 1 spaxel in each direction in the spectral fitting.  We correct the SFR maps for lensing magnification using the best-fit lensing magnification model from \cite{Klein.2024}.

\subsection{Measuring gas density}\label{sec:Methodsdensity}
At the R$\sim$1000 spectral resolution of our NIRSpec observations, the \oii\ doublet is mostly unresolved, except at the highest S/N where a slight asymmetry in the blended profile is apparent.  Therefore, we focus on the fainter yet spectrally-resolved \sii\ doublet to derive the spatial variation in the density of the nebular gas \citep{Osterbrock.2006}.  For completeness, we also include the global nebular density values inferred from both the \oii\ and \sii\ doublets.  

We use the \texttt{pyneb} code \citep{Luridiana.2015} to calculate the nebular gas density from the flux ratio of the \sii\ doublet.  As this doublet is fainter than \oii\ for the \waz, we fit the \sii\ doublet using a map spatially smoothed by a kernel of 5x5 pixels (such that every spaxel is combined with spaxels $\pm$ 2 spaxels in each direction, albeit at the cost of resolution). 
We use the \sii\ doublet as a diagnostic to infer the nebular gas density is valid over line flux ratio ranges of approximately 0.4--1.4 (with an appropriate range of 0.3--1.5 for \oii), which corresponds to densities of $\sim10-10^4~\text{cm}^{-3}$ (e.g., \citealt{Kewley.2019,Berg.2018,Sanders.2016}). We propagate the uncertainties by using \sii\ line flux ratio and uncertainty as the mean and standard deviation in a normal distribution, drawing 300 new values for each spaxel measurement. The resulting standard deviation of the randomly-drawn sample is used as the uncertainty for that density measurement.  We replaced spaxels that reported density values outside of 10--10$^4$ cm$^{-3}$ with their respective upper or lower bound to more conservatively fit within the recommended range defined by \citet{Kewley.2019}.

\subsection{Ionization parameter diagnostics}\label{sec:MethodslogU}
We use the classic ionization diagnostic O32 (= \oiii\ \lam5008 / \oii\ \lam3727,3730) ratio \citep[e.g.,][and many other works]{}.  Due to the faintness of the \oii\ emission, we spatially smooth our emission line maps using a 3x3 pixel kernel. Following the equation defined in \citet{Kewley.2019}, which uses both O32 and a representative gas-phase metallicity (12 + log$_{10}$O/H), we derive \logU\ for all spaxels that have dust-corrected O32 values.  For the representative metallicity, as previously described in \S\ref{subsec:metallicity}, we use the median spatially-resolved \citet{Marino.2013}-derived \niiha\ as an independent anchor from the cyclical nature of the \citet{Kewley.2019} metallicity and ionization equations.


\begin{figure*}
    \centering
    \includegraphics[width=0.87\linewidth]{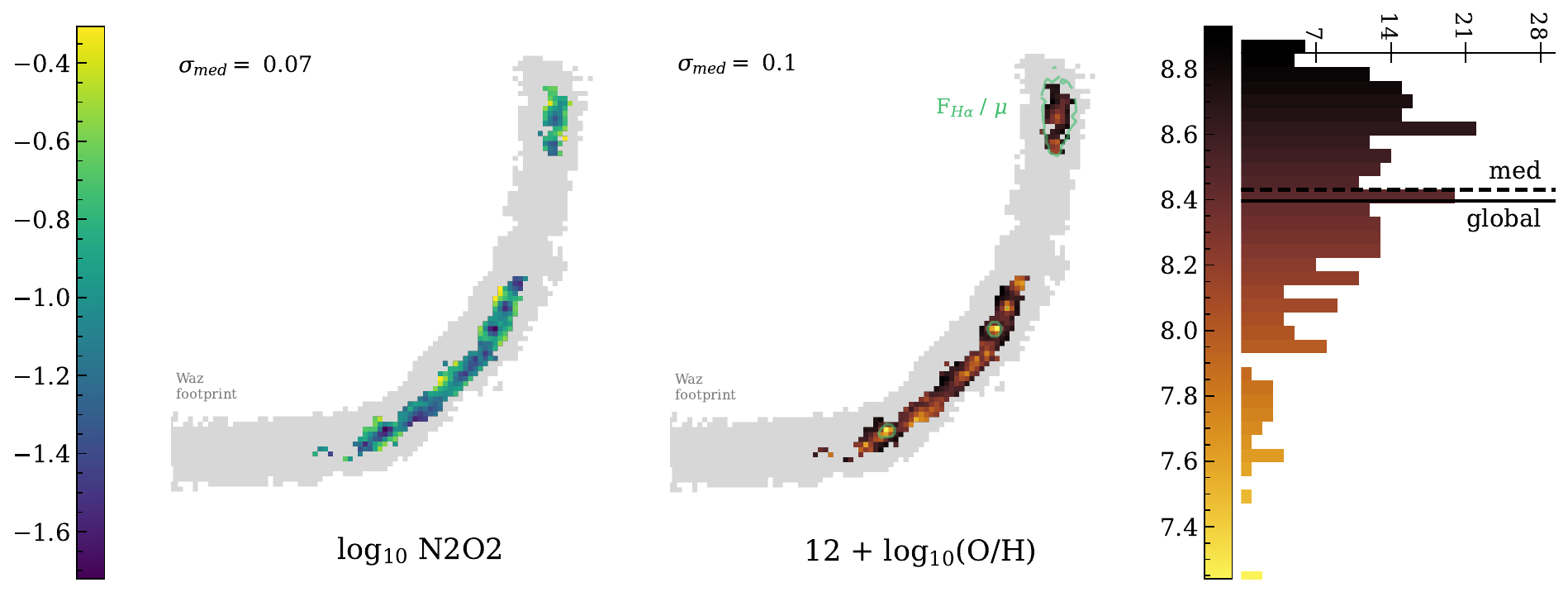}
    \includegraphics[width=0.87\linewidth]{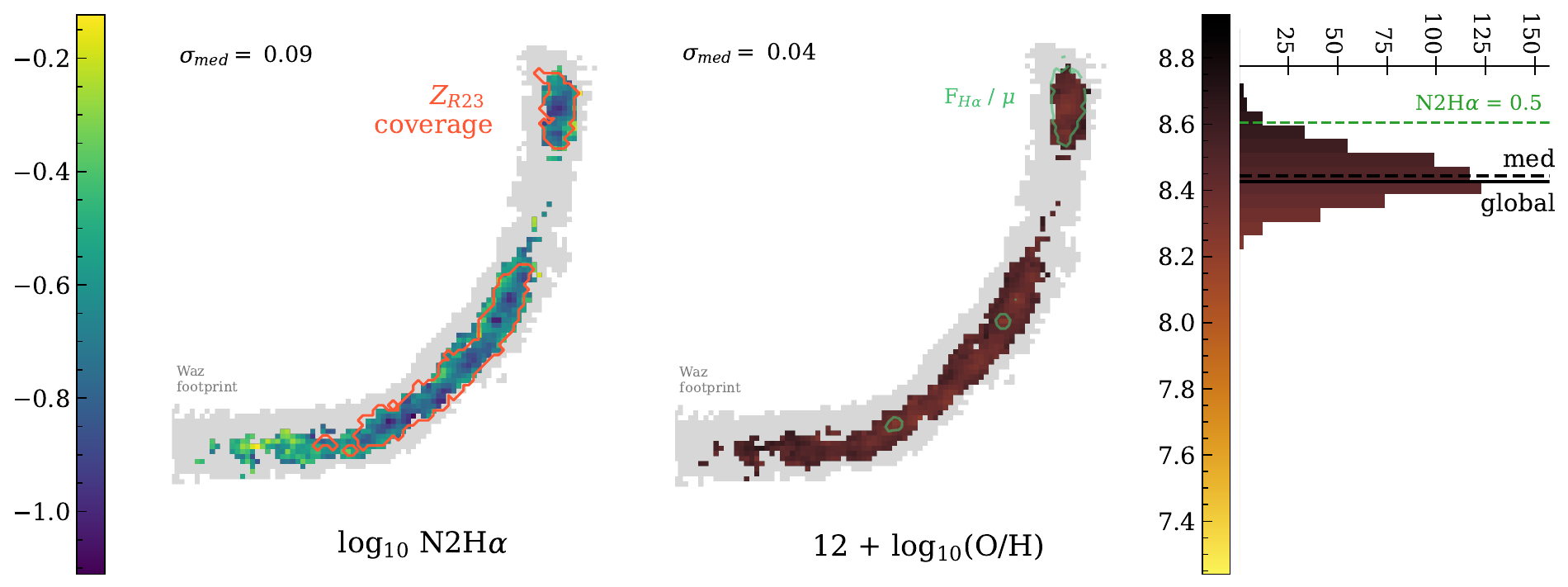}
    \includegraphics[width=0.87\linewidth]{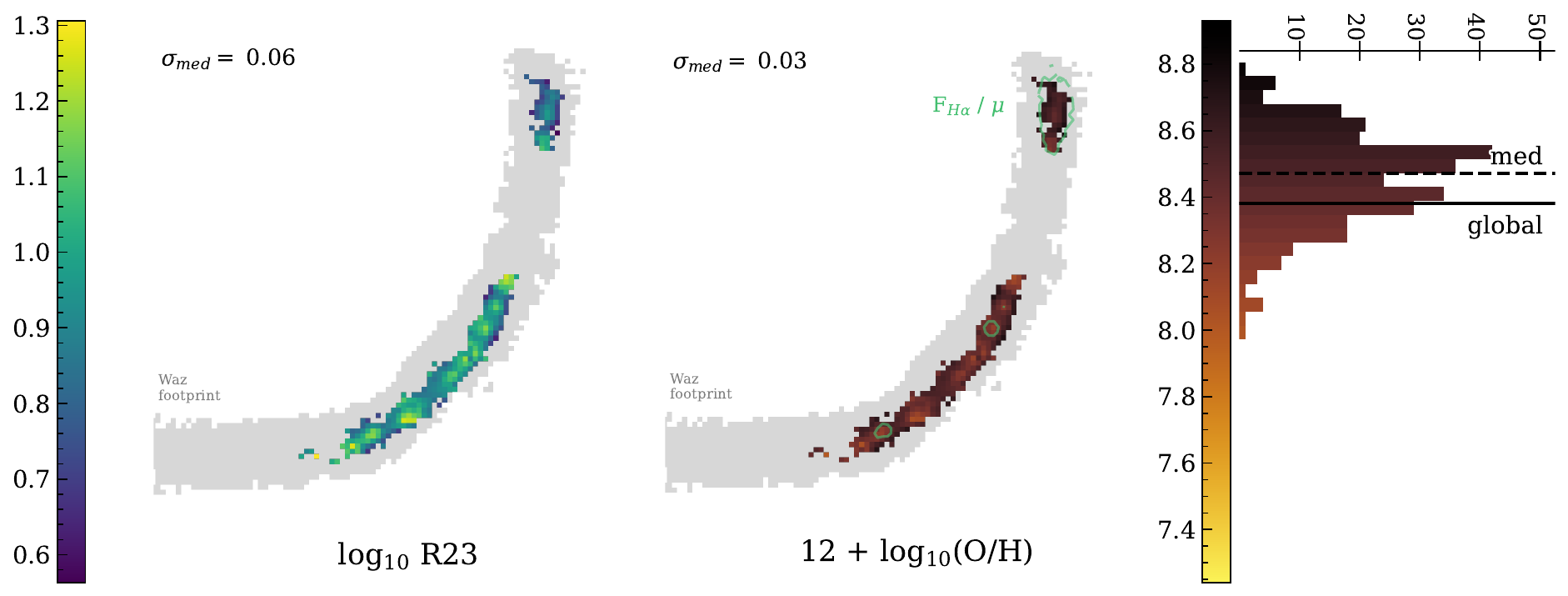}
    \caption{\textbf{Metallicity indicators suggest potential small-scale variations, yet show general global agreement.} Metallicity map using three different strong line diagnostics corrected for stellar absorption and dust attenuation, with the emission line flux ratio in the left panel on each row and the derived gas-phase metallicity in right panel.  The histogram attached to the right panel's colorbar show the dispersion of spatially-resolved values derived, as well as the median of spatial values (dashed line) and the global (solid line) values measured.  (\textit{top}) The \nii/\oii\ flux ratio and associated gas-phase metallicity derived using the equations from \cite{Kewley.2019}.  (\textit{middle}) The \nii/\ha\ flux ratio (left) and the associated gas-phase metallicity (right) derived using the equation from \citet{Marino.2013}. (\textit{bottom}) The R23 flux ratio and associated gas-phase metallicity derived using the equations from \cite{Kewley.2019}. The spaxel coverage of the R23 map is shown as an orange contour in the \nii/\ha\ panel, to highlight the difference in spatial coverage for a diagnostic that does not require dust correction.}
    \label{fig:metallicity}
\end{figure*}

\section{Results}\label{sec:Results}

In this section, we detail the various measurable physical conditions within the \waz, inferred from various strong nebular line fluxes and line ratio diagnostics.  In every diagnostic, we compare the spatially-resolved measurements to the global \waz\ spectrum, spatially integrated using all 1604 spaxels covering the galaxy (see \S\ref{subsec:first-global-ref} for more details), to serve as a reference for how the \waz\ would appear if it were not lensed (as a comparison to other high-z galaxies).

\subsection{Underlying Stellar Absorption}  

\label{subsec:Resultsstellarabs}
The depth of Balmer absorption in the stellar continuum is a sensitive diagnostic of the star formation history of the stellar population.  In Paper II we use this and other diagnostics to constrain the star formation histories of resolved and unresolved regions of the \waz.  However, for the purposes of the nebular line diagnostics discussed in this paper, Balmer absorption is a nuisance parameter.  

We find significant spatial variation in the amount of \hb\ absorption across the \waz, with some regions having no detectable \hb\ emission, and other regions having both \hb\ emission and signs of underlying stellar absorption. For all 30 apertures, we infer underlying stellar absorption with restframe equivalent widths of $-1$ to $-2$ \ang\ for \ha\ and $-5$ to $-6$ \ang\ for \hb.  As mentioned in \S\ref{subsec:Methodsstellarabs}, every spaxel is corrected based upon the inferred stellar absorption.


\subsection{Dust Attenuation} \label{subsubsec:Resultsdust}  

The right panel of Figure \ref{fig:dust} shows the derived map of dust attenuation for \waz, parameterized as $E(B-V)$, calculated as described in \S\ref{subsec:methodsdust}.  The global value measured from the spatially-integrated spectrum is indicated by the solid black line ($E(B-V) = 0.47 \pm 0.10$ mag), with the median of the spatially-resolved map represented by the black dashed line ($E(B-V) = 0.44 \pm 0.14$ mag).  While the dispersion of values covers $\sim$1 dex in magnitude, the median and global dust correction are generally consistent within errors.  We report both the median and global values in Table \ref{table:measurements}.
We note higher levels of attenuation for some of the most magnified regions, locations where bright clumps are identified in the NIRCam photometry (see Figure \ref{fig:rgb} and Paper II).
%


In certain locations along the arc, there is no detectable \hb\ emission; instead, we find either no discernible detection at all, or  \hb\ only in absorption. In these cases, we do not measure a Balmer decrement and do not estimate the dust attenuation. Therefore, these regions of the arc are not included in any of the following analyses where dust attenuation corrections are utilized and are represented by the gray shaded regions in every 2D map (which illustrates the full spaxel coverage/scale of the \waz).

\subsection{Gas-phase Metallicity, Z = 12 + log(O/H)}  
\label{subsec:metallicity}

Despite the high signal-to-noise of the NIRSpec spectra, optical auroral lines such as  \oiii\ \lam4364 emission are not detected in this galaxy, neither in the spatially resolved nor the global spectrum. It is not surprising that these ``gold standard'' temperature-sensitive metallicity diagnostics are not detected, as the \waz\ is a post-starburst galaxy (cf. Paper II) with relatively high metallicity -- a regime where it is notoriously challenging to detect auroral \oiii\ due to various factors including the intrinsic faintness of the emission line (which grows fainter as the hottest stars die), and potential contamination from Fe emission \citep[e.g.,][]{Curti.2020}. 

Instead, to infer the gas-phase metallicity of the \waz\ we use strong line diagnostic flux ratios such as R23 ($\equiv$ \oii+\oiii/\hb), N2O2 ($\equiv$ \nii/\oii), and N2H$\alpha$ ($\equiv$ \nii/\ha), following the relations described in \citet{Marino.2013} and \cite{Kewley.2019}.  Figure \ref{fig:metallicity} shows the metallicity maps inferred from each of these three metallicity diagnostics.  We report the spatial median and global values of both the line flux ratios and their derived metallicities in Table \ref{table:measurements}. 

\subsubsection{N2O2}

Many studies have observed an evolution in strong emission line ratios such as those in the traditional BPT diagram \citep[e.g.,][among other works]{Strom.2017,Kewley.2013,Kewley.2019,Curti.2020} in high-redshift galaxies. Various physical processes have been invoked to explain this evolution, including varying ionization and metallicities \citep[e.g.,][]{Feltre.2016,Steidel.2016,Strom.2017,Garg.2022}. This further emphasizes the need for strong line gas-phase metallicity diagnostics which are less sensitive to changing pressure and ionization.  One of the most consistent diagnostics is the N2O2 ratio \citep[$\equiv$ \nii/\oii; e.g.,][]{Kewley.2019}, as \nii\ brings a primary and secondary production mechanism while the \oii\ doublet is sensitive to temperature.  We use this ratio to provide an ionization- and pressure-insensitive gas-phase metallicity tracer. As an additional bonus, it is insensitive to alternate ionizing sources such as AGN or diffuse ionized gas \citep[DIG, ][]{Zhang.2017}.
The top panel of Figure \ref{fig:metallicity} shows a gas-phase metallicity map derived from the N2O2 ratio, with the median and integrated values overlaid on the histogram as described previously.

We note that this diagnostic has the largest dispersion of metallicites of any of the three diagnostics, covering almost 1.5 dex in range, even with the reduced spatial coverage driven by the dust correction map (see \S\ref{subsubsec:Resultsdust}).  As noted by \cite{Kewley.2019}, one important consideration with this diagnostic is the need for a well-defined dust correction due to the large wavelength separation of these two lines.  One could imagine slight changes in an assumed dust law could have significant effects on the perceived line flux ratio, and therefore derived metallicity.  

\subsubsection{N2H$\alpha$}\label{subsubsec:n2ha}

We derive the gas-phase metallicity from this ratio using the equation defined in \citet{Marino.2013}.  We choose this equation over others in the literature (including that in \citealt{Kewley.2019}) to remain more consistent with other works studying metal-enriched sources at higher redshift using this line diagnostic (e.g., \citealt{Birkin.2023}).  Additionally, as we describe below, the equations in \cite{Kewley.2019} require both metallicity and ionization values; therefore, we keep the N2H$\alpha$ diagnostic independent from the other strong line ratio metallicity and ionization diagnostics (that iteratively use the equations defined in \citealt{Kewley.2019}) to allow an anchor in those derivations. 
The middle panel of Figure \ref{fig:metallicity} shows a map of the \niiha\ ratio (left) and the derived gas-phase metallicity (right), with the derived metallicities shown both spatially and in a histogram of binned values. The median and global values using this diagnostic are shown in the colored histogram on the right as dashed and solid black lines, respectively.  

We find a remarkably narrow distribution of metallicities using the \niiha\ diagnostic -- implying a flat metallicity gradient in the source (unlensed) plane -- with a median value of \met\ = 8.44 $\pm$ 0.04 and a dispersion of $\sim$0.3 dex.  As the \niiha\ ratio can also be sensitive to other sources of ionization such as accreting supermassive black holes or shocks, we indicate the location of \niiha\ = 0.5 above which the source(s) powering the ratio could skew interpretation \citep[e.g.,][]{Curti.2020,Marino.2013}.  In Table \ref{table:measurements}, we list the line ratio measurements and derived gas-phase metallicity from \niiha\ where values above 0.5 are excluded or included, as well as the full global value.  However, while we keep this measurement for completeness, we note that we find no discernible difference between the two median measurements as only a small number of spaxels in the arc lie above the 0.5 threshold.

One advantage of using N2H$\alpha$ is the small separation in wavelength, removing the need to correct for dust attenuation.  This lets this diagnostic be used even in regions of the \waz\ where \hb\ is undetected and thus the dust attenuation cannot be measured.

\subsubsection{R23}

The R23 ratio ($\equiv$ \oii+\oiii/\hb) is one of the most common strong-line gas-phase metallicity tracers \citep[e.g.,][]{Pagel.1979,Kewley.2002,Tremonti.2004,Steidel.2016,Kewley.2019}.  Used as a tracer of the oxygen abundance of the nebular gas, R23 has a number of restrictions that limit its use except in specific conditions.  The ratio is known to be double-valued when compared to total oxygen abundance, with the peak of the curve hovering around 12 + log$_{10}$ $\sim$ 8.0.  Calibration studies of R23 suggest this ratio is most useful at the lowest and highest ends of the metallicity range --- or require inclusion of another line ratio to break the double-valued degeneracy --- however other line ratios are preferred if possible \citep[e.g.,][]{Kewley.2019,Sanders.2023,Sanders.2025}  We include R23 in this work for comparison to lower-z studies which focus primarily on R23 and O32 (used to trace the ionization parameter, log$_{10}$U; see \S\ref{subsec:logU}).
The bottom panel of Figure \ref{fig:metallicity} shows a gas-phase metallicity map derived from the R23 ratio.  Similar to the previous two panels, the median and integrated values using this diagnostic are overlaid as black lines in the colored histogram on the right.  The spatial coverage using this indicator is significantly less than that provided by the \niiha\ measurement, as the detectability of the \hb\ line impacted the dispersion of the R23 diagnostic (see \S\ref{subsubsec:Resultsdust}).
We note that while the dispersion of spatially-resolved metallicities shows significant variation between the three diagnostics, we find general overall agreement of the inferred gas-phase metallicities (both the global values and the median of the spatially-resolved maps) across the \waz\ within uncertainties, as shown in Table \ref{table:measurements}.  


\begin{figure*}[t!]
    \centering
    \includegraphics[width=0.98\linewidth]{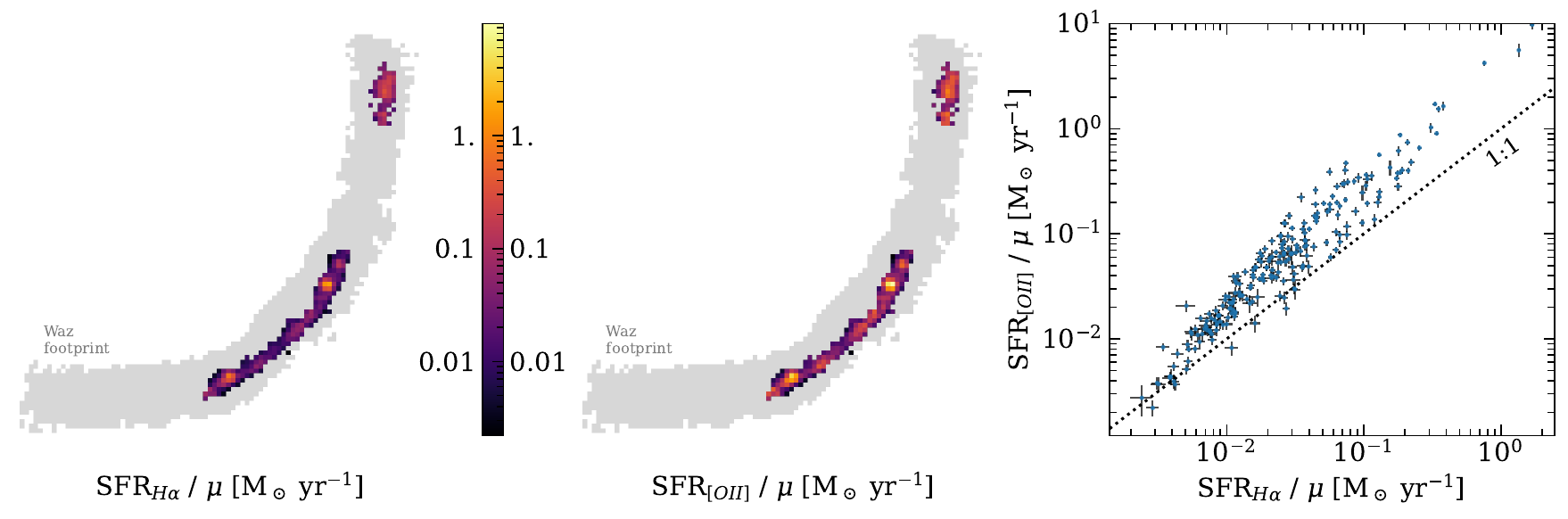}
    \caption{\textbf{Post-starburst nature of the \waz\ as seen by instantaneous SFR.} Spatially-resolved maps of dust- and lensing-corrected star formation rates, derived using (\textit{left}) \ha\ and (\textit{middle}) \oii\ emission, respectively. The brightest locations in both maps match to clumps visually identified via NIRCam photometry (see Paper II).  (\textit{right}) Both dust- and lensing-corrected SFR indicators compared against each other, with a 1:1 line underplotted.  While the SFRs for both indicators cover 1--2 dex across the Waz Arc, the majority of spaxels lie below $\sim$1 \Msolar\ yr$^{-1}$, reinforcing the post-starburst classification of this galaxy.}
    \label{fig:sfr} 
\end{figure*}

\subsection{Star Formation Rate} \label{subsec:ResultsSFR}

We estimate instantaneous star formation rates (SFRs, and the overall star formation history, SFH) of \waz\ using multiple indicators. In Paper II, we focus on stellar population synthesis models from \texttt{Prospector}'s Bayesian SED fitting framework, using nebular marginalization to infer the SFHs from the stellar continuum by avoiding physically-meaningful emission line strengths (as line ratio/strength calibrations in the early Universe are not well constrained, see Paper II).  The spatially-resolved work in this paper (Paper I) focuses on precisely the complementary approach -- what do nebular lines within \waz\ show us about SF activity (using oft-used indicators) within the galaxy?  

\begin{figure}[t]
    \centering
    \includegraphics[width=\linewidth,trim={4.6cm 7mm 5.4cm 0},clip]{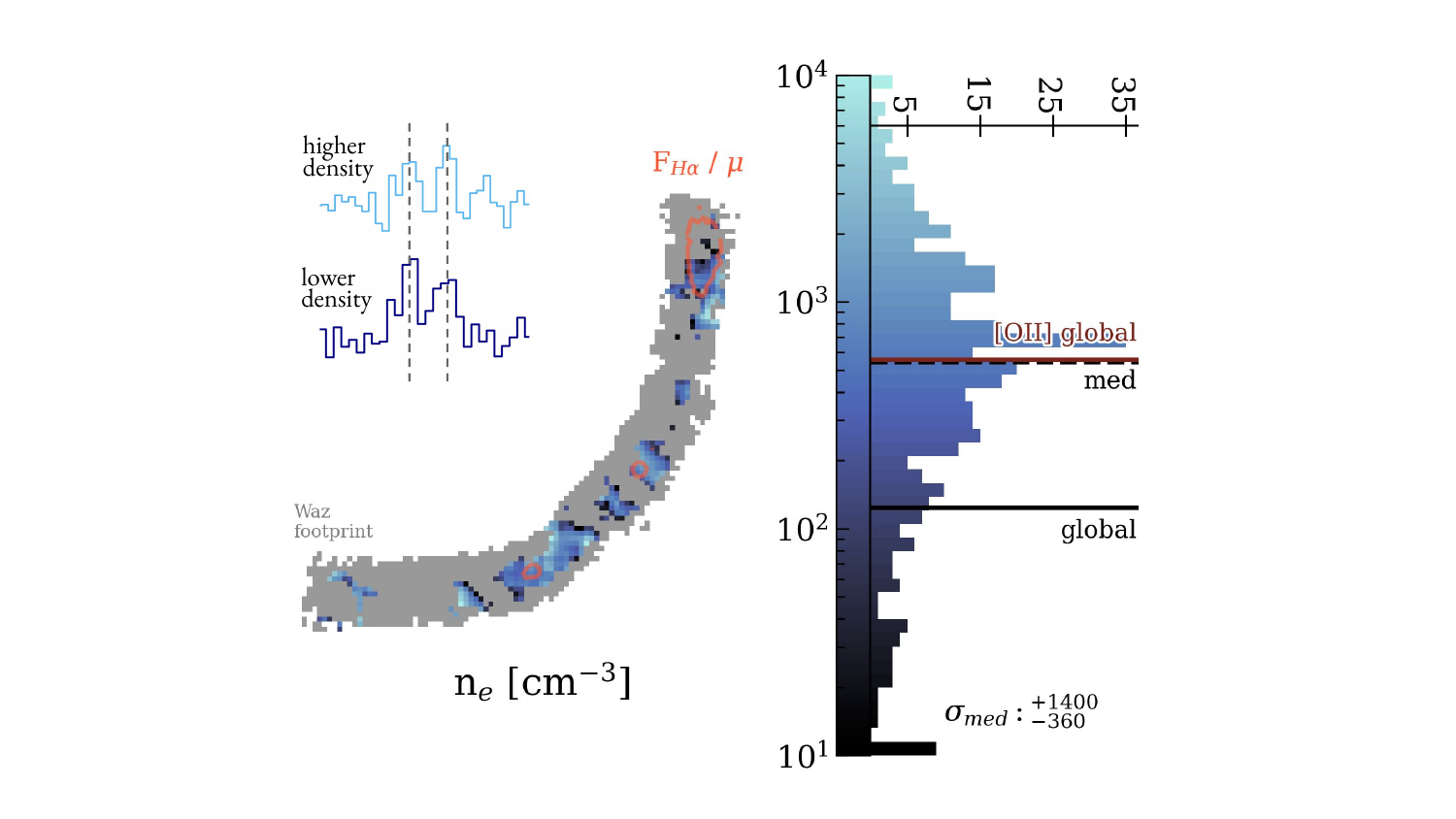}
    \caption{\textbf{Hints of higher nebular densities.}
    The spatially-resolved nebular density of the \waz\ using the doublet ratio \sii\lam6717/6731, where detectable above the noise.  Two spaxel examples of the doublet line strengths are shown in the top left.  The histogram on the right of the colorbar is the range of density values, with the median of spatial values and the global value marked as dashed and solid lines, respectively.  Additionally, the global density value derived from the mostly-unresolved \oii\ emission is marked as the red line.  Finally, the orange contours show peaks of lensing-corrected \ha\ flux, highlighting where the some of the brightest star-forming regions are located in the arc. \label{fig:density}}
\end{figure}

We map the instantaneous SFR of the \waz\ using dust-corrected \ha\ and \oii\ emission.  Both lines are frequently used to trace recent star formation, as they are powered by stars formed within the past $\lesssim$ 10 Myr \citep[such that massive stars are still present to produce ionizing photons; e.g.,][]{Calzetti.2013,Kennicutt.2012,Kewley.2004,Rosa-Gonzalez.2002,Kennicutt.1998}.  
Figure \ref{fig:sfr} shows the SFR maps derived from \ha\ (top) and \oii\ (bottom) emission.


We find general agreement between the two star formation rate tracers \ha\ and \oii, however we note an overall lower dispersion of SFR using \ha.  One possible reason for this is due to the corrections needed for the two lines.  Generally, after correcting \ha\ emission for any underlying stellar absorption, once a dust attenuation correction is applied, the SFR derived from both \ha\ and \oii\ should, in principle, be the same \citep[e.g.,][]{Kewley.2004}.  However, \oii\ emission is also sensitive to variations in the amount of diffuse gas present, with the emission found to be strong in diffuse ionized gas (DIG) in starburst galaxies \citep[e.g.,][we discuss the possibility of DIG in \S\ref{subsec:dig}]{Kennicutt.1998,Hunter1990}. It is possible that the enhancement we see in the SFR(\oii) map may be influenced by this effect.
Separately, as previously mentioned, the only other strong effect would be related to the extinction correction of the two emission lines: one could envision that if an incorrect dust attenuation law is used, the resulting discrepancies would propagate across wavelengths.  Further analysis of such implications are explored in Sarmiento et al.\ (in prep).

\begin{figure*}
    \centering
    \includegraphics[width=1.0\linewidth]{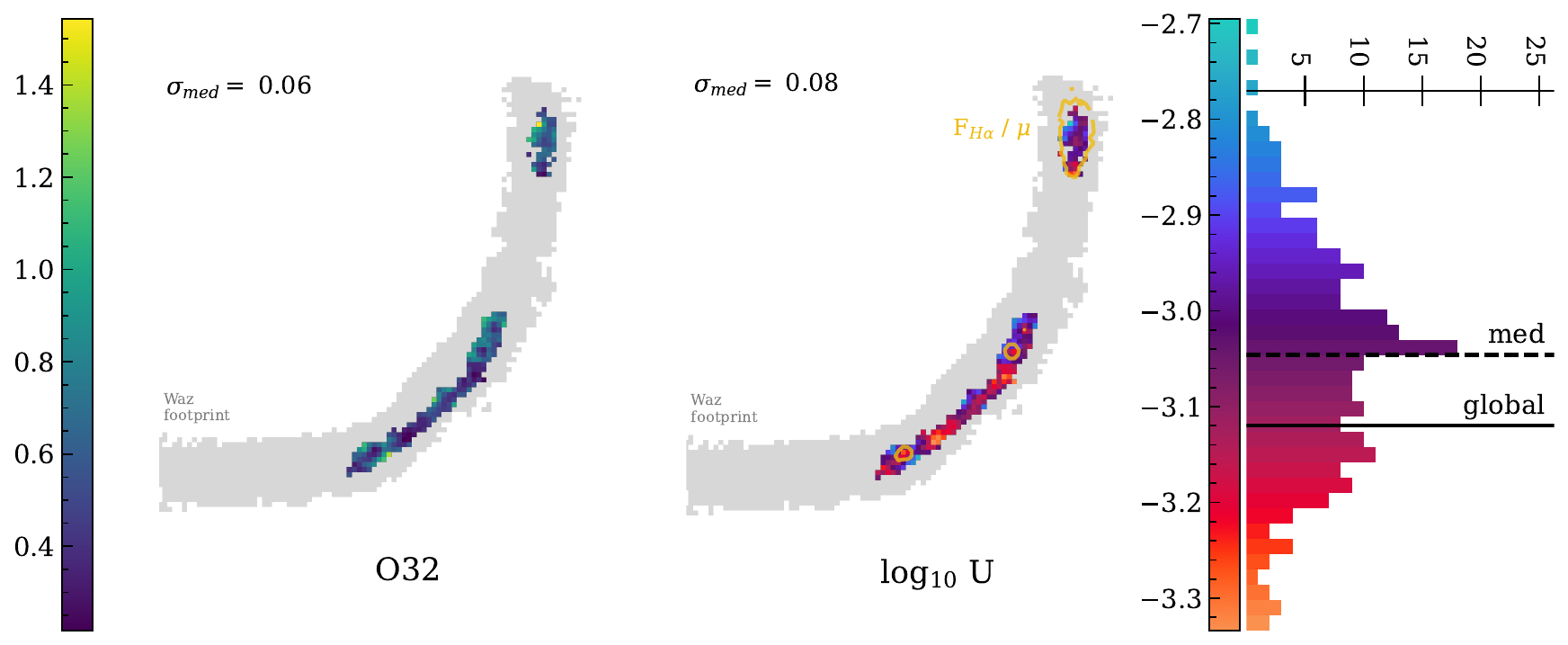}
    \caption{\textbf{Ionization parameter changing across the \waz.}
    Ionization map using the O32 strong line diagnostic, corrected for stellar absorption and dust attenuation.  As with previous figures, the emission line flux ratio is shown in the left panel and the derived ionization parameter (log$_{10}$U) in the right panel. The histogram attached to the right panel's colorbar shows the dispersion of spatially-resolved values derived, as well as the median of spatial values (dashed line) and the global (solid line) values measured. Similar to previous figures, the gold contours overlaid on the right panel indicate higher lensing-corrected \ha\ regions in the arc.}
    \label{fig:ionization}\end{figure*}

\subsection{Nebular gas density}\label{subsec:results-density}

Figure \ref{fig:density} shows the nebular gas density map using the \sii\ ratio, with the coral contour highlighting regions of higher lensing-corrected \ha\ emission (the same contour regions are highlighted in green in other figures).  In the histogram on the right, we note the global and median values for the \sii-derived densities, as well as the global \oii-derived density value.  
While there are not many spaxels in the field of view that make our S/N cuts, we see a distribution of density values covering over 2 orders of magnitude. The median density derived from the spatial map of \sii\ is $540^{+1400}_{-360}$ cm$^{-3}$, which is offset from the global \sii\ density of $120^{+90}_{-90}$ cm$^{-3}$ but generally within uncertainties.  While the overlap of the median \sii-derived density and the global \oii-derived density values is curious, they are consistent within 1$\sigma$ based upon their respective uncertainties -- suggesting this is likely more coincidence than driven by any physical processes.
Both these measurements and the global \oii\ density are shown in Table \ref{table:measurements}.

\begin{figure}
    \centering
    \includegraphics[width=\linewidth]{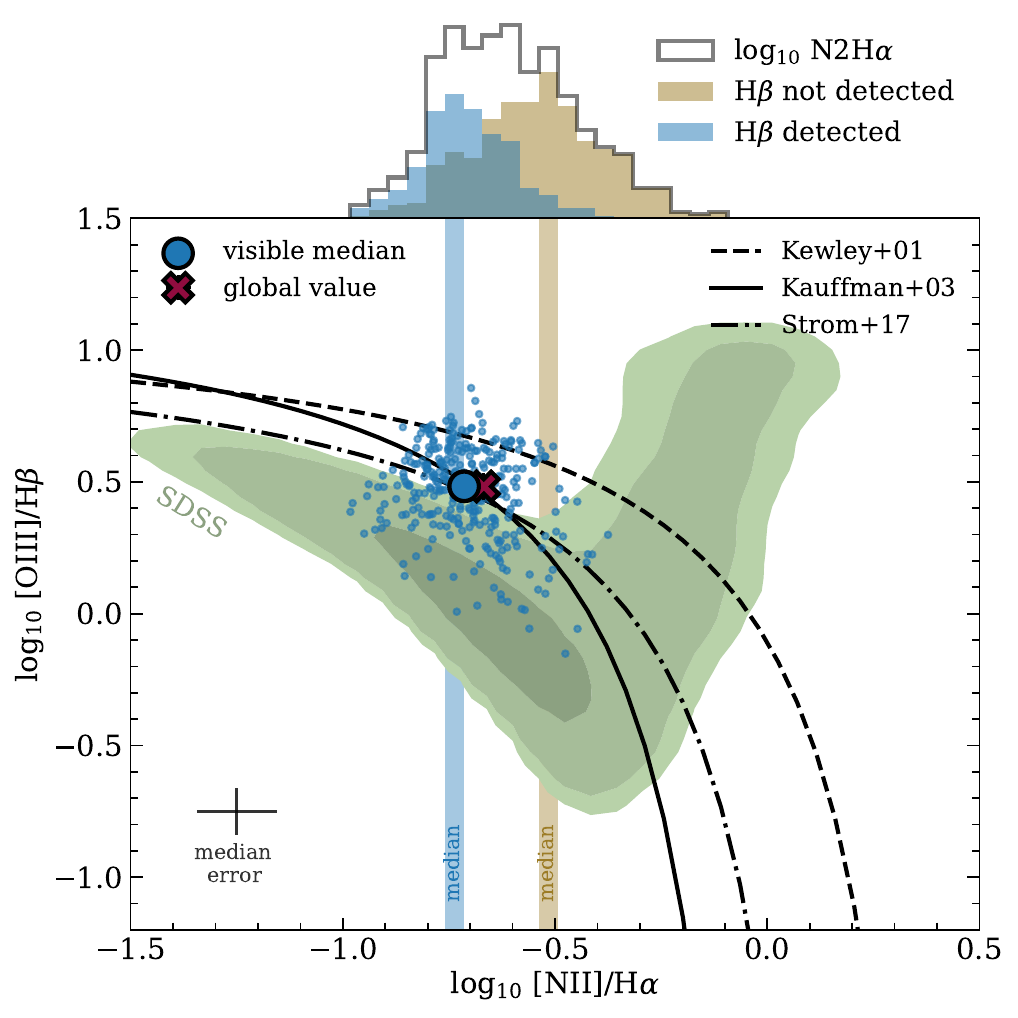}
    \caption{\textbf{Missing H$\mathbf{\beta}$ in context with BPT.}
    The \nii\ BPT diagram, with flux ratios of \oiii/\hb\ versus \nii/\ha.  The sloped black lines denote the ``maximum starburst'' threshold (dashed; \citealt{Kewley.2001}), the local SF-AGN theoretical line (solid; \citealt{Kauffmann.2003}), and the $z\sim2-3$ SF-AGN observational threshold (dot-dash; \citealt{Strom.2017}), respectively.  The values of individual spaxels are shown as blue circles.  Only spaxels where \hb\ was detected are shown in the scatter plot. The larger blue circle indicated the median value of these spaxels, while the red cross shows the global value for the \waz. On top of the panel is a histogram showing the total distribution of \nii/\ha\ spaxels, highlighting those that have \hb\ detected (blue, matching the scatter points) and those that do not have \hb\ detected (tan).
    }
    \label{fig:bpt}
\end{figure}

\subsection{Ionization Parameter}\label{subsec:logU}


The ionization parameter is a dimensionless quantity defined as the ratio of the number density of ionizing photons to the number density of gas particles: U $\equiv n_\gamma$/$n_H$.  The logarithm of the quantity (\logU) is commonly quoted in studies.
\logU\ is a powerful tool, used to describe the intensity of the ionizing continuum in a galaxy. There have been numerous strong (and faint) line diagnostics explored both in observations and in simulations to accurately probe this parameter --- while accounting for how ionization-sensitive line ratios compare to changes in electron temperature, density, elemental abundances, and other key physics --- both pre-\jwst\ \citep[e.g.,][]{Hainline.2009,Bian.2010,Bian.2016,Siana.2010,Kewley.2013,Madau.2014,Shapley.2015,Shapley.2019,Shivaei.2015,Shivaei.2018,Sanders.2015,Sanders.2016,Sanders.2018,Sanders.2020,Steidel.2014,Steidel.2016,Strom.2017,Strom.2018,Strom.2022,Suzuki.2017,Kaasinen.2017,Kashino.2017,Gburek.2019,Maiolino.2019,Simons.2021,Papovich.2022,Backhaus.2022} and since \jwst\ began observations \citep[e.g.,][]{Isobe.2023,Sanders.2023,Sanders.2024,Abdurrouf.2024,Backhaus.2024,Cameron.2023,Christensen.2023,Curti.2023,Fujimoto.2023,Heintz.2023,Hsiao.2024a,Hsiao.2024b,Nakajima.2025}.

Figure \ref{fig:ionization} shows the O32 ratio (left) and the resulting \logU\ map (right), with the global and median spatially-resolved values denoted in black following previous figures.  Similar to the other diagnostic maps, the gold contours indicate higher lensing-corrected \ha\ regions in the arc.  Overall, while we see a significant dispersion of \logU\ across 0.5 dex, we note that there may possibly be even more diversity in the regions along the \waz\ that do not have \hb\ detections -- such regions were not dust-corrected, and therefore not included in this derivation -- but deeper data is needed to better constrain the stellar continuum and possible presence of weak \hb\ in those regions.  In the more \ha-bright regions, such as those denoted by the gold contours, we see evidence of a \textit{decrease} in ionization. This is intriguing as one may expect to see a correlation between higher ionization and star formation. However, as the difference between the cores of these clumps of \ha\ emission and their immediate surroundings is generally no more than $\sim0.2$ dex,  this does not appear too dramatic a difference.  We discuss the possible cause of these dips in ionization and their implications in more detail in \S\ref{subsec:sfr-spatial}.


\subsection{The BPT diagram}

A common diagnostic used to separate the dominant source of ionization in a galaxy, the \nii\ BPT diagram (\nii/\ha\ vs \oiii/\hb; \citealt{Baldwin.1981}) has been frequently used due to the ready availability of the chosen emission lines at lower redshift ($z<2$; see contextual discussion in \citealt{Steidel.2016}, among countless other works). This diagnostic, a comparison of forbidden lines against nearby hydrogen lines, leverages the lack of need for dust attenuation corrections. With theoretical lines demarcating regions of the parameter space into star-forming, composite, accreting black holes, etc., the \nii\ BPT is generally effective for low-z studies while often complicated to decipher when invoked in higher-z studies \citep{Cleri.2025}.  We employ this diagnostic diagram in this work due to the spatially-resolved nature of the \waz\ (in effect, looking at individual ``HII-dominated'' regions), as well as a tool to investigate the lack of detectable \hb\ emission in some regions of the arc (see \S\ref{subsec:sfr-spatial} and \S\ref{subsec:dig} for further discussion on this point).


The top panel of Figure \ref{fig:bpt} shows the \nii\ BPT diagram, with the green shaded regions representing SDSS galaxies and the black lines separating the star-forming and accreting black hole relative locations on the diagram (based upon theoretical and observational samples at $z\sim0$, \citealt{Kewley.2001,Kauffmann.2003}, and $z\sim2-3$, \citealt{Strom.2017}).
Leveraging the spatial resolution provided by NIRSpec/IFS, we show the location of individual spaxels on the BPT diagram. The median value of both line ratios (from the entire sample of spaxels, which differs in coverage between the two axes for reasons we discuss below) is shown as blue dotted lines, and the median of the spaxels plotted is shown as a larger blue circle.  Finally, we include the global (spatially-integrated) value for the \waz\ as the red cross.  The spaxels hover around the edge of the ``star-forming'' region of the diagram, with the median error shown as the black plus on the bottom left corner.

The full coverage of \niiha\ spaxels are not shown in the BPT diagram, as there are regions in the \waz\ where \hb\ was not detected (see \S\ref{subsubsec:Resultsdust}).  We show the spaxels that have detected \niiha\ but no \hb\ in the histogram of \niiha\ values on above the top panel of Figure \ref{fig:bpt}. The full span of \niiha\ values covers $\sim$0.8 dex, however we find a bimodal distribution when separating the \niiha\ spaxels into those with \hb\ detected (and therefore an \oiii/\hb\ value, blue points) and those where \hb\ is not detected (gold distribution).  The spaxels where \hb\ is not detected skew to larger \niiha\ values, while those with \hb\ detected are located on the lower end of the measured \niiha\ distribution.  

We discuss the implications from this result in \S\ref{subsec:sfr-spatial} and \S\ref{subsec:dig}.

\begin{deluxetable*}{lcclcc}[!ht]
\tablecaption{\label{table:measurements}Spectroscopically Measured \& Derived Physical Conditions of the \waz}
\tablecolumns{6}
\tablehead{ \colhead{Measurement} & \colhead{Median} & \colhead{Global} & \colhead{Derived Property} & \colhead{Median} & \colhead{Global} }
\startdata
    \niiha\ & 0.24 $\pm$ 0.05 & 0.215 $\pm$ 0.008 &
        12 $+$ log(O/H) (N2\ha) & 8.45 $\pm$ 0.04 & 8.4272 $\pm$ 0.008 \\
    \niiha\ ($<0.5$)\tablenotemark{a} & 0.23 $\pm$ 0.05 & ------ &
        12 $+$ log(O/H) (N2\ha$<$0.5) & 8.44 $\pm$ 0.04 & ------ \\
    R23 & 8.5 $\pm$ 1.3 & 10.2 $\pm$ 1.1 &
        12 $+$ log(O/H) (R23)\tablenotemark{b} & 8.47 $\pm$ 0.03 & 8.38 $\pm$ 0.02 \\ 
    N2O2 & 0.09 $\pm$ 0.02 & 0.092 $\pm$ 0.006 &
        12 $+$ log(O/H) (N2O2)\tablenotemark{b} & 8.43 $\pm$ 0.10 & 8.40 $\pm$ 0.02 \\ 
    \\[-1mm] 
    %
    $L_{\ha} / \mu$ [erg s$^{-1}$] & (4.5 $\pm$ 0.2)\,$\times10^{39}$ & [$\bar{\mu}$] (8.05 $\pm$ 0.08)\,$\times10^{39}$ &
        SFR (\ha$/\mu$) [\Msolar/yr] & 0.024 $\pm$ 0.001 & [$\bar{\mu}$]  0.0443 $\pm$ 0.0005 \\
    $L_{\ha} / \bar{\mu}_L$ [erg s$^{-1}$] & ------ & (6.07 $\pm$ 0.06)\,$\times10^{39}$ &
        SFR (\ha$/\bar{\mu}_L$) [\Msolar/yr] & ------ & 0.0657 $\pm$ 0.0007 \\
    $L_{\ha} / \bar{\mu}_H$ [erg s$^{-1}$] & ------ & (1.2 $\pm$ 0.1)\,$\times10^{40}$ &
        SFR (\ha$/\bar{\mu}_H$) [\Msolar/yr] & ------ & 0.0334 $\pm$ 0.0003 \\ 
    \\[-2mm]
    $L_{\oii} / \mu$ [erg s$^{-1}$] & (8.2 $\pm$ 0.6)\,$\times10^{39}$ & [$\bar{\mu}$] (5.8 $\pm$ 0.3)\,$\times10^{40}$ &
        SFR (\oii$/\mu$) [\Msolar/yr] & 0.053 $\pm$ 0.004 & [$\bar{\mu}$] 0.38 $\pm$ 0.02 \\
    $L_{\oii} / \bar{\mu}_L$ [erg s$^{-1}$] & ------ & (8.6 $\pm$ 0.4)\,$\times10^{40}$ &
        SFR (\oii$/\bar{\mu}_L$) [\Msolar/yr] & ------ & 0.56 $\pm$ 0.03 \\
    $L_{\oii} / \bar{\mu}_H$ [erg s$^{-1}$] & ------ & (4.3 $\pm$ 0.2)\,$\times10^{40}$ &
        SFR (\oii$/\bar{\mu}_H$) [\Msolar/yr] & ------ & 0.28 $\pm$ 0.01 \\
    \\[-1mm] 
    %
    R$_{\sii}$ & 1.16 $\pm$ 0.39 & 1.33 $\pm$ 0.14 &
        $n_e$ (\sii) [cm$^{-3}$] & 540$^{+1400}_{-360}$ & 120$^{+90}_{-90}$  \\
    R$_{\oii}$ & ------ & 0.9 $\pm$ 0.4 &
        $n_e$ (\oii) [cm$^{-3}$] & ------ & 560$^{+790}_{-360}$ \\
    \\[-1mm] 
    %
    O32 & 0.57 $\pm$ 0.08 & 0.46 $\pm$ 0.04 & 
        log\,U (O32)\tablenotemark{c} & $-$3.05 $\pm$ 0.08 & $-$3.12 $\pm$ 0.02 \\
    \\[-1mm] 
    %
    %
    \ha/\hb\ & 3.93 $\pm$ 0.64 & 4.57 $\pm$ 0.48 & 
        $E(B-V)$ [mag] & 0.44 $\pm$ 0.14 & 0.47 +/- 0.10
\enddata 
\vspace{2mm}
\tablecomments{The \ha\ and \oii\ luminosities and their respective SFRs have been corrected for lensing magnification.  For the median values, the spatially-resolved maps themselves were corrected for lensing magnification prior to measurement (see also \S\ref{subsec:ResultsSFR}).  For the global values, we apply the average magnification from \citet{Klein.2024}: $\bar{\mu} = 92^{+37 (H)}_{-31 (L)}$ (see also Table \ref{table:properties}). All other values in this table are lensing insensitive, as they are line flux ratios.}
\tablenotetext{a}{For \niiha\ ($<0.5$), we masked out any spaxels with \niiha\ above 0.5 and then calculated the median of the remaining spaxels (see \S\ref{subsubsec:n2ha} for more details).}
\tablenotetext{b}{Derived from \citet{Kewley.2019} using median O32 ionization parameter.}
\tablenotetext{c}{Derived from \citet{Kewley.2019} using median \niiha\ ($<0.5$) gas-phase metallicity.} 
\end{deluxetable*}

\section{Discussion} \label{sec:discuss}

Before \jwst{}, the general consensus in the literature was that galaxies in the early Universe housed relatively young ($<100$ Myr), actively star-forming stellar populations that provided intense radiation fields for very low metallicity nebular gas (for reviews of the pre-\jwst\ census of galaxies in the early Universe, see \citealt{Stark.2016,Finkelstein.2016}; for pre-\jwst\ expectations of the discoveries to come from \jwst, see \citealt{Robertson.2022}). Such galaxies were known for their strong nebular emission lines and very negative UV continuum slopes ($\beta\lesssim-2$; \citealt{Dunlop.2012,Larson.2023}).  The star formation histories (SFHs) of these early sources were inferred to be bursty, continually rising with time to assemble the star forming main sequence at lower redshifts \citep{Leja2022,ciesla2024}. These assumptions matched well with observational evidence of the universal star formation rate surface density as a function of time, rising from the earliest times to peak around cosmic noon ($z\sim2$) and exponentially declining at later times \citep{Madau.2014}.  This is not to say that there were not sources at higher redshifts beginning to cease or decline in star formation. Indeed, before \jwst\ there were already spectroscopic detections of quenching and/or quiescent galaxies at redshifts as high as $z\sim3-4$ \citep[e.g.,][among others]{Glazebrook.2017,dugenio2021,Antwi-Danso.2023}. However, such sources were rare and difficult to observe.


\jwst\ has revolutionized what we understand of the history of star formation in galaxies in the early Universe ($z>5$), discovering $z>5$ galaxies housing weak nebular emission that suggest actively declining SFHs, or those which have recently had a burst of star formation and then quickly lost their fuel \citep{Strait2023,carnall2024,Weibel.2024,Setton.2024,Looser.2024,covelo2025}.
%
%
Despite these advances, two challenges remain -- (a) high-redshift quenched galaxies are hard to spatially resolve, even with \jwst, and therefore only provide access to mostly globally-averaged properties, and (b) the properties of low-redshift PSBs and quenched systems are challenging to extrapolate to higher redshifts, given the limitations of scaling relations and calibrations in the methodologies used to measure these properties. Despite these challenges, we aim here to describe the current status of some of these studies, to bring into context the novelty of the \waz\ and its physical conditions.

\subsection{Contextualizing Star Formation (or lack thereof) in a Post-Starburst Galaxy at z=5 \\ with Existing Studies} \label{subsec:sfr-spatial}

Studies of post-starburst galaxies at lower redshifts (by definition) indicate that, at their early epochs, star formation is ongoing in the early epochs of their star formation histories, both globally and locally \citep{chen2019,french2021}, and at the epoch of observation, such processes have been either suppressed or have ceased altogether. 

\cite{french2021} investigate recent SFHs of PSBs (which they label as ``E+A'' systems based upon historical classification) to find evidence for multiple bursts in low-redshift systems,  and characteristic molecular gas depletion times corresponding to AGN feedback. Spatially resolved studies via SDSS/MaNGA have taken a step further to identify particular regions of star formation and recent quenching within relatively low-redshift galaxies \citep{chen2019,cheng2024,Leung2025}. \cite{chen2019} identify for the first time new classes of PSBs -- ring-like and irregular PSBs -- with clear evidence of patchy quenching with a central starburst, potentially caused by gas fueling disruptions or mergers. On studying these systems globally, their average mass-weighted ages are on the younger side. \cite{Nielsen2025} also find three distinct classes of PSBs, albeit using a UMAP (Uniform Manifold Approximation and Projection) and clustering approach training on stellar population signatures in PSB spectra. The \waz, from both morphological and stellar populations perspective, is situated in the ring-like/irregular PSB category -- with weak emission lines and Balmer absorption features, with B star populations quite likely still present. In the parameter space defined by \cite{Nielsen2025}, the \waz\ is a ``Group 1'' galaxy, with younger stellar population ages ($<200-300$ Myr) and evolution due to a merger-driven scenario. The most star-forming region within \waz, close to the center, hints towards a rejuvenating phase -- we explore this further in Paper II.  

While gravitationally lensed PSBs are rare in literature and almost all at lower redshifts \citep{Fassnacht,Shin2000}, studies like \cite{Cabanac2005} find (globally) fading stellar populations post-burst in a $z=3.78$ lensed PSB. One of the most comprehensive studies of the physical conditions within a fast rotator PSB at $z=3$ was conducted by \cite{dugenio2024}, where weak emission line characterization shows evidence for quenching due to supermassive black hole feedback. Compared to the \waz\ where quenching occured 20-100 Myr ago, this system (dubbed GS-10578) quenched 100-300 Myr ago, and shows evidence for ionized and neutral gas outflows, which are challenging to characterize in a lensed system with the spectral resolution our study possesses. Despite this, there are several hypotheses regarding the physical conditions of the \waz\ that we have strong evidence for.

\begin{figure*}
    \centering
    \includegraphics[width=\linewidth]{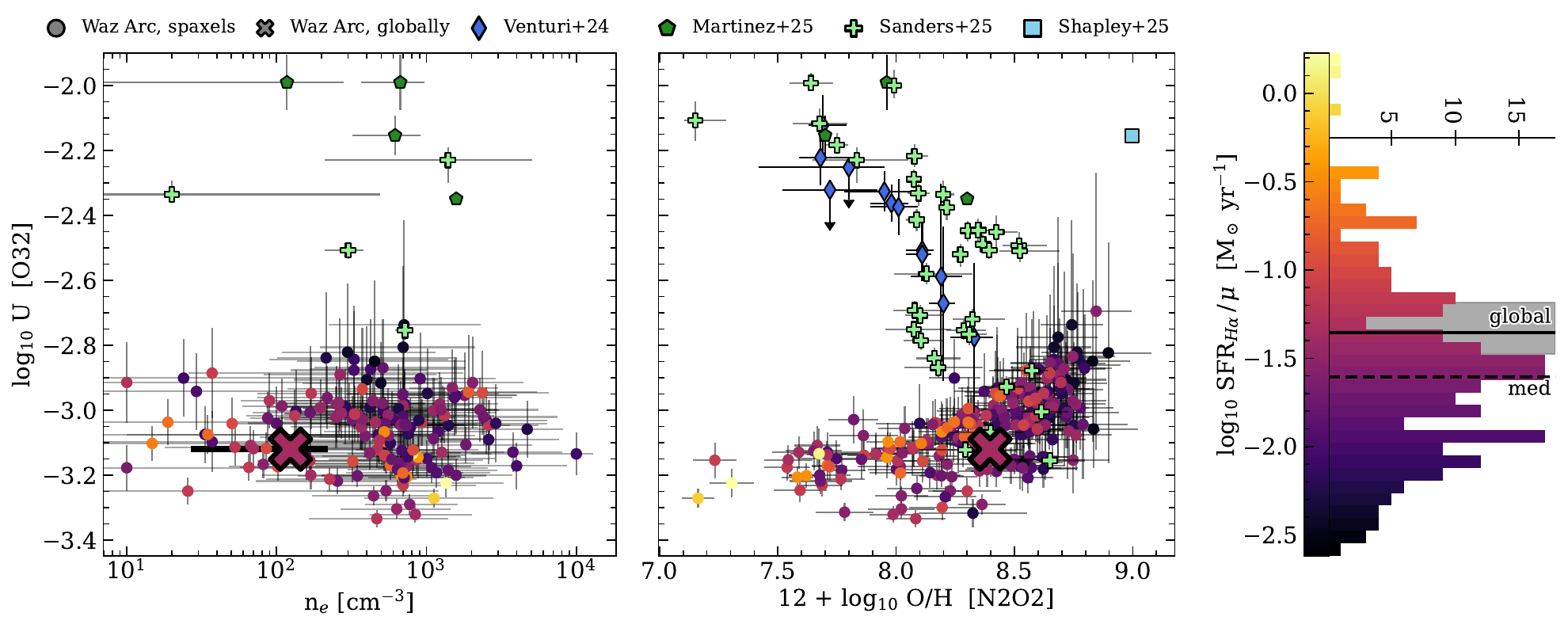}
    \caption{\textbf{The diversity of ISM conditions at smaller scales compared to a global view.}
    Ionization parameter versus (\textit{left}) electron density and (\textit{middle}) gas-phase metallicity for detected spaxels in the \waz.  The spaxels (circles) are colored by star formation rate derived from dust- and lensing-corrected \ha\ emission.  The global, spatially-integrated value for the \waz\ (large X) is colored to match the global \ha-derived star formation rate (see Table \ref{table:measurements}, also the rightmost panel). (\textit{right}) Histogram of the SFR, with the median (dashed) and global (solid) shown, including a shaded gray region denoting the uncertainty range in the average lensing magnification correction used for the global SFR ($\bar{\mu}=92^{+37}_{-31}$; \citealt{Klein.2024}).  For comparison, global measurements from high-z studies by \citet[diamond, $z\sim6.4-7.9$]{Venturi.2024}, \citet[pentagon, $z\sim2-9$]{Martinez.2025}, \citet[plus, $z\sim2-7$]{Sanders.2025}, \citet[square, $z=6.731$]{Shapley.2025} are included in the electron density and gas-phase metallicity panels.  These studies highlight the diversity of high-z galaxy properties, but also how globally the \waz\ fits well in this parameter space -- while the dispersion of \waz\ properties from individual spaxels illustrates the value of spatially-resolution and what we may be missing in other high-z galaxy studies. 
    }
    \label{fig:comparison}
\end{figure*}

\subsection{What may be missing from a global view \\ of a high-redshift PSB?}  \label{subsec:u-z-ne-sfr}

%
%



Even with the introduction of \jwst\ to high-redshift galaxy evolution studies, we are still limited by the apparent size and brightness of sources in the early Universe.  Gravitational lensing is a boon to these studies, magnifying the light from distant galaxies and allowing us to peer within the depths of individual sources.  While \jwst/NIRSpec IFS is a powerful tool, due to its small field of view (3\arcsec$\times$3\arcsec) observations are limited to measuring one source (or very nearby sources) at a time.  There have been advances on larger scales through clever techniques such as ``slit-stepping'' (which utilize \jwst/NIRSpec MSA as a massive IFS, e.g., \citealt{Barisic.2025}; see also programs \citealt{piKassin.2023,piJones.2023}) in order to target several sources at once. However, the number of $z\gtrsim5$ sources that have spatially-resolved spectroscopy is still in the low dozens, while the number of sources at these redshifts with spatially-integrated (``slitted'') spectroscopy using \jwst/NIRSpec MSA is well into the many hundreds, if not thousands \citep[e.g.,][]{Adamo.2025}.
Therefore, much of what we now know spectroscopically of galaxies in the early Universe is via their global properties -- the spatially-averaged (``global'') information about their stellar populations, ISM properties, accreting black holes, etc.

%
%
The \waz\ provides an excellent illustration of what may be hiding at smaller spatial scales within high-z galaxies. 
In the arc, the locations of higher SFR appear to relate to the relatively higher nebular densities.  As noted previously, there appears to be a dip inside the brightest star-forming knots in both ionization and metallicity -- such that the very centers of these regions are lower in both gas-phase metallicity (using N2O2 and R23) and ionization compared to both SFR and nebular density.
In Figure \ref{fig:comparison}, nebular density ($n_e$; left panel) and gas-phase oxygen abundance (``metallicity'' in this context; middle panel) are compared against ionization parameter (log$_{10}$\,U).  The \waz\ is shown both in individual spaxel measurements (circles) and the globally-averaged measurement (large X), colored by their respective SFR value (see right panel of Figure \ref{fig:comparison}). Global measurements from a modest number of high-z sources from the literature ($z>3$, with nearly 30\% at $z>6$; \citealt{Venturi.2024,Martinez.2025,Sanders.2025,Shapley.2025}) are shown in both panels.  

The nebular density of the high-z sources from the literature, as well as the global \waz\ measurement, span over two order of magnitude.  This is not unexpected, as studies at higher redshifts have found a range of nebular densities \citep[$n_e\sim10^2-10^4$ cm$^{-3}$, e.g.,][]{Topping.2025,Isobe.2023}.  However, what is not measurable with the global data is the dispersion of densities within a given galaxy.  Low-z studies and simulations show a broad diversity of nebular densities driven by and/or influencing stellar population ages, stellar feedback, relative abundances, and ambient ISM temperatures, among other factors (\citealt{Kewley.2019}).  While the \waz\ spaxels show a broad range of nebular densities, with very little trend in SFR, the majority cluster near $\sim10^3$ cm$^{-3}$.  As shown in Figure \ref{fig:density} as well, the median of the spatially-resolved measurements is $\sim$0.5 dex larger than the global \waz\ measurement. This difference highlights the value in spatially-resolved studies, where the magnitude and extent of such physical properties can directly inform studies investigating the history and location of star formation within individual high-z galaxies.  

\begin{figure*}[th]
    \centering \hspace*{-0.5cm}
    \includegraphics[width=1.02\linewidth]{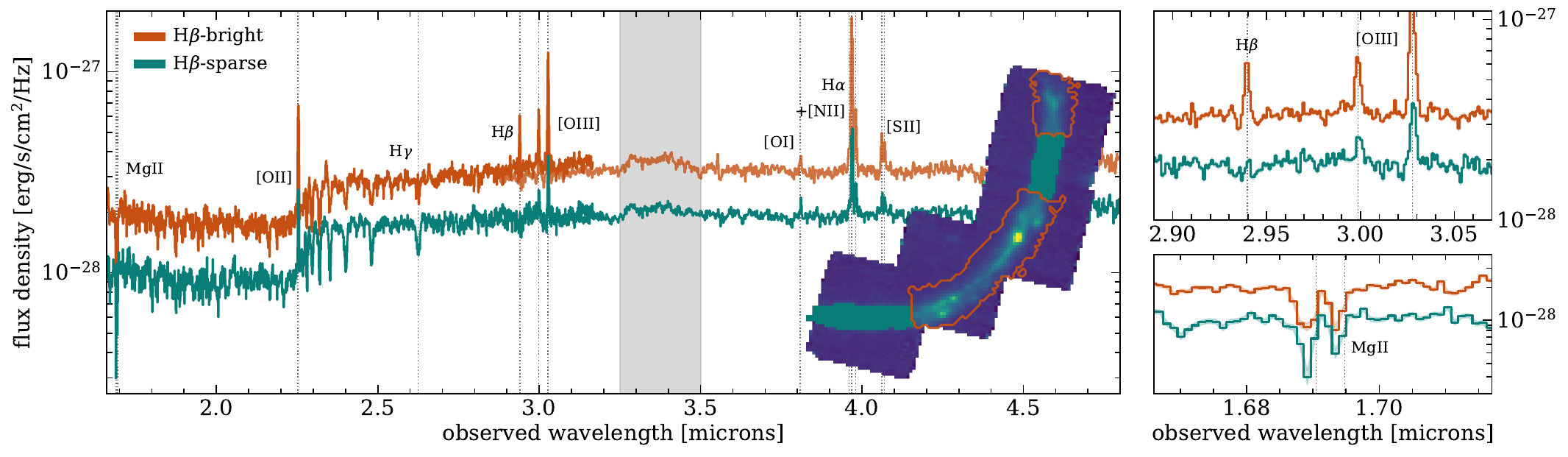}
    \caption{\textbf{Non-detected H$\mathbf{\beta}$ regions connected to potential first indication of DIG at $\mathbf{z>5}$.}
    (\textit{left}) Integrated 1D spectroscopy of the \hb-bright (orange) and \hb-sparse (teal) regions of the \waz, comparing the differences in emission and absorption line strengths.  The map of the arc is in the bottom right of this panel, with the two regions colored to match the spectra.  (\textit{right}) Zoom-ins of i) the \hb\ and \oiii\ profiles reinforcing the general lack of \hb\ emission in the \hb-sparse regions of the arc (as well as overall weaker nebular emission), and ii) the \mgii\ absorption feature showing varying absorption strengths suggesting different optical depths, both blue-shifted from the systemic redshift (dotted lines) due to stellar outflows.}
    \label{fig:dig}
\end{figure*}

Comparing gas-phase metallicity to ionization parameter, we find the dispersion of \waz\ spaxel measurements distributing nearly perpendicular to the direction of higher ionization and lower metallicities that the high-z sources from the literature trace.  Indeed, even the global \waz\ measurement overlaps with some of the \citet{Sanders.2025} sources that reach down into the same parameter space.  What is intriguing is the notion of what each of these high-z sources would show, if similar spatial resolution were possible.  In the \waz, the points with the highest SFR appear to follow the upper edge of the distribution, traveling to both lower metallicities and ionization parameter.  Conversely, the points with lower SFR appear to predominantly reside in the bottom and right edges of the distribution.  The relationship of lower SFR with lower ionization and higher metallicities logically tracks with global (spatially-integrated) trends measured in the literature (such as quenching/post-starburst characteristics; e.g., \citealt{Leung2025} and references therein) as well as the path traced by the global measurements in this figure.  However, there also exist many points with very low SFR that occupy both high metallicity and high ionization parameter space.  The diversity of physical properties implied here point to the various mechanisms that can simultaneously be at play within a galaxy -- both in star formation-dominated regions as well as in regions where there is very little or no active star formation (such as diffuse ionized gas, see \S\ref{subsec:dig} for more discussion on this mechanism).

Similar IFS studies using \jwst\ at lower redshifts have found comparable nearly-perpendicular dispersion in certain diagnostics when compared to global spatially-averaged values.  In Olivier et al.\ (in prep), they show spatially-resolved measurements of auroral lines in a galaxy at $z=1.3$, comparing the direct metallicity method (using the auroral lines) to strong line ratio diagnostics that trace the general trends of metallicity, such as those in this work (see \S\ref{subsec:metallicity}).  In their calibration figures of strong line ratios versus direct measurements via auroral lines, they find similar scatter around the global spatially-averaged value of their source, where the spaxels with the brightest auroral line detections were more centered while the lower luminosity spaxels extended further out.  Separately, \citet{Khoram.2025} measured stacked IFS spectra in radial bins for $>4000$ MaNGA IFS galaxies, with their results also showing a broader dispersion in individual spaxel bins versus their globally-averaged measurements.  These studies reinforce the dispersion measured in this work, with all of these studies involving IFS data where each spaxel could contain both star-forming and non-star-forming dominated regions. Thus contributing to the variety and dispersion in derived physical properties and directly informing studies following the growth and evolution of individual galaxies.

\subsection{The Potential Presence of Diffuse Ionized Gas}\label{subsec:dig}

Another consideration for spatially-resolved studies of highly magnified galaxies is the effect of other smaller-scale galaxy properties otherwise impossible to discern at higher redshifts. This includes the potential to separate HII regions from diffuse ionized gas \citep[DIG; e.g.,][]{Zhang.2017}. Locally DIG emission has been found to contribute over 50\% of the total light in face-on galaxies (mean fraction of warm ionized medium [DIG], $f_{WIM}$ = 0.59 $\pm$ 0.19; Oey+2007). For lower surface brightness sources, this fraction grows even larger.
Spatially-resolved studies at $z\sim0$ from surveys such as MaNGA have found evidence of DIG interspersed throughout galaxies, identified by regions with lower \ha\ surface brightness ($\Sigma$\ha) and enhanced line ratios in a few DIG-sensitive diagnostics \citep[e.g.,][]{Zhang.2017,Bracci.2025}. From these studies, DIG have been shown to contaminate key strong line ratio diagnostics that are frequently used to infer the physical conditions of star-forming (HII) regions within a galaxy. Namely, physical conditions estimated via metallicity tracers such as \niiha\ and R23 and ionization indicators such at O32, leading lower-z studies to caution extragalactic programs to consider the possible contribution of DIG in such diagnostics \citep[e.g.,][among others]{Garg.2024,McClymont.2024,Bracci.2025,Clarendon.2025}. Such contamination is thought to come from DIG emission containing stronger nebular metal lines (\nii, \sii, \oii, \oiii) while simultaneously weaker recombination lines (\hb, \ha) -- likely driven by higher relative abundances of those ionized elements in increasingly lower nebular densities compared to HII regions (due to previous star formation and stellar feedback; e.g., \citealt{McClymont.2024,Bracci.2025}) coupled with higher temperatures in DIG-dominated regions.
Some studies suggest that DIGs may be powered in part by hot, low-mass, evolved stars (HOLMES, generally $>100$ Myr stars; e.g., \citealt{Zhang.2017,Postnikova.2023}) emitting very hard ionizing photons, which may make DIGs even more prominent in post-starburst or quiescent galaxies, where the majority (or all) of the OB stars have died out \citep[]{Postnikova.2023,Boardman.2023}. 
However, other studies note that while HOLMES may play a part in ionizing DIGs, a more likely primary contributor is leaking radiation from stellar populations aged $\sim$10--25 Myr.  This radiation is also hard (albeit less so than HOLMES and other 100 Myr and older stars) but more critically also has a much higher intrinsic ionizing luminosity (see discussion in \citealt{McClymont.2024}).  Both of these scenarios favor sources with aging stellar populations, where one has the spatial resolution to separate more DIG-dominated regions from HII-dominated regions within a galaxy.

We investigate the potential for DIG in the spatially-resolved \waz\ by comparing the regions of strong and non-detected \hb\ emission, respectively. The location of these regions is clearly defined by the spaxels that have dust attenuation corrections versus those which not used (i.e., where we were able to spectrally measure a Balmer decrement), as shown in the gray footprints outlining the full extent of the arc in Figures \ref{fig:dust}--\ref{fig:ionization}.  We highlight the two regions in the middle of Figure \ref{fig:dig}, overplotted on a continuum map of the four NIRSpec/IFS pointings.  We denote these two regions as \hb-bright (orange contour) and \hb-sparse (filled teal). The \hb-sparse regions represent where \hb\ emission is not detected above the local continuum, where even correcting for infill from stellar absorption is not enough to generate reliable detection (\S\ref{subsec:Resultsstellarabs}).  Integrating the spaxels in each region, we observe some key differences in their 1D spectra. The left panel of Figure \ref{fig:dig} shows the two spectra, where the differences in the various strong emission lines are stark. However, this is most notable in the shape of the \hb\ line (further highlighted in the top right panel), where in the \hb-sparse spectrum the profile is dominated by stellar absorption.  

Additionally, we note the difference between the profiles of the blue-shifted \mgii\ absorption (from systemic, dotted lines; also shown in the bottom right panel).  The relative strengths of the \mgii\ absorption lines can shed light on the relative opacity of the nebular gas through which it is measured, including any indications of optically-thick scenarios (where measuring exact absorption line strength is no longer feasible; e.g., \citealt{Finley.2017}).  
For \mgii, the oscillator strengths of the lines result in a line ratio of 2:1 in an optically-thin scenario and, as the optical depth increases, the absorption lines saturate and the line ratio approaches 1:1 \citep[e.g.,][]{Rubin.2010}.

In the bottom right panel of Figure \ref{fig:dig}, we zoom in on the \mgii\ spectral feature.  In the \hb-bright spectrum, the ratio of the \mgii\ absorption line strengths are nearly unity, suggesting a higher optical depth in the regions with higher star formation.  Conversely, in the \hb-spares spectrum, the blue line is stronger suggesting more optically-thin gas in the regions with little to no star formation.  To quantify this, in forcing a single gaussian fit to both lines in each spectrum, we find line flux ratios of 1.07 $\pm$ 0.19 and 1.47 $\pm$ 0.27 for the \hb-bright and \hb-sparse spectra, respectively.  This could suggest that the \hb-sparse regions are dominated by less dense gas, while the \hb-bright regions have higher densities of gas (which generally mirror the results shown in the nebular density map, \S\ref{subsec:results-density}).
However, we note that these measurements are lower limits, as blue-shifted absorption profiles are better fit with multiple components (especially in the bluer \mgii\ line) which may lead to a slightly larger ratio in both; however a deeper analysis of the stellar outflows and precise absorption features across the \waz\ requires a combination of higher-resolution spectroscopy, bluer gratings, and even deeper integrations.  In this work, we merely note the presence of these lines and their support of the potential identification of DIG-dominated regions.

\begin{figure}[h!]
    \centering 
    \includegraphics[width=\linewidth]{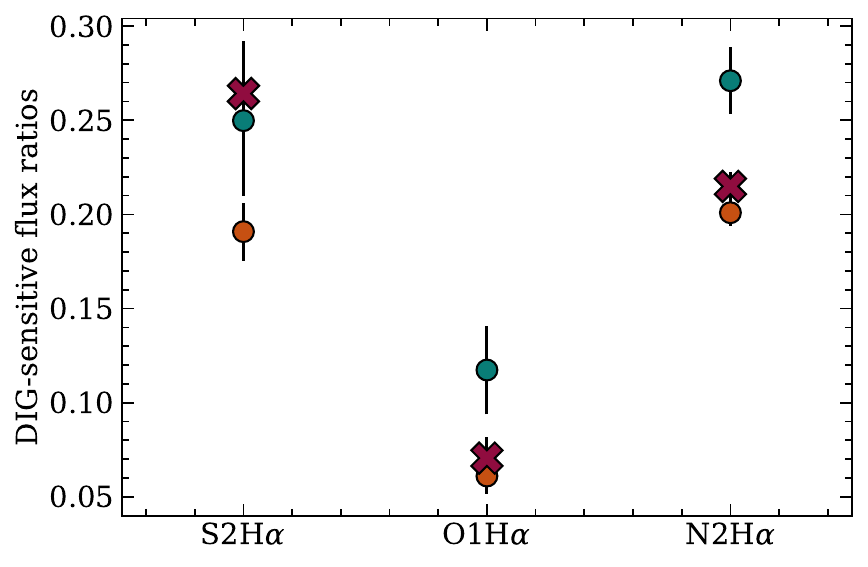}
    \caption{\textbf{\hb-sparse regions in the \waz\ higher in DIG-sensitive line ratios.} Three line ratio diagnostics sensitive to DIG regions within a galaxy: \sii/\ha, \oi/\ha, and \nii/\ha. The three diagnostics shown for the spatially-integrated diffuse regions, with the globally-integrated (all spaxels) value for each ratio included as well (red cross).}\label{fig:dig-ratio}
\end{figure}

\begin{table}[h!]
\centering
\caption{DIG-sensitive Line Flux Ratios} ~\\ [-3mm]
\begin{tabular}{cccc}
    \hline \hline \\ [-3ex] 
    Ratio & \hb-sparse & \hb-bright & Waz Arc, globally \\ [0.5ex] 
    \hline \\ [-3ex] 
    S2H$\alpha$ & 0.25 $\pm$ 0.04 & 0.19 $\pm$ 0.02 & 0.26 $\pm$ 0.03 \\
    O1H$\alpha$ & 0.12 $\pm$ 0.02 & 0.06 $\pm$ 0.01 & 0.07 $\pm$ 0.01 \\
    N2H$\alpha$ & 0.27 $\pm$ 0.02 & 0.20 $\pm$ 0.01 & 0.21 $\pm$ 0.01 \\
    \hline
\end{tabular}
\end{table}

Creating a spatially-integrated spectrum for both regions, we measure the DIG-sensitive diagnostics as shown in Figure \ref{fig:dig-ratio}.  Such diagnostics include \sii/\ha\, \nii/\ha\, and \oi/\ha, which are more sensitive to lower ionization and are not impacted by dust attenuation due to their small wavelength separation (especially valuable when looking at spatially-resolved regions where a Balmer decrement cannot be reliably measured). 
The DIG-sensitive line diagnostics show intriguing differences between the \hb-sparse and \hb-bright regions. All three diagnostics show higher ratios for the \hb-sparse regions in the arc, with the strongest evidence in the \oiha\ and \niiha\ diagnostic ratios where the \hb-sparse ratios are $\sim2--3\sigma$ higher. This may indicate the presence of DIG in the regions of the \waz\ where \hb\ emission is either not detected or only seen in absorption. This result aligns well with spatially-resolved studies of DIGs in nearby galaxies from the MaNGA survey, where similar enhancement has been linked to harder radiation yet lower ionizing luminosities powered by older stellar populations \citep[e.g.,][]{McClymont.2024}, and HOLMES specifically \citep[e.g.,][]{Postnikova.2023,Zhang.2017}.

The global values for the three DIG diagnostics are plotted in Figure \ref{fig:dig-ratio} as a large red X.  While the global \niiha\ and \oiha\ values agree with those derived from the \hb-bright regions, the \siiha\ ratio overlaps more in the \hb-sparse/DIG-like regime.  In a series of local MaNGA sources, \citet{Zhang.2017} used \siiha\ to verify that $\Sigma$\ha\ was an accurate tracer in separating DIG regions from HII regions.  While all three of the DIG line ratio diagnostics in Figure \ref{fig:dig-ratio} were explored in their work, the \siiha\ had the strongest, clearest separation of the two regions.  Therefore, it is possible that the enhanced global \siiha\ in the \waz\ may be influenced by the presence of DIG in part due to the overall expected rise in densities at high-redshift -- driving ions with low critical densities to collisionally de-excite in higher-density star-forming regions (such as \oii\ and \sii\ $\sim1-5\times10^3$ cm$^{-3}$, respectively), allowing DIG to shine through more prominently. For another potential factor, we also note the discussion in \citet{Bracci.2025} regarding the prevalence of \siiha\ globally in galaxies and studies that show the fainter \oiha\ ratio provides more reliable diagnostic in separating DIGs from HII regions. 

This discussion reinforces the caution supplied by lower-z studies regarding the influence of DIG in physical conditions derived from line ratio diagnostics.  Somewhat serendipitously, the spatially-resolved physical conditions probed in this work represent only the \hb-bright regions (barring the broader coverage of \niiha\ and \sii\ ratios, due to their lack of correction for dust attenuation).  Therefore, the results from this work are representing results from the HII-dominated regions of the \waz, likely minimizing the influence of DIG in those spatially-resolved measurements.  However, any DIG-dominated regions would be included in a global spatially-averaged spectrum, therein potentially influencing any global measurements presented in this work.  Additionally, as noted such an effect from DIGs is anticipated to be more significant in older, more evolved galaxies.  While this analysis is the highest redshift thus far to trace DIG-dominated regions in an individual galaxy, such work is truly only possible at these epochs through the combination of sensitive NIR technology and strong gravitational lensing.

\begin{figure*}
    \centering
    \includegraphics[width=1.0\linewidth]{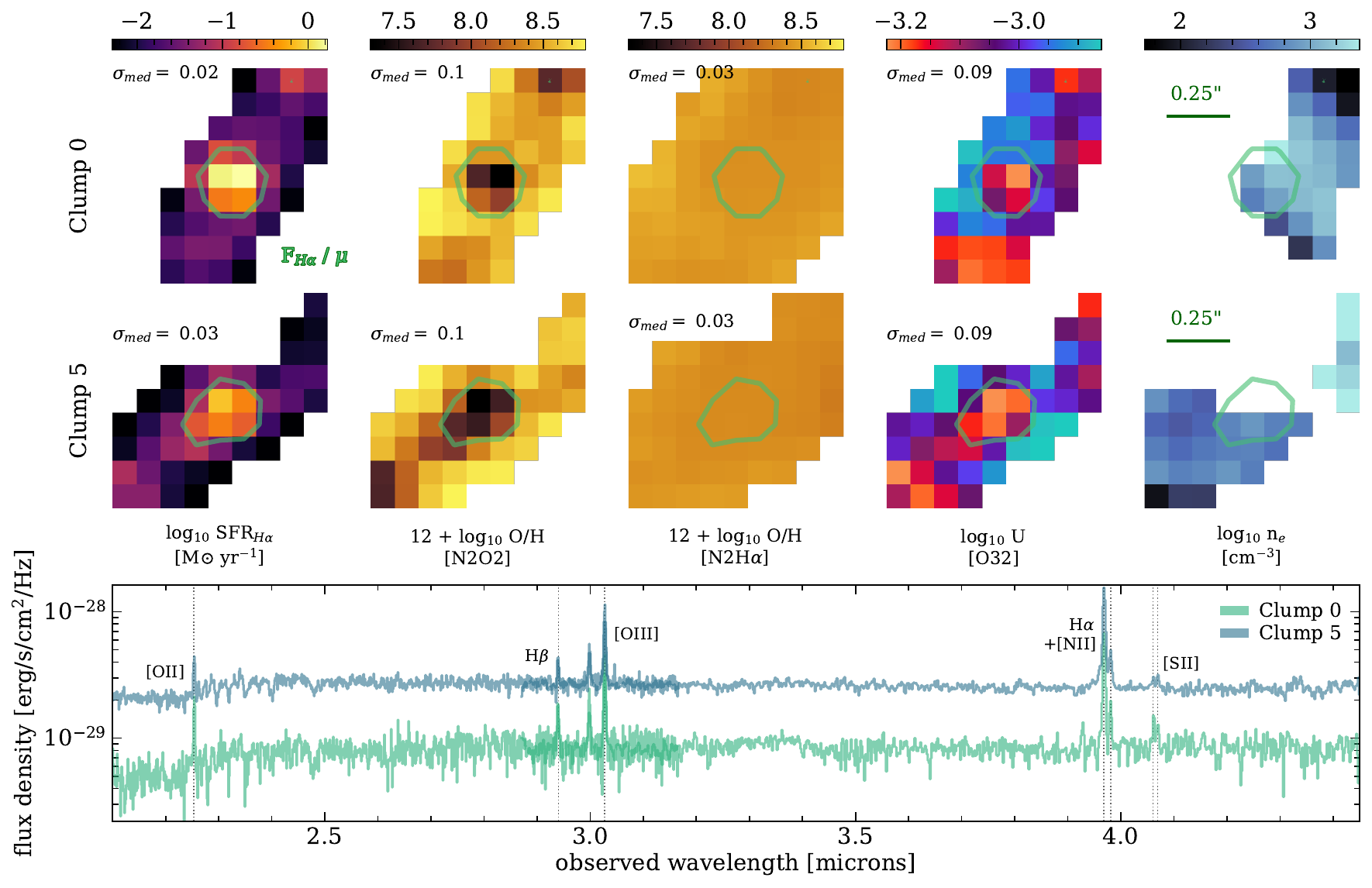}
    \caption{\textbf{Spatially-resolved spectroscopy allows study of individual clumps at high-z.}
    (\textit{top}) Zoom-in panels for two clumps identified in Paper II (Clumps 0 and 5; Khullar et al. in prep), showing SFR and a variety of ISM physical conditions (with the color-coding of each map matching previous figures). The green contours in each subpanel show the lensing-corrected \ha\ flux for each clump.  In the rightmost panel of each clump, the green horizontal line indicates 0\farcs25 in this scaling. 
    (\textit{bottom}) The integrated 1D spectrum of Clumps 0 and 5, with the strong lines used in deriving the physical conditions labeled.  We note that these spectra are in the image plane (i.e., have not been corrected for magnification).
    }
    \label{fig:zoomin}
\end{figure*}

\subsection{Metallicity ``Gradients'' within the Waz Arc}\label{subsec:z-gradient}

Some hydrodynamical simulations have suggested that as galaxies evolve and build their stellar mass, the degree of abundance gradients from inside-out compact growth tend to flatten out with time \citep[e.g., FIRE-2;][]{Graf.2025}. Recent work studying spatially-resolved star-forming galaxies around $z\sim1$ using \jwst\ have found a correlation between stellar mass and negative metallicity gradient \citep{Ju.2025}, such that more massive sources appear to have metal enrichment in their cores and decreasing metallicity radially outward. 
Simulations like TNG50 match these observations \citep[e.g.,][]{Hemler.2021}, often predicting less starburst-like feedback driving negative gradients that grow steeper with increasing redshift -- while simulations such as FIRE disagree in some scenarios, at times predicting a more flattened metallicity gradient due to mergers and galactic outflows driven by supernovae (SNe) bursts \citep[e.g.,][]{Ma.2017}. This difference in predictions is not uncommon, as different hydrodynamical simulations in the literature show a broad variety of effects and gradients which largely depend upon star formation histories and feedback, as well as other galaxy properties \citep[e.g.,][]{Garcia.2025b,Hemler.2021,Tissera.2019,Ma.2017}.

We leverage the lensing magnification of the \waz\ to discuss individual clumps of star formation that we see within the extended arc.  For the purpose of this discussion section, we treat the star-forming clumps as individual HII regions.  The top two rows of Figure \ref{fig:zoomin} show zoom-in panels on two distinct star-forming clumps colored by various ISM properties such as gas-phase metallicity (derived from the N2O2 and \niiha\ diagnostics), ionization, and nebular density.
Contrary to most predictions of high-redshift metallicity gradients following negative, ``inside-out'' growth \citep[e.g.,][among many others, beginning with \citealt{Chiosi.1980} and contemporaries]{Acharyya.2025,Garcia.2025a,Ibrahim.2025}, we find what appears to be the opposite effect in the sub-kpc star-forming regions of the \waz.  The brightest star-forming knots appear to have some of the lowest metallicities in the arc, with the local metallicity gradient rapidly increasing outwards from their centers by over 0.5 dex (when using our fiducial gas-phase metallicity indicator, N2O2). 

The existence of a relatively metal-poor center increasing to higher metallicities radially outward is less common in simulations of lower-z galaxies, where generally the reverse is found due to standard cycles of star formation and mechanical feedback (see references above). However, at higher redshifts ($z>5$) some simulations suggest a broad diversity of metallicity gradients due to the chaotic nature of the earlier Universe (many mergers, various drivers of feedback, re-accretion of enriched material, accretion of metal-poor gas, mixing with the circumgalactic medium, etc.; e.g., \citealt{Tapia-Contreras.2025,Ibrahim.2025}).  We note that this scenario is likely the case on a global scale -- where all of the various regions within a galaxy are averaged together -- yet it is worth exploring the diversity of gas-phase metallicity probed around individual star-forming clumps at small spatial scales within a single source.

The observed positive ``gradient'' measured for the star-forming clumps in Figure \ref{fig:zoomin} provide clues of the clumps' local SFHs. In Paper II, we show the global SFH of the \waz\ includes a recent burst of star-formation ($\sim$20 Myr in lookback time) followed by abrupt quenching, leaving enough existing massive stars to power the nebular line emission seen in certain parts of the \waz.
A potential source of this gradient includes the possibility of a recent in-fall of relatively pristine gas (i.e., very low metallicity gas; compared to the median Waz gas-phase metallicity) onto the star-forming regions (see discussion in e.g., \citealt{Venturi.2024}).  Such an event in the recent history of the \waz\ would fuel new star-formation in the clumpy regions, generating slightly more metal-poor stellar populations, while the outskirts and non-star-forming regions of the \waz\ remain broadly unaffected.  

Alternatively, focusing upon the \niiha\ map, we see very little variation across the entire arc.  Using this diagnostic alone would suggest that the \waz\ may have undergone a merger or experienced strong outflows powered by SNe winds in its recent history. Such a history is not unexpected for massive galaxies at these early redshifts \citep{Ibrahim.2025}, especially as the \waz\ is among the more massive sources for its redshift (M$_{*} \sim 10^{10}$ \Msolar; \citealt{Khullar.2021}, Paper II).  However, this possibility is partially testable by inspecting the spectroscopy for indications of asymmetry or broad components that would hint towards other galaxy effects at play. Upon refitting the stronger emission lines in the arc (i.e., \oiii\ and \ha) we found no evidence of additional broad components in their profiles, except at very low confidence in the brightest star-forming clump located in the visual center of arc (Clump 5 in Figures \ref{fig:rgb} \& \ref{fig:zoomin}). The lack of a (significant) broad component anywhere in the arc does not disprove the presence of merger activity impacting the gradients derived, however this does suggest that such activity likely happened far enough in the past to not have a visible effect at present \citep[e.g.,][and references therein]{Venturi.2024}.

However, there exists a third possibility, one that leverages all metallicity diagnostics probed. In the most star-forming regions (clumps 0 and 5; Figure \ref{fig:zoomin}), depending upon the timescale of the most recent burst of star-formation, there 1) may still exist massive stars, and 2) the stellar population is young enough that the significant production of N by low- to intermediate-mass stars has not begun.  If this is the case, the more massive stars would be steadily building the oxygen abundance of the local gas while contributing some of the nitrogen abundance, but not all.  If we assume that the low- and intermediate-mass stars in the same young stellar populations have not lived long enough to begin the secondary enrichment of nitrogen, this could explain the difference measured via the N2O2 diagnostic.  This behavior, with low N/O at high O/H, can be produced with bursts of high-efficiency star formation \citep[e.g.,][]{Henry.2000, Vincenzo.2016}. Some simulations of high-redshift galaxies have found local regions of high star formation efficiency can be triggered following a global starburst \citep{Wang.2025}, similar to what we observe for the Waz Arc. Additionally, low N/O at high O/H has been observed in other high-redshift galaxies \citep[e.g.,][]{Rogers.2025}.  Thus, rather than describing lower gas-phase metallicities in these regions, the N2O2 ratio rather may be indicating a lower N/O abundance.  This possibility is further supported by the \niiha\ diagnostic, where we see very little variation -- suggesting that there may be a decrease of N/O in the star-forming regions at fixed N/H.  Testing this possibility requires bluer spectroscopy and deep higher-resolution spectroscopic observations, in order to tease out weaker signal and measure line profiles.

~\\
\section{Summary and Future Directions} \label{sec:conclude}

In this work, we present \jwst\ NIRSpec/IFS observations of the \waz\ (n\'e COOL J1241+2219), a highly magnified galaxy at $z=5.04534$, right after the end of the reionization era ($z\sim13-6$). Originally identified via ground-based imaging \citep{Khullar.2021}, this restframe UV-bright ($z_{AB}=20.5$) massive ($M_\star\sim10^{10}$ \Msolar) galaxy was thought to be filled with low-metallicity gas and house ample star formation.  Using four pointings of \jwst\ NIRSpec/IFS (to cover the full $\sim$12\arcsec\ arc), we measure the spatially-resolved spectral features across the galaxy as well as the globally-integrated features (measured from a single, integrated spectrum; to serve as a global comparison to the spatially-resolved results).  We summarize the key findings below.

\begin{itemize}
    \item We find the \waz\ to be globally post-starburst in nature, with strong absorption features indicating a recent burst, a fast quenching episode, older stellar populations, and relatively weak nebular emission. The \waz\, in its source plane modeling, shows a ring-like/irregular PSB morphology -- a central SF region surrounded by quenched pockets.  
    
    \item Based on nebular line SFR indicators, we measure very low star formation rates (SFR) across the \waz\ (SFR$_{H\alpha} \lesssim$1 \Msolar\ yr$^{-1}$), with locations of distinct ``star-forming'' clumps matching some of the clumps identified in the \jwst/NIRCam imaging (see Paper II; Khullar et al, in prep).

    \item We use the Balmer decrement of \ha/\hb\ emission to estimate and correct for dust attenuation across the arc.  However, due to the nature of the \waz\ and the presence of underlying stellar absorption, among other factors, we do not measure \hb\ in emission consistently throughout the galaxy.  Therefore, the majority of the spatially-resolved maps of the physical conditions of the \waz\ show only the spatial regions where \hb\ was detected in emission.  We explore the possible mechanisms driving the lack of detectable \hb\ emission in \S\ref{subsec:dig}, investigating the potential discovery of an exciting not-yet-seen-at-high-z component from low-z galaxy studies (see summary below).
    
    \item We estimate the gas-phase metallicity of the \waz\ using the strong emission line ratio diagnostics \nii/\oii\ (\niioii), \nii/\ha\ (\niiha), and R23 ($\equiv$ (\oiii\ + \oii) / \hb).  Our results show median gas-phase metallicities of 12 + log$_{10}$ O/H $\sim$ 50--54\%  (8.43 $\pm$ 0.1, \niioii; 8.45 $\pm$ 0.04, \niiha; and 8.47 $\pm$ 0.03, R23), with a dispersion of values up to 1 dex for our fiducial diagnostic \niioii.  We note a much more narrow distribution measured by \niiha\ -- suggesting less spatial variation -- and discuss the possible implications in \S\ref{subsec:z-gradient}, including the possibility of the fiducial \niioii\ diagnostic potentially tracing decreased N/O abundances compared to O/H.
    
    \item We measure the nebular density across the \waz\ by the weak \sii\ doublet, finding a broad range of densities spanning $\sim10^2-10^{3.5}$ cm$^{-3}$.  Leveraging the spatial resolution provided by the IFS data, we find the median nebular density higher than the global density by $\sim0.5$ dex ($n_e\sim500$ vs $n_e\sim100$ cm$^{-3}$), with the highest densities overlapping with the locations of the largest SFR.
    
    \item We estimate the ionization parameter, log$_{10}$U, for the \waz\ using the strong line ratio diagnostic O32 ($\equiv$ \oiii5008 / \oii). While covering a range of nearly 0.5 dex in spatially-resolved measurements, we find general agreement between the median ($-3.05\pm0.08$) and the globally-integrated value ($-3.12\pm0.02$), within uncertainties.  We further discuss the \waz\ in context of its ionization, density, and gas-phase metallicities relative to other high-z galaxies in \S\ref{subsec:u-z-ne-sfr}, highlighting the extra layer of information provided by spatially-resolved data and what may be missing from the majority of high-z extragalactic globally-integrated (i.e., single slit) studies.

    \item Zooming into two of the most ``star-forming'' clumps in the \waz, we find lower metallicities within the clump that quickly enriches radially outward.  We compare these lower-metallicity clumps against their SFR, ionization parameter, and densities and find possible hints of either a recent infall of pristine gas onto the star-forming clumps, or perhaps evidence of lower N/O abundances compared to O/H in a region with higher densities.
    
    \item We investigate the regions of the \waz\ with no \hb\ observed in emission and find potential detection of diffuse ionized gas (DIG) for the first time at such high epochs.  Spatially-integrating the \hb-bright and \hb-sparse (i.e., not detected) regions along the arc, we find varied \mgii\ absorption, strong \hb\ absorption in the \hb-sparse spectrum, and DIG-sensitive line ratios showing higher values for the \hb-sparse regions (with the \hb-sparse region $\sim2-3\sigma$ higher in two of the three diagnostics).  All of this evidence supports the possibility of directly-observed DIG at $z=5$, observations which are only feasible at such high redshifts due to the combination of \jwst\ and strong gravitational lensing.

    \item We find that the locations of higher SFR appear to relate to the relatively higher nebular densities, and the brightest star forming knots have lower ionization and metallicity, which would have otherwise blended/averaged out in globally characterized observations.  Such results could indicate different SF/mass assembly paths, including a likely recent infall of pristine (low-metallicity) gas onto the star-forming regions.
\end{itemize}

This JWST-based study of the \waz\ is a clear example of the improvements over extant studies of PSBs at high redshift \citep{Strait2023,dugenio2024, Looser.2024} -- the combination of strong gravitational lensing and JWST provides a robust interpretation of the physical conditions within SF and quenched pockets within this system, analogous to low-redshift MaNGA studies of PSBs. This study is also a comprehensive demonstration of the power of IFS studies of high-redshift lensed galaxies, and a look into the physical conditions within a system with a diversity of star formation histories, as well as nebular emission. What powers this nebular emission requires higher resolution spectra with NIRSpec/IFS, as well as a detailed characterization of systematics -- any obscured AGN contribution to the minimal SF within \waz, feedback processes, outflows. Moreover, the final piece of the puzzle is the content and distribution of molecular gas (or lack thereof) in this quenched galaxy, well within the purview of ALMA observations in the near future.

\begin{acknowledgements}
TAH \& GK acknowledge that a significant part of our work is done on stolen land, and we support the efforts of Land Back and true stewardship to those same peoples whose land is occupied. We also note that we do not use the full name of \jwst\ due to the person after whom this telescope is named and their role as NASA administrator during the ``Lavender Scare'', as per the $\#RenameJWST$ protest movement.

TAH's research is supported by an appointment to the NASA Postdoctoral Program at the NASA Goddard Space, administered by Oak Ridge Associated Universities under contract with NASA, as well as the University of Maryland Baltimore County and the Center for Space Sciences and Technology.
GK would like to thank the Baum Grant and Fellowship at the University of Washington for support during this work, as well as the ALMA Ambassador Program (administered by NAASC and NRAO). 
RLL appreciates support from a Giacconi Fellowship at the Space Telescope Science Institute, which is operated by the Association of Universities for Research in Astronomy, Inc., under NASA contracts NAS 5-26555 and NAS5-03127.

This work is on observations made with the NASA/ESA/CSA \emph{JWST}. The data were obtained from the Mikulski Archive for Space Telescopes at the Space Telescope Science Institute, which is operated by the Association of Universities for Research in Astronomy, Inc., under NASA contract NAS 5-03127 for \JWST. These observations are associated with \JWST\ Cycle 1 GO program 2566. Support for the program JWST-GO-2566 was provided by NASA through a grant from the Space Telescope Science Institute, which is operated by the Associations of Universities for Research in Astronomy, Incorporated, under NASA contract NAS5-26555.

This research was supported in part by the University of Pittsburgh Center for Research Computing, RRID:SCR\_022735, through the resources provided. Specifically, this work used the H2P/MPI cluster, which is supported by NSF award number OAC-2117681. Discovery of the \waz\ was supported by a program initiated by The College Undergraduate program at the University of Chicago, and the Department of Astronomy and Astrophysics at the University of Chicago.
\end{acknowledgements}

\facilities{\JWST, \HST}

\software{\texttt{Python 3.6 — Prospector, python-FSPS, SEDpy, pygtc, Matplotlib, Numpy, Scipy, Astropy, LENSTOOL, Jupyter, IPython Notebooks, GALFIT, SAO Image DS9, IRAF, IDL}}

\appendix

\section{Uncertainty Maps}
In this section, for completeness we include the two-dimensional (2D) uncertainty maps for every spatially-resolved 2D map shown in this work.  The median of each of these maps has been represented in the main text via the $\sigma_{med}$ annotation in every 2D panel.

\begin{figure*}
    \centering
    \hspace*{-3mm}
    \includegraphics[width=1.03\linewidth]{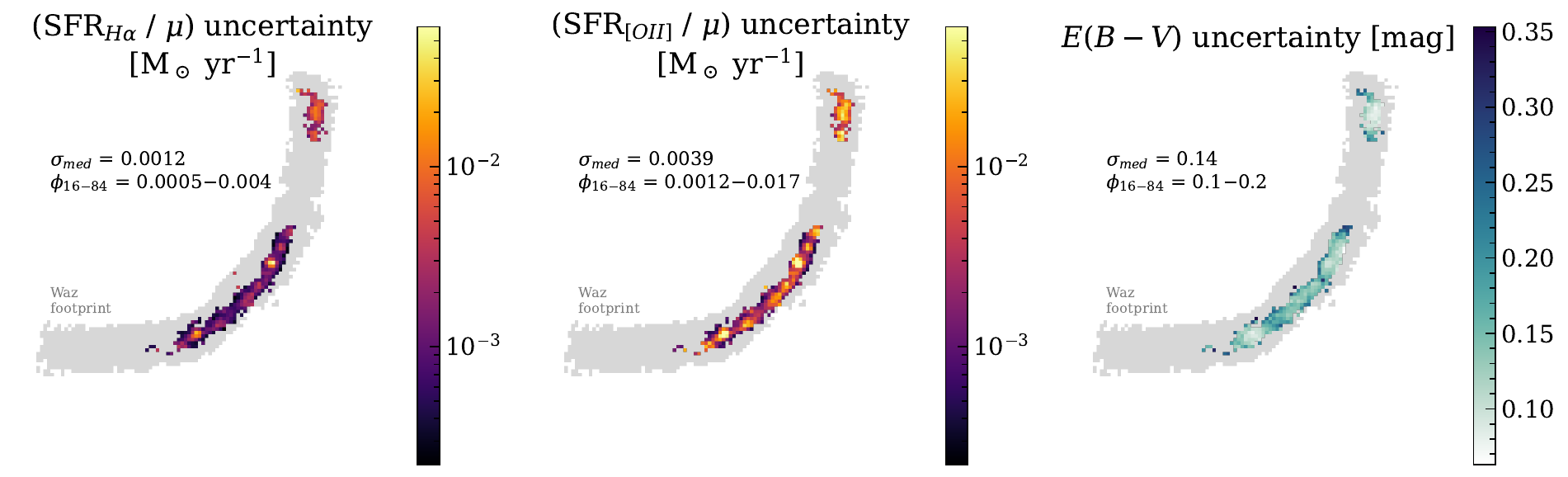}
    \caption{\textbf{Uncertainty maps}. The 2D uncertainty maps for each diagnostic and line ratio shown in this work. In each map, we include the median of the uncertainties (also shown in the main text in their respective figures) as well as the 16$^{th}$ and 84$^{th}$ percentiles, represented by $\phi_{16{-}84}$.}
\end{figure*}

\begin{figure*}
    \centering
    \hspace*{-3mm}
    \includegraphics[width=1.03\linewidth]{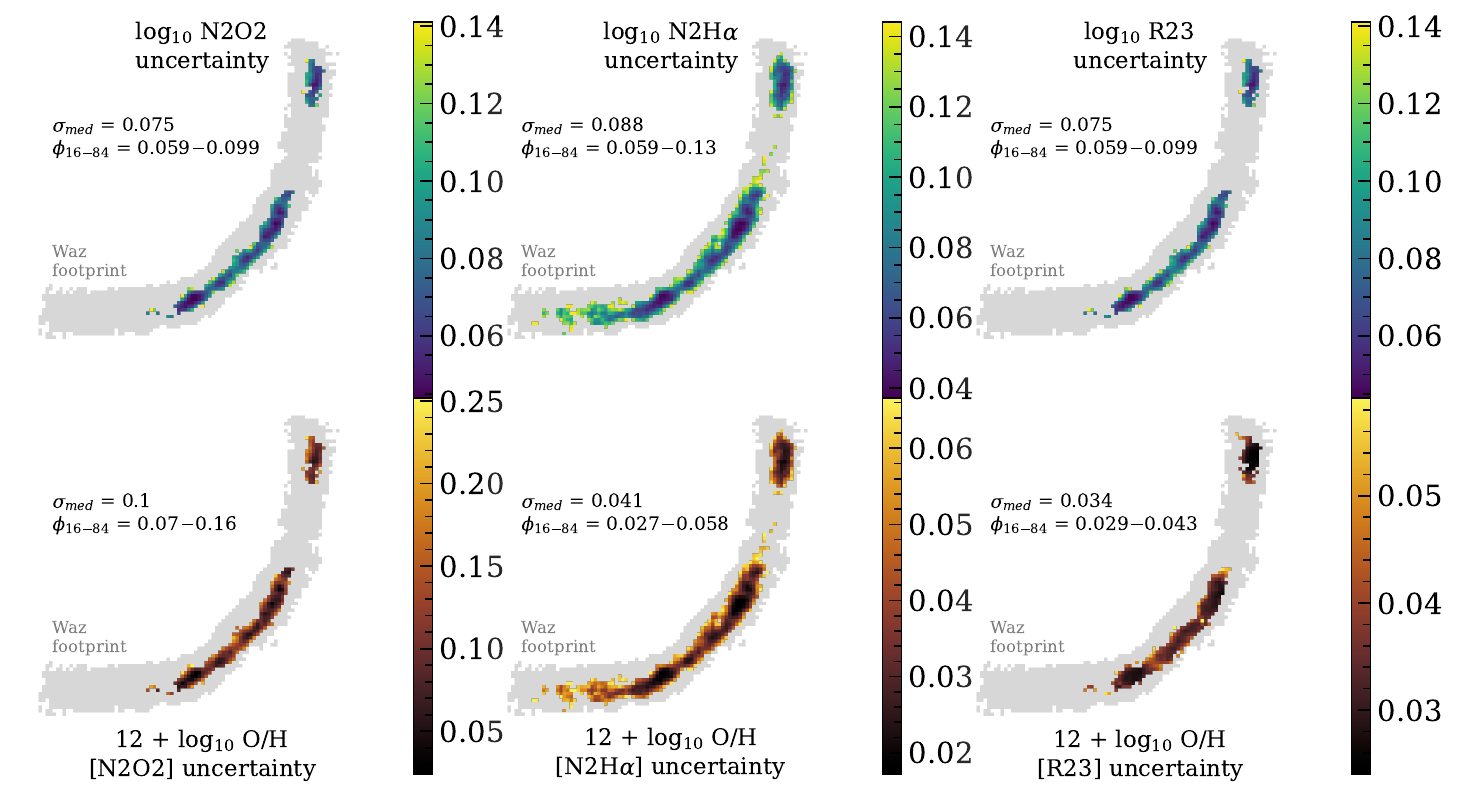}
    \caption{\textbf{Uncertainty maps, continued}. The 2D uncertainty maps for each diagnostic and line ratio shown in this work.  In each map, we include the median of the uncertainties (also shown in the main text in their respective figures) as well as the 16$^{th}$ and 84$^{th}$ percentiles, represented by $\phi_{16{-}84}$.}
\end{figure*}

\begin{figure*}
    \centering
    \includegraphics[width=0.9\linewidth]{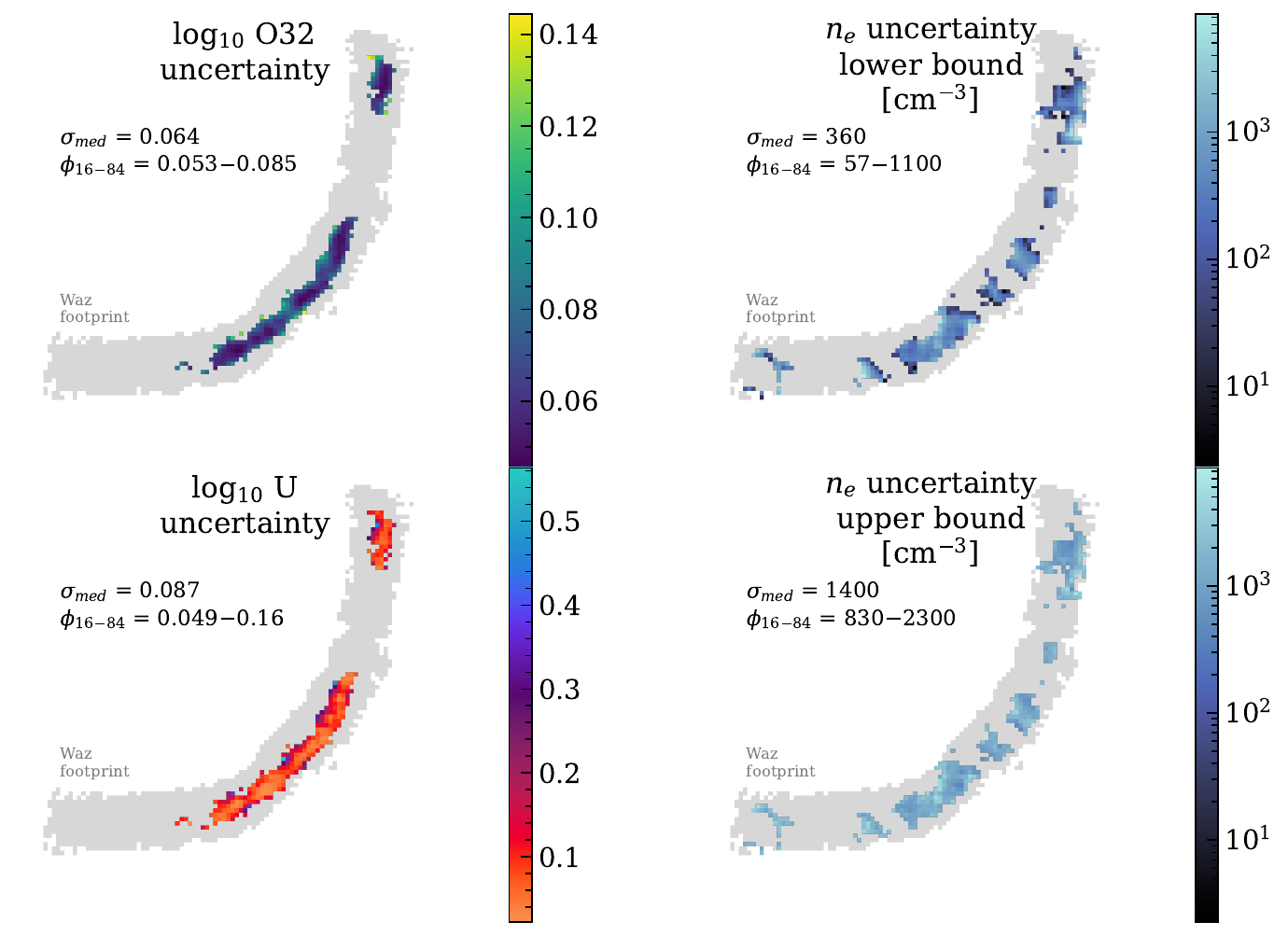}
    \caption{\textbf{Uncertainty maps, continued}. The 2D uncertainty maps for each diagnostic and line ratio shown in this work.   In each map, we include the median of the uncertainties (also shown in the main text in their respective figures) as well as the 16$^{th}$ and 84$^{th}$ percentiles, represented by $\phi_{16{-}84}$.}
\end{figure*}

\pagebreak

\bibliography{papers}{}

\begin{thebibliography}{}
\expandafter\ifx\csname natexlab\endcsname\relax\def\natexlab#1{#1}\fi
\providecommand{\url}[1]{\href{#1}{#1}}
\providecommand{\dodoi}[1]{doi:~\href{http://doi.org/#1}{\nolinkurl{#1}}}
\providecommand{\doeprint}[1]{\href{http://ascl.net/#1}{\nolinkurl{http://ascl.net/#1}}}
\providecommand{\doarXiv}[1]{\href{https://arxiv.org/abs/#1}{\nolinkurl{https://arxiv.org/abs/#1}}}

\bibitem[{{Abdurro'uf} {et~al.}(2024){Abdurro'uf}, {Larson}, {Coe}, {Hsiao}, {{\'A}lvarez-M{\'a}rquez}, {G{\'o}mez}, {Adamo}, {Bhatawdekar}, {Bik}, {Bradley}, {Conselice}, {Dayal}, {Diego}, {Fujimoto}, {Furtak}, {Hutchison}, {Jung}, {Killi}, {Kokorev}, {Mingozzi}, {Norman}, {Resseguier}, {Ricotti}, {Rigby}, {Vanzella}, {Welch}, {Windhorst}, {Xu}, \& {Zitrin}}]{Abdurrouf.2024}
{Abdurro'uf}, {Larson}, R.~L., {Coe}, D., {et~al.} 2024, \apj, 973, 47, \dodoi{10.3847/1538-4357/ad6001}

\bibitem[{{Acharyya} {et~al.}(2025){Acharyya}, {Peeples}, {Tumlinson}, {O'Shea}, {Lochhaas}, {Wright}, {Simons}, {Augustin}, {Smith}, \& {Lee}}]{Acharyya.2025}
{Acharyya}, A., {Peeples}, M.~S., {Tumlinson}, J., {et~al.} 2025, \apj, 979, 129, \dodoi{10.3847/1538-4357/ad9dd8}

\bibitem[{{Adamo} {et~al.}(2025){Adamo}, {Atek}, {Bagley}, {Ba{\~n}ados}, {Barrow}, {Berg}, {Bezanson}, {Brada{\v{c}}}, {Brammer}, {Carnall}, {Chisholm}, {Coe}, {Dayal}, {Eisenstein}, {Eldridge}, {Ferrara}, {Fujimoto}, {Graaff}, {Habouzit}, {Hutchison}, {Kartaltepe}, {Kassin}, {Kriek}, {Labb{\'e}}, {Maiolino}, {Marques-Chaves}, {Maseda}, {Mason}, {Matthee}, {McQuinn}, {Meynet}, {Naidu}, {Oesch}, {Pentericci}, {P{\'e}rez-Gonz{\'a}lez}, {Rigby}, {Roberts-Borsani}, {Schaerer}, {Shapley}, {Stark}, {Stiavelli}, {Strom}, {Vanzella}, {Wang}, {Wilkins}, {Williams}, {Willott}, {Wylezalek}, \& {Nota}}]{Adamo.2025}
{Adamo}, A., {Atek}, H., {Bagley}, M.~B., {et~al.} 2025, Nature Astronomy, 9, 1134, \dodoi{10.1038/s41550-025-02624-5}

\bibitem[{{Antwi-Danso} {et~al.}(2023){Antwi-Danso}, {Papovich}, {Leja}, {Marchesini}, {Marsan}, {Martis}, {Labb{\'e}}, {Muzzin}, {Glazebrook}, {Straatman}, \& {Tran}}]{Antwi-Danso.2023}
{Antwi-Danso}, J., {Papovich}, C., {Leja}, J., {et~al.} 2023, \apj, 943, 166, \dodoi{10.3847/1538-4357/aca294}

\bibitem[{{Backhaus} {et~al.}(2022){Backhaus}, {Trump}, {Cleri}, {Simons}, {Momcheva}, {Papovich}, {Estrada-Carpenter}, {Finkelstein}, {Matharu}, {Ji}, {Weiner}, {Giavalisco}, \& {Jung}}]{Backhaus.2022}
{Backhaus}, B.~E., {Trump}, J.~R., {Cleri}, N.~J., {et~al.} 2022, \apj, 926, 161, \dodoi{10.3847/1538-4357/ac3919}

\bibitem[{{Backhaus} {et~al.}(2024){Backhaus}, {Trump}, {Pirzkal}, {Barro}, {Finkelstein}, {Arrabal Haro}, {Simons}, {Wessner}, {Cleri}, {Bagley}, {Hirschmann}, {Nicholls}, {Dickinson}, {Kartaltepe}, {Papovich}, {Kocevski}, {Koekemoer}, {Bisigello}, {Jaskot}, {Lucas}, {Jung}, {Wilkins}, {Yung}, {Ferguson}, {Fontana}, {Grazian}, {Grogin}, {Kewley}, {Kirkpatrick}, {Lotz}, {Pentericci}, {P{\'e}rez-Gonz{\'a}lez}, {Ravindranath}, {Somerville}, {Yang}, {Holwerda}, {Kurczynski}, {Hathi}, {Rose}, \& {Davis}}]{Backhaus.2024}
{Backhaus}, B.~E., {Trump}, J.~R., {Pirzkal}, N., {et~al.} 2024, \apj, 962, 195, \dodoi{10.3847/1538-4357/ad1520}

\bibitem[{{Baldwin} {et~al.}(1981){Baldwin}, {Phillips}, \& {Terlevich}}]{Baldwin.1981}
{Baldwin}, J.~A., {Phillips}, M.~M., \& {Terlevich}, R. 1981, \pasp, 93, 5, \dodoi{10.1086/130766}

\bibitem[{{Bari{\v{s}}i{\'c}} {et~al.}(2025){Bari{\v{s}}i{\'c}}, {Jones}, {Mortensen}, {Nanayakkara}, {Chen}, {Sanders}, {Bullock}, {Bundy}, {Faucher-Gigu{\`e}re}, {Glazebrook}, {Henry}, {Ju}, {Malkan}, {Morishita}, {Obreschkow}, {Roy}, {Espejo Salcedo}, {Shapley}, {Treu}, {Wang}, \& {Westfall}}]{Barisic.2025}
{Bari{\v{s}}i{\'c}}, I., {Jones}, T., {Mortensen}, K., {et~al.} 2025, \apj, 983, 139, \dodoi{10.3847/1538-4357/ada617}

\bibitem[{{Berg} {et~al.}(2018){Berg}, {Erb}, {Auger}, {Pettini}, \& {Brammer}}]{Berg.2018}
{Berg}, D.~A., {Erb}, D.~K., {Auger}, M.~W., {Pettini}, M., \& {Brammer}, G.~B. 2018, \apj, 859, 164, \dodoi{10.3847/1538-4357/aab7fa}

\bibitem[{{Bezanson} {et~al.}(2022){Bezanson}, {Spilker}, {Suess}, {Setton}, {Feldmann}, {Greene}, {Kriek}, {Narayanan}, \& {Verrico}}]{bezanson2022}
{Bezanson}, R., {Spilker}, J.~S., {Suess}, K.~A., {et~al.} 2022, \apj, 925, 153, \dodoi{10.3847/1538-4357/ac3dfa}

\bibitem[{{Bian} {et~al.}(2016){Bian}, {Kewley}, {Dopita}, \& {Juneau}}]{Bian.2016}
{Bian}, F., {Kewley}, L.~J., {Dopita}, M.~A., \& {Juneau}, S. 2016, \apj, 822, 62, \dodoi{10.3847/0004-637X/822/2/62}

\bibitem[{{Bian} {et~al.}(2010){Bian}, {Fan}, {Bechtold}, {McGreer}, {Just}, {Sand}, {Green}, {Thompson}, {Peng}, {Seifert}, {Ageorges}, {Juette}, {Knierim}, \& {Buschkamp}}]{Bian.2010}
{Bian}, F., {Fan}, X., {Bechtold}, J., {et~al.} 2010, \apj, 725, 1877, \dodoi{10.1088/0004-637X/725/2/1877}

\bibitem[{{Birkin} {et~al.}(2023){Birkin}, {Hutchison}, {Welch}, {Spilker}, {Aravena}, {Bayliss}, {Cathey}, {Chapman}, {Gonzalez}, {Gururajan}, {Hayward}, {Khullar}, {Kim}, {Mahler}, {Malkan}, {Narayanan}, {Olivier}, {Phadke}, {Reuter}, {Rigby}, {Smith}, {Solimano}, {Sulzenauer}, {Vieira}, {Vizgan}, \& {Weiss}}]{Birkin.2023}
{Birkin}, J.~E., {Hutchison}, T.~A., {Welch}, B., {et~al.} 2023, \apj, 958, 64, \dodoi{10.3847/1538-4357/acf712}

\bibitem[{{Boardman} {et~al.}(2023){Boardman}, {Wild}, {Heckman}, {Sanchez}, {Riffel}, {Riffel}, \& {Zasowski}}]{Boardman.2023}
{Boardman}, N., {Wild}, V., {Heckman}, T., {et~al.} 2023, \mnras, 520, 4301, \dodoi{10.1093/mnras/stad277}

\bibitem[{{B{\"o}ker} {et~al.}(2023){B{\"o}ker}, {Beck}, {Birkmann}, {Giardino}, {Keyes}, {Kumari}, {Muzerolle}, {Rawle}, {Zeidler}, {Abul-Huda}, {Alves de Oliveira}, {Arribas}, {Bechtold}, {Bhatawdekar}, {Bonaventura}, {Bunker}, {Cameron}, {Carniani}, {Charlot}, {Curti}, {Espinoza}, {Ferruit}, {Franx}, {Jakobsen}, {Karakla}, {L{\'o}pez-Caniego}, {L{\"u}tzgendorf}, {Maiolino}, {Manjavacas}, {Marston}, {Moseley}, {Ogle}, {Perna}, {Pe{\~n}a-Guerrero}, {Pirzkal}, {Plesha}, {Proffitt}, {Rauscher}, {Rix}, {Rodr{\'\i}guez del Pino}, {Rustamkulov}, {Sabbi}, {Sing}, {Sirianni}, {te Plate}, {{\'U}beda}, {Wahlgren}, {Wislowski}, {Wu}, \& {Willott}}]{Boker.2023}
{B{\"o}ker}, T., {Beck}, T.~L., {Birkmann}, S.~M., {et~al.} 2023, \pasp, 135, 038001, \dodoi{10.1088/1538-3873/acb846}

\bibitem[{{Bracci} {et~al.}(2025){Bracci}, {Belfiore}, {Ginolfi}, {Feltre}, {Mannucci}, {Marconi}, {Cresci}, {Bertola}, {Bombini}, {Ceci}, {Marconcini}, {Moreschini}, {Scialpi}, {Tozzi}, {Ulivi}, \& {Venturi}}]{Bracci.2025}
{Bracci}, C., {Belfiore}, F., {Ginolfi}, M., {et~al.} 2025, \aap, 697, A148, \dodoi{10.1051/0004-6361/202554196}

\bibitem[{{Cabanac} {et~al.}(2005){Cabanac}, {Valls-Gabaud}, {Jaunsen}, {Lidman}, \& {Jerjen}}]{Cabanac2005}
{Cabanac}, R.~A., {Valls-Gabaud}, D., {Jaunsen}, A.~O., {Lidman}, C., \& {Jerjen}, H. 2005, \aap, 436, L21, \dodoi{10.1051/0004-6361:200500115}

\bibitem[{{Calzetti}(2013)}]{Calzetti.2013}
{Calzetti}, D. 2013, in Secular Evolution of Galaxies, ed. J.~{Falc{\'o}n-Barroso} \& J.~H. {Knapen}, 419, \dodoi{10.48550/arXiv.1208.2997}

\bibitem[{{Cameron} {et~al.}(2023){Cameron}, {Katz}, \& {Rey}}]{Cameron.2023}
{Cameron}, A.~J., {Katz}, H., \& {Rey}, M.~P. 2023, \mnras, 522, L89, \dodoi{10.1093/mnrasl/slad046}

\bibitem[{Carnall {et~al.}(2024)Carnall, Cullen, McLure, McLeod, Begley, Donnan, Dunlop, Shapley, Rowlands, Almaini, Arellano-Córdova, Barrufet, Cimatti, Ellis, Grogin, Hamadouche, Illingworth, Koekemoer, Leung, Lovell, Pérez-González, Santini, Stanton, \& Wild}]{carnall2024}
Carnall, A.~C., Cullen, F., McLure, R.~J., {et~al.} 2024, Monthly Notices of the Royal Astronomical Society, 534, 325, \dodoi{10.1093/mnras/stae2092}

\bibitem[{{Cerny} {et~al.}(2025){Cerny}, {Mahler}, {Sharon}, {Jauzac}, {Khullar}, {Beauchesne}, {Diego}, {Lagattuta}, {Limousin}, {Patel}, {Richard}, {Cornil-Baiotto}, {Gladders}, {Werner}, {Doppel}, {Floyd}, {Gonzalez}, {Massey}, {Montes}, {Bayliss}, {Bleem}, {Canning}, {Edge}, {McDonald}, {Natarjan}, {Stark}, \& {Gassis}}]{cerny2025}
{Cerny}, C., {Mahler}, G., {Sharon}, K., {et~al.} 2025, arXiv e-prints, arXiv:2503.17498, \dodoi{10.48550/arXiv.2503.17498}

\bibitem[{{Chabrier}(2003)}]{Chabrier2003}
{Chabrier}, G. 2003, \pasp, 115, 763, \dodoi{10.1086/376392}

\bibitem[{{Chen} {et~al.}(2019){Chen}, {Shi}, {Wild}, {Tremonti}, {Rowlands}, {Bizyaev}, {Yan}, {Lin}, \& {Riffel}}]{chen2019}
{Chen}, Y.-M., {Shi}, Y., {Wild}, V., {et~al.} 2019, \mnras, 489, 5709, \dodoi{10.1093/mnras/stz2494}

\bibitem[{{Cheng} {et~al.}(2024){Cheng}, {Li}, {Li}, {Yan}, \& {Mo}}]{cheng2024}
{Cheng}, Z., {Li}, C., {Li}, N., {Yan}, R., \& {Mo}, H. 2024, \apj, 961, 216, \dodoi{10.3847/1538-4357/ad1510}

\bibitem[{{Chiosi}(1980)}]{Chiosi.1980}
{Chiosi}, C. 1980, \aap, 83, 206

\bibitem[{{Choe} {et~al.}(2025){Choe}, {Emil Rivera-Thorsen}, {Dahle}, {Sharon}, {Owens}, {Rigby}, {Bayliss}, {Hayes}, {Hutchison}, {Welch}, {Chisholm}, {Gladders}, {Khullar}, \& {Kim}}]{Choe.2025}
{Choe}, S., {Emil Rivera-Thorsen}, T., {Dahle}, H., {et~al.} 2025, \aap, 698, A16, \dodoi{10.1051/0004-6361/202450685}

\bibitem[{{Choi} {et~al.}(2016){Choi}, {Dotter}, {Conroy}, {Cantiello}, {Paxton}, \& {Johnson}}]{Choi2016}
{Choi}, J., {Dotter}, A., {Conroy}, C., {et~al.} 2016, \apj, 823, 102, \dodoi{10.3847/0004-637X/823/2/102}

\bibitem[{{Christensen} {et~al.}(2023){Christensen}, {Jakobsen}, {Willott}, {Arribas}, {Bunker}, {Charlot}, {Maiolino}, {Marshall}, {Perna}, \& {{\"U}bler}}]{Christensen.2023}
{Christensen}, L., {Jakobsen}, P., {Willott}, C., {et~al.} 2023, \aap, 680, A82, \dodoi{10.1051/0004-6361/202347943}

\bibitem[{{Ciesla} {et~al.}(2024){Ciesla}, {Elbaz}, {Ilbert}, {Buat}, {Magnelli}, {Narayanan}, {Daddi}, {G{\'o}mez-Guijarro}, \& {Arango-Toro}}]{ciesla2024}
{Ciesla}, L., {Elbaz}, D., {Ilbert}, O., {et~al.} 2024, \aap, 686, A128, \dodoi{10.1051/0004-6361/202348091}

\bibitem[{{Clarendon} {et~al.}(2025){Clarendon}, {Strom}, \& {von Raesfeld}}]{Clarendon.2025}
{Clarendon}, A., {Strom}, A., \& {von Raesfeld}, C. 2025, Research Notes of the American Astronomical Society, 9, 206, \dodoi{10.3847/2515-5172/adf4c5}

\bibitem[{{Cleri} {et~al.}(2025){Cleri}, {Olivier}, {Backhaus}, {Leja}, {Papovich}, {Trump}, {Arrabal Haro}, {Buat}, {Burgarella}, {Burnham}, {Calabro}, {Cohn}, {Cole}, {Davis}, {Dickinson}, {Finkelstein}, {Garner}, {Hirschmann}, {Hu}, {Hutchison}, {Kocevski}, {Koekemoer}, {Larson}, {Lewis}, {Maseda}, {Seille}, \& {Simons}}]{Cleri.2025}
{Cleri}, N.~J., {Olivier}, G.~M., {Backhaus}, B.~E., {et~al.} 2025, arXiv e-prints, arXiv:2506.21660, \dodoi{10.48550/arXiv.2506.21660}

\bibitem[{{Covelo-Paz} {et~al.}(2025){Covelo-Paz}, {Meuwly}, {Oesch}, {Witten}, {Weibel}, {Carvajal-Bohorquez}, {Ciesla}, {Giovinazzo}, \& {Brammer}}]{covelo2025}
{Covelo-Paz}, A., {Meuwly}, C., {Oesch}, P.~A., {et~al.} 2025, arXiv e-prints, arXiv:2506.22540, \dodoi{10.48550/arXiv.2506.22540}

\bibitem[{{Curti} {et~al.}(2020){Curti}, {Maiolino}, {Cirasuolo}, {Mannucci}, {Williams}, {Auger}, {Mercurio}, {Hayden-Pawson}, {Cresci}, {Marconi}, {Belfiore}, {Cappellari}, {Cicone}, {Cullen}, {Meneghetti}, {Ota}, {Peng}, {Pettini}, {Swinbank}, \& {Troncoso}}]{Curti.2020}
{Curti}, M., {Maiolino}, R., {Cirasuolo}, M., {et~al.} 2020, \mnras, 492, 821, \dodoi{10.1093/mnras/stz3379}

\bibitem[{{Curti} {et~al.}(2023){Curti}, {D'Eugenio}, {Carniani}, {Maiolino}, {Sandles}, {Witstok}, {Baker}, {Bennett}, {Piotrowska}, {Tacchella}, {Charlot}, {Nakajima}, {Maheson}, {Mannucci}, {Amiri}, {Arribas}, {Belfiore}, {Bonaventura}, {Bunker}, {Chevallard}, {Cresci}, {Curtis-Lake}, {Hayden-Pawson}, {Jones}, {Kumari}, {Laseter}, {Looser}, {Marconi}, {Maseda}, {Scholtz}, {Smit}, {{\"U}bler}, \& {Wallace}}]{Curti.2023}
{Curti}, M., {D'Eugenio}, F., {Carniani}, S., {et~al.} 2023, \mnras, 518, 425, \dodoi{10.1093/mnras/stac2737}

\bibitem[{{de Graaff} {et~al.}(2025){de Graaff}, {Setton}, {Brammer}, {Cutler}, {Suess}, {Labb{\'e}}, {Leja}, {Weibel}, {Maseda}, {Whitaker}, {Bezanson}, {Boogaard}, {Cleri}, {De Lucia}, {Franx}, {Greene}, {Hirschmann}, {Matthee}, {McConachie}, {Naidu}, {Oesch}, {Price}, {Rix}, {Valentino}, {Wang}, \& {Williams}}]{degraaf2025}
{de Graaff}, A., {Setton}, D.~J., {Brammer}, G., {et~al.} 2025, Nature Astronomy, 9, 280, \dodoi{10.1038/s41550-024-02424-3}

\bibitem[{{D'Eugenio} {et~al.}(2021){D'Eugenio}, {Daddi}, {Gobat}, {Strazzullo}, {Lustig}, {Delvecchio}, {Jin}, {Cimatti}, \& {Onodera}}]{dugenio2021}
{D'Eugenio}, C., {Daddi}, E., {Gobat}, R., {et~al.} 2021, \aap, 653, A32, \dodoi{10.1051/0004-6361/202040067}

\bibitem[{{D'Eugenio} {et~al.}(2024){D'Eugenio}, {P{\'e}rez-Gonz{\'a}lez}, {Maiolino}, {Scholtz}, {Perna}, {Circosta}, {{\"U}bler}, {Arribas}, {B{\"o}ker}, {Bunker}, {Carniani}, {Charlot}, {Chevallard}, {Cresci}, {Curtis-Lake}, {Jones}, {Kumari}, {Lamperti}, {Looser}, {Parlanti}, {Rix}, {Robertson}, {Rodr{\'\i}guez Del Pino}, {Tacchella}, {Venturi}, \& {Willott}}]{dugenio2024}
{D'Eugenio}, F., {P{\'e}rez-Gonz{\'a}lez}, P.~G., {Maiolino}, R., {et~al.} 2024, Nature Astronomy, 8, 1443, \dodoi{10.1038/s41550-024-02345-1}

\bibitem[{{Dey} {et~al.}(2019){Dey}, {Schlegel}, {Lang}, {Blum}, {Burleigh}, {Fan}, {Findlay}, {Finkbeiner}, {Herrera}, {Juneau}, {Landriau}, {Levi}, {McGreer}, {Meisner}, {Myers}, {Moustakas}, {Nugent}, {Patej}, {Schlafly}, {Walker}, {Valdes}, {Weaver}, {Y{\`e}che}, {Zou}, {Zhou}, {Abareshi}, {Abbott}, {Abolfathi}, {Aguilera}, {Alam}, {Allen}, {Alvarez}, {Annis}, {Ansarinejad}, {Aubert}, {Beechert}, {Bell}, {BenZvi}, {Beutler}, {Bielby}, {Bolton}, {Brice{\~n}o}, {Buckley-Geer}, {Butler}, {Calamida}, {Carlberg}, {Carter}, {Casas}, {Castander}, {Choi}, {Comparat}, {Cukanovaite}, {Delubac}, {DeVries}, {Dey}, {Dhungana}, {Dickinson}, {Ding}, {Donaldson}, {Duan}, {Duckworth}, {Eftekharzadeh}, {Eisenstein}, {Etourneau}, {Fagrelius}, {Farihi}, {Fitzpatrick}, {Font-Ribera}, {Fulmer}, {G{\"a}nsicke}, {Gaztanaga}, {George}, {Gerdes}, {Gontcho}, {Gorgoni}, {Green}, {Guy}, {Harmer}, {Hernandez}, {Honscheid}, {Huang}, {James}, {Jannuzi}, {Jiang}, {Joyce}, {Karcher}, {Karkar}, {Kehoe}, {Kneib}, {Kueter-Young}, {Lan},
  {Lauer}, {Le Guillou}, {Le Van Suu}, {Lee}, {Lesser}, {Perreault Levasseur}, {Li}, {Mann}, {Marshall}, {Mart{\'\i}nez-V{\'a}zquez}, {Martini}, {du Mas des Bourboux}, {McManus}, {Meier}, {M{\'e}nard}, {Metcalfe}, {Mu{\~n}oz-Guti{\'e}rrez}, {Najita}, {Napier}, {Narayan}, {Newman}, {Nie}, {Nord}, {Norman}, {Olsen}, {Paat}, {Palanque-Delabrouille}, {Peng}, {Poppett}, {Poremba}, {Prakash}, {Rabinowitz}, {Raichoor}, {Rezaie}, {Robertson}, {Roe}, {Ross}, {Ross}, {Rudnick}, {Safonova}, {Saha}, {S{\'a}nchez}, {Savary}, {Schweiker}, {Scott}, {Seo}, {Shan}, {Silva}, {Slepian}, {Soto}, {Sprayberry}, {Staten}, {Stillman}, {Stupak}, {Summers}, {Sien Tie}, {Tirado}, {Vargas-Maga{\~n}a}, {Vivas}, {Wechsler}, {Williams}, {Yang}, {Yang}, {Yapici}, {Zaritsky}, {Zenteno}, {Zhang}, {Zhang}, {Zhou}, \& {Zhou}}]{2019AJ....157..168D}
{Dey}, A., {Schlegel}, D.~J., {Lang}, D., {et~al.} 2019, \aj, 157, 168, \dodoi{10.3847/1538-3881/ab089d}

\bibitem[{{Dressler} \& {Gunn}(1983)}]{dressler1983}
{Dressler}, A., \& {Gunn}, J.~E. 1983, \apj, 270, 7, \dodoi{10.1086/161093}

\bibitem[{{Dunlop} {et~al.}(2012){Dunlop}, {McLure}, {Robertson}, {Ellis}, {Stark}, {Cirasuolo}, \& {de Ravel}}]{Dunlop.2012}
{Dunlop}, J.~S., {McLure}, R.~J., {Robertson}, B.~E., {et~al.} 2012, \mnras, 420, 901, \dodoi{10.1111/j.1365-2966.2011.20102.x}

\bibitem[{{Fassnacht} {et~al.}(1996){Fassnacht}, {Womble}, {Neugebauer}, {Browne}, {Readhead}, {Matthews}, \& {Pearson}}]{Fassnacht}
{Fassnacht}, C.~D., {Womble}, D.~S., {Neugebauer}, G., {et~al.} 1996, \apjl, 460, L103, \dodoi{10.1086/309984}

\bibitem[{{Feltre} {et~al.}(2016){Feltre}, {Charlot}, \& {Gutkin}}]{Feltre.2016}
{Feltre}, A., {Charlot}, S., \& {Gutkin}, J. 2016, \mnras, 456, 3354, \dodoi{10.1093/mnras/stv2794}

\bibitem[{{Finkelstein}(2016)}]{Finkelstein.2016}
{Finkelstein}, S.~L. 2016, \pasa, 33, e037, \dodoi{10.1017/pasa.2016.26}

\bibitem[{{Finley} {et~al.}(2017){Finley}, {Bouch{\'e}}, {Contini}, {Paalvast}, {Boogaard}, {Maseda}, {Bacon}, {Blaizot}, {Brinchmann}, {Epinat}, {Feltre}, {Marino}, {Muzahid}, {Richard}, {Schaye}, {Verhamme}, {Weilbacher}, \& {Wisotzki}}]{Finley.2017}
{Finley}, H., {Bouch{\'e}}, N., {Contini}, T., {et~al.} 2017, \aap, 608, A7, \dodoi{10.1051/0004-6361/201731499}

\bibitem[{{Florian} {et~al.}(2025){Florian}, {Gladders}, {Khullar}, {Sharon}, {Cloonan}, {Solhaug}, {Welch}, {Bayliss}, {Dahle}, {Hutchison}, \& {Rigby}}]{Florian.2025}
{Florian}, M.~K., {Gladders}, M.~D., {Khullar}, G., {et~al.} 2025, arXiv e-prints, arXiv:2510.10376, \dodoi{10.48550/arXiv.2510.10376}

\bibitem[{{French}(2021)}]{french2021}
{French}, K.~D. 2021, \pasp, 133, 072001, \dodoi{10.1088/1538-3873/ac0a59}

\bibitem[{{Fujimoto} {et~al.}(2023){Fujimoto}, {Arrabal Haro}, {Dickinson}, {Finkelstein}, {Kartaltepe}, {Larson}, {Burgarella}, {Bagley}, {Behroozi}, {Chworowsky}, {Hirschmann}, {Trump}, {Wilkins}, {Yung}, {Koekemoer}, {Papovich}, {Pirzkal}, {Ferguson}, {Fontana}, {Grogin}, {Grazian}, {Kewley}, {Kocevski}, {Lotz}, {Pentericci}, {Ravindranath}, {Somerville}, {Wilkins}, {Amor{\'\i}n}, {Backhaus}, {Calabr{\`o}}, {Casey}, {Cooper}, {Fern{\'a}ndez}, {Franco}, {Giavalisco}, {Hathi}, {Harish}, {Hutchison}, {Iyer}, {Jung}, {Lucas}, \& {Zavala}}]{Fujimoto.2023}
{Fujimoto}, S., {Arrabal Haro}, P., {Dickinson}, M., {et~al.} 2023, \apjl, 949, L25, \dodoi{10.3847/2041-8213/acd2d9}

\bibitem[{{Garcia} {et~al.}(2025a){Garcia}, {Torrey}, {Bhagwat}, {Wright}, {Chen}, {Grasha}, {Ridolfo}, {Hemler}, {Sarkar}, {Chakraborty}, {Nelson}, {Sanders}, {Costa}, {Vogelsberger}, {Kewley}, {Ellison}, \& {Hernquist}}]{Garcia.2025a}
{Garcia}, A.~M., {Torrey}, P., {Bhagwat}, A., {et~al.} 2025a, \apj, 989, 147, \dodoi{10.3847/1538-4357/adea51}

\bibitem[{{Garcia} {et~al.}(2025b){Garcia}, {Torrey}, {Bhagwat}, {Shen}, {Vogelsberger}, {McClymont}, {Nagarajan-Swenson}, {Ridolfo}, {Zhu}, {Zimmerman}, {Zier}, {Biddle}, {Sarkar}, {Chakraborty}, {Wright}, {Grasha}, {Costa}, {Keating}, {Kannan}, {Smith}, {Garaldi}, {Puchwein}, {Ciardi}, {Hernquist}, \& {Kewley}}]{Garcia.2025b}
---. 2025b, arXiv e-prints, arXiv:2510.26877, \dodoi{10.48550/arXiv.2510.26877}

\bibitem[{{Garg} {et~al.}(2024){Garg}, {Narayanan}, {Sanders}, {Dav{\'e}}, {Popping}, {Shapley}, {Stark}, \& {Trump}}]{Garg.2024}
{Garg}, P., {Narayanan}, D., {Sanders}, R.~L., {et~al.} 2024, \apj, 972, 113, \dodoi{10.3847/1538-4357/ad5ae1}

\bibitem[{{Garg} {et~al.}(2022){Garg}, {Narayanan}, {Byler}, {Sanders}, {Shapley}, {Strom}, {Dav{\'e}}, {Hirschmann}, {Lovell}, {Otter}, {Popping}, \& {Privon}}]{Garg.2022}
{Garg}, P., {Narayanan}, D., {Byler}, N., {et~al.} 2022, \apj, 926, 80, \dodoi{10.3847/1538-4357/ac43b8}

\bibitem[{{Gburek} {et~al.}(2019){Gburek}, {Siana}, {Alavi}, {Emami}, {Richard}, {Freeman}, {Stark}, {Snapp-Kolas}, \& {Lucero}}]{Gburek.2019}
{Gburek}, T., {Siana}, B., {Alavi}, A., {et~al.} 2019, \apj, 887, 168, \dodoi{10.3847/1538-4357/ab5713}

\bibitem[{{Glazebrook} {et~al.}(2017){Glazebrook}, {Schreiber}, {Labb{\'e}}, {Nanayakkara}, {Kacprzak}, {Oesch}, {Papovich}, {Spitler}, {Straatman}, {Tran}, \& {Yuan}}]{Glazebrook.2017}
{Glazebrook}, K., {Schreiber}, C., {Labb{\'e}}, I., {et~al.} 2017, \nat, 544, 71, \dodoi{10.1038/nature21680}

\bibitem[{{Goto}(2004)}]{goto2004}
{Goto}, T. 2004, in Proceedings of the 6th RESCEU International Symposium, 1, \dodoi{10.48550/arXiv.astro-ph/0411519}

\bibitem[{{Graf} {et~al.}(2025){Graf}, {Wetzel}, {Bailin}, \& {Orr}}]{Graf.2025}
{Graf}, R.~L., {Wetzel}, A., {Bailin}, J., \& {Orr}, M.~E. 2025, \apj, 991, 139, \dodoi{10.3847/1538-4357/adfa07}

\bibitem[{{Hainline} {et~al.}(2009){Hainline}, {Shapley}, {Kornei}, {Pettini}, {Buckley-Geer}, {Allam}, \& {Tucker}}]{Hainline.2009}
{Hainline}, K.~N., {Shapley}, A.~E., {Kornei}, K.~A., {et~al.} 2009, \apj, 701, 52, \dodoi{10.1088/0004-637X/701/1/52}

\bibitem[{{Heintz} {et~al.}(2023){Heintz}, {Gim{\'e}nez-Arteaga}, {Fujimoto}, {Brammer}, {Espada}, {Gillman}, {Gonz{\'a}lez-L{\'o}pez}, {Greve}, {Harikane}, {Hatsukade}, {Knudsen}, {Koekemoer}, {Kohno}, {Kokorev}, {Lee}, {Magdis}, {Nelson}, {Rizzo}, {Sanders}, {Schaerer}, {Shapley}, {Strait}, {Toft}, {Valentino}, {van der Wel}, {Vijayan}, {Watson}, {Bauer}, {Christiansen}, \& {Wilson}}]{Heintz.2023}
{Heintz}, K.~E., {Gim{\'e}nez-Arteaga}, C., {Fujimoto}, S., {et~al.} 2023, \apjl, 944, L30, \dodoi{10.3847/2041-8213/acb2cf}

\bibitem[{{Hemler} {et~al.}(2021){Hemler}, {Torrey}, {Qi}, {Hernquist}, {Vogelsberger}, {Ma}, {Kewley}, {Nelson}, {Pillepich}, {Pakmor}, \& {Marinacci}}]{Hemler.2021}
{Hemler}, Z.~S., {Torrey}, P., {Qi}, J., {et~al.} 2021, \mnras, 506, 3024, \dodoi{10.1093/mnras/stab1803}

\bibitem[{{Henry} {et~al.}(2000){Henry}, {Edmunds}, \& {K{\"o}ppen}}]{Henry.2000}
{Henry}, R.~B.~C., {Edmunds}, M.~G., \& {K{\"o}ppen}, J. 2000, \apj, 541, 660, \dodoi{10.1086/309471}

\bibitem[{{Hinshaw} {et~al.}(2013){Hinshaw}, {Larson}, {Komatsu}, {Spergel}, {Bennett}, {Dunkley}, {Nolta}, {Halpern}, {Hill}, {Odegard}, {Page}, {Smith}, {Weiland}, {Gold}, {Jarosik}, {Kogut}, {Limon}, {Meyer}, {Tucker}, {Wollack}, \& {Wright}}]{2013ApJS..208...19H}
{Hinshaw}, G., {Larson}, D., {Komatsu}, E., {et~al.} 2013, \apjs, 208, 19, \dodoi{10.1088/0067-0049/208/2/19}

\bibitem[{{Hsiao} {et~al.}(2024a){Hsiao}, {Abdurro'uf}, {Coe}, {Larson}, {Jung}, {Mingozzi}, {Dayal}, {Kumari}, {Kokorev}, {Vikaeus}, {Brammer}, {Furtak}, {Adamo}, {Andrade-Santos}, {Antwi-Danso}, {Brada{\v{c}}}, {Bradley}, {Broadhurst}, {Carnall}, {Conselice}, {Diego}, {Donahue}, {Eldridge}, {Fujimoto}, {Henry}, {Hernandez}, {Hutchison}, {James}, {Norman}, {Park}, {Pirzkal}, {Postman}, {Ricotti}, {Rigby}, {Vanzella}, {Welch}, {Wilkins}, {Windhorst}, {Xu}, {Zackrisson}, \& {Zitrin}}]{Hsiao.2024a}
{Hsiao}, T. Y.-Y., {Abdurro'uf}, {Coe}, D., {et~al.} 2024a, \apj, 973, 8, \dodoi{10.3847/1538-4357/ad5da8}

\bibitem[{{Hsiao} {et~al.}(2024b){Hsiao}, {{\'A}lvarez-M{\'a}rquez}, {Coe}, {Crespo G{\'o}mez}, {Abdurro'uf}, {Dayal}, {Larson}, {Bik}, {Blanco-Prieto}, {Colina}, {P{\'e}rez-Gonz{\'a}lez}, {Costantin}, {Prieto-Jim{\'e}nez}, {Adamo}, {Bradley}, {Conselice}, {Fujimoto}, {Furtak}, {Hutchison}, {James}, {Jim{\'e}nez-Teja}, {Jung}, {Kokorev}, {Mingozzi}, {Norman}, {Ricotti}, {Rigby}, {Sharon}, {Vanzella}, {Welch}, {Xu}, {Zackrisson}, \& {Zitrin}}]{Hsiao.2024b}
{Hsiao}, T. Y.-Y., {{\'A}lvarez-M{\'a}rquez}, J., {Coe}, D., {et~al.} 2024b, \apj, 973, 81, \dodoi{10.3847/1538-4357/ad6562}

\bibitem[{{Hunter} \& {Gallagher}(1990)}]{Hunter1990}
{Hunter}, D.~A., \& {Gallagher}, III, J.~S. 1990, \apj, 362, 480, \dodoi{10.1086/169286}

\bibitem[{{Hutchison} {et~al.}(2024){Hutchison}, {Welch}, {Rigby}, {Olivier}, {Birkin}, {Phadke}, {Khullar}, {Rauscher}, {Sharon}, {Aravena}, {Bayliss}, {Elicker}, {Kim}, {Solimano}, {Vieira}, {Vizgan}, \& {Jwst Templates Early Release Science Team}}]{Hutchison.2024}
{Hutchison}, T.~A., {Welch}, B.~D., {Rigby}, J.~R., {et~al.} 2024, \pasp, 136, 044503, \dodoi{10.1088/1538-3873/ad34fd}

\bibitem[{{Ibrahim} \& {Kobayashi}(2025)}]{Ibrahim.2025}
{Ibrahim}, D., \& {Kobayashi}, C. 2025, \mnras, 544, 815, \dodoi{10.1093/mnras/staf1727}

\bibitem[{{Isobe} {et~al.}(2023){Isobe}, {Ouchi}, {Nakajima}, {Harikane}, {Ono}, {Xu}, {Zhang}, \& {Umeda}}]{Isobe.2023}
{Isobe}, Y., {Ouchi}, M., {Nakajima}, K., {et~al.} 2023, \apj, 956, 139, \dodoi{10.3847/1538-4357/acf376}

\bibitem[{{Jones} {et~al.}(2025){Jones}, {Bunker}, {Telikova}, {Arribas}, {Carniani}, {Charlot}, {D'Eugenio}, {Maiolino}, {Perna}, {Rodr{\'\i}guez Del Pino}, {{\"U}bler}, {Willott}, {Aravena}, {B{\"o}ker}, {Cresci}, {Curti}, {Gonz{\'a}lez-L{\'o}pez}, {Herrera-Camus}, {Lamperti}, {Parlanti}, {P{\'e}rez-Gonz{\'a}lez}, \& {Villanueva}}]{Jones.2025}
{Jones}, G.~C., {Bunker}, A.~J., {Telikova}, K., {et~al.} 2025, \mnras, 540, 3311, \dodoi{10.1093/mnras/staf899}

\bibitem[{{Jones} {et~al.}(2023){Jones}, {Barisic}, {Bullock}, {Faucher-Giguere}, {Glazebrook}, {Henry}, {Malkan}, {Morishita}, {Mortensen}, {Nanayakkara}, {Sanders}, {Shapley}, {Treu}, \& {Wang}}]{piJones.2023}
{Jones}, T., {Barisic}, I., {Bullock}, J.~S., {et~al.} 2023, {Confirming the population of disk galaxies at z>3}, JWST Proposal. Cycle 2, ID. \#3426

\bibitem[{{Ju} {et~al.}(2025){Ju}, {Wang}, {Jones}, {Bari{\v{s}}i{\'c}}, {Nanayakkara}, {Bundy}, {Faucher-Gigu{\`e}re}, {Feng}, {Glazebrook}, {Henry}, {Malkan}, {Obreschkow}, {Roy}, {Sanders}, {Sun}, {Treu}, \& {Zhou}}]{Ju.2025}
{Ju}, M., {Wang}, X., {Jones}, T., {et~al.} 2025, \apjl, 978, L39, \dodoi{10.3847/2041-8213/ada150}

\bibitem[{{Kaasinen} {et~al.}(2017){Kaasinen}, {Bian}, {Groves}, {Kewley}, \& {Gupta}}]{Kaasinen.2017}
{Kaasinen}, M., {Bian}, F., {Groves}, B., {Kewley}, L.~J., \& {Gupta}, A. 2017, \mnras, 465, 3220, \dodoi{10.1093/mnras/stw2827}

\bibitem[{{Kashino} {et~al.}(2017){Kashino}, {More}, {Silverman}, {Daddi}, {Renzini}, {Sanders}, {Rodighiero}, {Puglisi}, {Kajisawa}, {Valentino}, {Kartaltepe}, {Le F{\`e}vre}, {Nagao}, {Arimoto}, \& {Sugiyama}}]{Kashino.2017}
{Kashino}, D., {More}, S., {Silverman}, J.~D., {et~al.} 2017, \apj, 843, 138, \dodoi{10.3847/1538-4357/aa789d}

\bibitem[{{Kassin} {et~al.}(2023){Kassin}, {Pacifici}, {Amorin}, {Arrabal Haro}, {Barro}, {Bovill}, {Dave}, {Dekel}, {Devriendt}, {Ferguson}, {Gardner}, {Giavalisco}, {Hathi}, {Holwerda}, {Huertas-Company}, {Iyer}, {Jogee}, {Kartaltepe}, {Kocevski}, {Koekemoer}, {Lucas}, {Madhani}, {Mulcahey}, {Ogle}, {Pandya}, {Perez-Gonzalez}, {Pichon}, {Pirzkal}, {Primack}, {Qin}, {Rafelski}, {Slyz}, {Sukay}, {Wang}, {Weiner}, {Welker}, {Willmer}, {de la Vega}, {Laigle}, \& {Kraljic}}]{piKassin.2023}
{Kassin}, S., {Pacifici}, C., {Amorin}, R., {et~al.} 2023, {Galaxy angular momentum alignment with filaments at z {\ensuremath{\sim}} 3: The effect of large scale structure on galaxies}, JWST Proposal. Cycle 2, ID. \#4291

\bibitem[{{Kauffmann} {et~al.}(2003){Kauffmann}, {Heckman}, {Tremonti}, {Brinchmann}, {Charlot}, {White}, {Ridgway}, {Brinkmann}, {Fukugita}, {Hall}, {Ivezi{\'c}}, {Richards}, \& {Schneider}}]{Kauffmann.2003}
{Kauffmann}, G., {Heckman}, T.~M., {Tremonti}, C., {et~al.} 2003, \mnras, 346, 1055, \dodoi{10.1111/j.1365-2966.2003.07154.x}

\bibitem[{{Kennicutt} \& {Evans}(2012)}]{Kennicutt.2012}
{Kennicutt}, R.~C., \& {Evans}, N.~J. 2012, \araa, 50, 531, \dodoi{10.1146/annurev-astro-081811-125610}

\bibitem[{{Kennicutt}(1998)}]{Kennicutt.1998}
{Kennicutt}, Jr., R.~C. 1998, \araa, 36, 189, \dodoi{10.1146/annurev.astro.36.1.189}

\bibitem[{{Kewley} \& {Dopita}(2002)}]{Kewley.2002}
{Kewley}, L.~J., \& {Dopita}, M.~A. 2002, \apjs, 142, 35, \dodoi{10.1086/341326}

\bibitem[{{Kewley} {et~al.}(2001){Kewley}, {Dopita}, {Sutherland}, {Heisler}, \& {Trevena}}]{Kewley.2001}
{Kewley}, L.~J., {Dopita}, M.~A., {Sutherland}, R.~S., {Heisler}, C.~A., \& {Trevena}, J. 2001, \apj, 556, 121, \dodoi{10.1086/321545}

\bibitem[{{Kewley} {et~al.}(2004){Kewley}, {Geller}, \& {Jansen}}]{Kewley.2004}
{Kewley}, L.~J., {Geller}, M.~J., \& {Jansen}, R.~A. 2004, \aj, 127, 2002, \dodoi{10.1086/382723}

\bibitem[{{Kewley} {et~al.}(2013){Kewley}, {Maier}, {Yabe}, {Ohta}, {Akiyama}, {Dopita}, \& {Yuan}}]{Kewley.2013}
{Kewley}, L.~J., {Maier}, C., {Yabe}, K., {et~al.} 2013, \apjl, 774, L10, \dodoi{10.1088/2041-8205/774/1/L10}

\bibitem[{{Kewley} {et~al.}(2019){Kewley}, {Nicholls}, \& {Sutherland}}]{Kewley.2019}
{Kewley}, L.~J., {Nicholls}, D.~C., \& {Sutherland}, R.~S. 2019, \araa, 57, 511, \dodoi{10.1146/annurev-astro-081817-051832}

\bibitem[{{Khoram} \& {Belfiore}(2025)}]{Khoram.2025}
{Khoram}, A.~H., \& {Belfiore}, F. 2025, \aap, 693, A150, \dodoi{10.1051/0004-6361/202451980}

\bibitem[{{Khullar} {et~al.}(2021){Khullar}, {Gozman}, {Lin}, {Martinez}, {Matthews Acu{\~n}a}, {Medina}, {Merz}, {Sanchez}, {Sisco}, {Kavin Stein}, {Sukay}, {Tavangar}, {Bayliss}, {Bleem}, {Brownsberger}, {Dahle}, {Florian}, {Gladders}, {Mahler}, {Rigby}, {Sharon}, \& {Stark}}]{Khullar.2021}
{Khullar}, G., {Gozman}, K., {Lin}, J.~J., {et~al.} 2021, \apj, 906, 107, \dodoi{10.3847/1538-4357/abcb86}

\bibitem[{{Klein} {et~al.}(2024){Klein}, {Sharon}, {Napier}, {Gladders}, {Khullar}, {Bayliss}, {Dahle}, {Owens}, {Stark}, {Brownsberger}, {Kim}, {Kuchta}, {Mahler}, {Smith}, {Walker}, {Gozman}, {Martinez}, {Matthews Acu{\~n}a}, {Merz}, {Sanchez}, {Stein}, {Sukay}, \& {Tavangar}}]{Klein.2024}
{Klein}, M., {Sharon}, K., {Napier}, K., {et~al.} 2024, \apj, 963, 44, \dodoi{10.3847/1538-4357/ad22de}

\bibitem[{{Kriek} \& {Conroy}(2013)}]{Kriek2013}
{Kriek}, M., \& {Conroy}, C. 2013, \apjl, 775, L16, \dodoi{10.1088/2041-8205/775/1/L16}

\bibitem[{{Larson} {et~al.}(2023){Larson}, {Hutchison}, {Bagley}, {Finkelstein}, {Yung}, {Somerville}, {Hirschmann}, {Brammer}, {Holwerda}, {Papovich}, {Morales}, \& {Wilkins}}]{Larson.2023}
{Larson}, R.~L., {Hutchison}, T.~A., {Bagley}, M., {et~al.} 2023, \apj, 958, 141, \dodoi{10.3847/1538-4357/acfed4}

\bibitem[{{Law} {et~al.}(2023){Law}, {E. Morrison}, {Argyriou}, {Patapis}, {{\'A}lvarez-M{\'a}rquez}, {Labiano}, \& {Vandenbussche}}]{law2023}
{Law}, D.~R., {E. Morrison}, J., {Argyriou}, I., {et~al.} 2023, \aj, 166, 45, \dodoi{10.3847/1538-3881/acdddc}

\bibitem[{{Leja} {et~al.}(2019){Leja}, {Carnall}, {Johnson}, {Conroy}, \& {Speagle}}]{leja2019}
{Leja}, J., {Carnall}, A.~C., {Johnson}, B.~D., {Conroy}, C., \& {Speagle}, J.~S. 2019, \apj, 876, 3, \dodoi{10.3847/1538-4357/ab133c}

\bibitem[{{Leja} {et~al.}(2022){Leja}, {Speagle}, {Ting}, {Johnson}, {Conroy}, {Whitaker}, {Nelson}, {van Dokkum}, \& {Franx}}]{Leja2022}
{Leja}, J., {Speagle}, J.~S., {Ting}, Y.-S., {et~al.} 2022, \apj, 936, 165, \dodoi{10.3847/1538-4357/ac887d}

\bibitem[{{Leung} {et~al.}(2025){Leung}, {Wild}, {Papathomas}, {Carnall}, \& {Chen}}]{Leung2025}
{Leung}, H.-H., {Wild}, V., {Papathomas}, M., {Carnall}, A.~C., \& {Chen}, Y. 2025, \mnras, 543, 738, \dodoi{10.1093/mnras/staf1493}

\bibitem[{{Long} {et~al.}(2024){Long}, {Antwi-Danso}, {Lambrides}, {Lovell}, {de la Vega}, {Valentino}, {Zavala}, {Casey}, {Wilkins}, {Yung}, {Arrabal Haro}, {Bagley}, {Bisigello}, {Chworowsky}, {Cooper}, {Cooper}, {Cooray}, {Croton}, {Dickinson}, {Finkelstein}, {Franco}, {Gould}, {Hirschmann}, {Hutchison}, {Kartaltepe}, {Kocevski}, {Koekemoer}, {Lucas}, {McKinney}, {Nere}, {Papovich}, {P{\'e}rez-Gonz{\'a}lez}, {Pirzkal}, \& {Santini}}]{long2024}
{Long}, A.~S., {Antwi-Danso}, J., {Lambrides}, E.~L., {et~al.} 2024, \apj, 970, 68, \dodoi{10.3847/1538-4357/ad4cea}

\bibitem[{{Looser} {et~al.}(2023{\natexlab{a}}){Looser}, {D'Eugenio}, {Maiolino}, {Tacchella}, {Curti}, {Arribas}, {Baker}, {Baum}, {Bonaventura}, {Boyett}, {Bunker}, {Carniani}, {Charlot}, {Chevallard}, {Curtis-Lake}, {Danhaive}, {Eisenstein}, {de Graaff}, {Hainline}, {Ji}, {Johnson}, {Kumari}, {Nelson}, {Parlanti}, {Rix}, {Robertson}, {Rodr{\'\i}guez Del Pino}, {Sandles}, {Scholtz}, {Smit}, {Stark}, {{\"U}bler}, {Williams}, {Willott}, \& {Witstok}}]{Looser2023b}
{Looser}, T.~J., {D'Eugenio}, F., {Maiolino}, R., {et~al.} 2023{\natexlab{a}}, arXiv e-prints, arXiv:2306.02470, \dodoi{10.48550/arXiv.2306.02470}

\bibitem[{{Looser} {et~al.}(2023{\natexlab{b}}){Looser}, {D'Eugenio}, {Maiolino}, {Witstok}, {Sandles}, {Curtis-Lake}, {Chevallard}, {Tacchella}, {Johnson}, {Baker}, {Suess}, {Carniani}, {Ferruit}, {Arribas}, {Bonaventura}, {Bunker}, {Cameron}, {Charlot}, {Curti}, {de Graaff}, {Maseda}, {Rawle}, {Rix}, {Rodriguez Del Pino}, {Smit}, {{\"U}bler}, {Willott}, {Alberts}, {Egami}, {Eisenstein}, {Endsley}, {Hausen}, {Rieke}, {Robertson}, {Shivaei}, {Williams}, {Boyett}, {Chen}, {Ji}, {Jones}, {Kumari}, {Nelson}, {Perna}, {Saxena}, \& {Scholtz}}]{Looser2023a}
---. 2023{\natexlab{b}}, arXiv e-prints, arXiv:2302.14155, \dodoi{10.48550/arXiv.2302.14155}

\bibitem[{{Looser} {et~al.}(2024){Looser}, {D'Eugenio}, {Maiolino}, {Witstok}, {Sandles}, {Curtis-Lake}, {Chevallard}, {Tacchella}, {Johnson}, {Baker}, {Suess}, {Carniani}, {Ferruit}, {Arribas}, {Bonaventura}, {Bunker}, {Cameron}, {Charlot}, {Curti}, {de Graaff}, {Maseda}, {Rawle}, {Rix}, {Del Pino}, {Smit}, {{\"U}bler}, {Willott}, {Alberts}, {Egami}, {Eisenstein}, {Endsley}, {Hausen}, {Rieke}, {Robertson}, {Shivaei}, {Williams}, {Boyett}, {Chen}, {Ji}, {Jones}, {Kumari}, {Nelson}, {Perna}, {Saxena}, \& {Scholtz}}]{Looser.2024}
---. 2024, \nat, 629, 53, \dodoi{10.1038/s41586-024-07227-0}

\bibitem[{{Luridiana} {et~al.}(2015){Luridiana}, {Morisset}, \& {Shaw}}]{Luridiana.2015}
{Luridiana}, V., {Morisset}, C., \& {Shaw}, R.~A. 2015, \aap, 573, A42, \dodoi{10.1051/0004-6361/201323152}

\bibitem[{{Ma} {et~al.}(2017){Ma}, {Hopkins}, {Feldmann}, {Torrey}, {Faucher-Gigu{\`e}re}, \& {Kere{\v{s}}}}]{Ma.2017}
{Ma}, X., {Hopkins}, P.~F., {Feldmann}, R., {et~al.} 2017, \mnras, 466, 4780, \dodoi{10.1093/mnras/stx034}

\bibitem[{{Madau} \& {Dickinson}(2014)}]{Madau.2014}
{Madau}, P., \& {Dickinson}, M. 2014, \araa, 52, 415, \dodoi{10.1146/annurev-astro-081811-125615}

\bibitem[{{Maiolino} \& {Mannucci}(2019)}]{Maiolino.2019}
{Maiolino}, R., \& {Mannucci}, F. 2019, \aapr, 27, 3, \dodoi{10.1007/s00159-018-0112-2}

\bibitem[{{Marconcini} {et~al.}(2024){Marconcini}, {D'Eugenio}, {Maiolino}, {Arribas}, {Bunker}, {Carniani}, {Charlot}, {Perna}, {Rodr{\'\i}guez Del Pino}, {{\"U}bler}, {Willott}, {B{\"o}ker}, {Cresci}, {Curti}, {Jones}, {Lamperti}, {Parlanti}, \& {Venturi}}]{Marconcini.2024}
{Marconcini}, C., {D'Eugenio}, F., {Maiolino}, R., {et~al.} 2024, \mnras, 533, 2488, \dodoi{10.1093/mnras/stae1971}

\bibitem[{{Marino} {et~al.}(2013){Marino}, {Rosales-Ortega}, {S{\'a}nchez}, {Gil de Paz}, {V{\'\i}lchez}, {Miralles-Caballero}, {Kehrig}, {P{\'e}rez-Montero}, {Stanishev}, {Iglesias-P{\'a}ramo}, {D{\'\i}az}, {Castillo-Morales}, {Kennicutt}, {L{\'o}pez-S{\'a}nchez}, {Galbany}, {Garc{\'\i}a-Benito}, {Mast}, {Mendez-Abreu}, {Monreal-Ibero}, {Husemann}, {Walcher}, {Garc{\'\i}a-Lorenzo}, {Masegosa}, {Del Olmo Orozco}, {Mour{\~a}o}, {Ziegler}, {Moll{\'a}}, {Papaderos}, {S{\'a}nchez-Bl{\'a}zquez}, {Gonz{\'a}lez Delgado}, {Falc{\'o}n-Barroso}, {Roth}, {van de Ven}, \& {CALIFA Team}}]{Marino.2013}
{Marino}, R.~A., {Rosales-Ortega}, F.~F., {S{\'a}nchez}, S.~F., {et~al.} 2013, \aap, 559, A114, \dodoi{10.1051/0004-6361/201321956}

\bibitem[{{Martinez} {et~al.}(2025){Martinez}, {Berg}, {James}, {Arellano-C{\'o}rdova}, {Stark}, {Senchyna}, {Skillman}, {Rogers}, \& {Chisholm}}]{Martinez.2025}
{Martinez}, Z., {Berg}, D.~A., {James}, B.~L., {et~al.} 2025, arXiv e-prints, arXiv:2510.21960, \dodoi{10.48550/arXiv.2510.21960}

\bibitem[{{McClymont} {et~al.}(2024){McClymont}, {Tacchella}, {Smith}, {Kannan}, {Maiolino}, {Belfiore}, {Hernquist}, {Li}, \& {Vogelsberger}}]{McClymont.2024}
{McClymont}, W., {Tacchella}, S., {Smith}, A., {et~al.} 2024, \mnras, 532, 2016, \dodoi{10.1093/mnras/stae1587}

\bibitem[{{Messa} {et~al.}(2025){Messa}, {Vanzella}, {Loiacono}, {Adamo}, {Oguri}, {Sharon}, {Bradley}, {Christensen}, {Claeyssens}, {Richard}, {Abdurro'uf}, {Bauer}, {Bergamini}, {Bolamperti}, {Brada{\v{c}}}, {Calura}, {Coe}, {Diego}, {Grillo}, {Y-Y. Hsiao}, {Inoue}, {Fujimoto}, {Lombardi}, {Meneghetti}, {Resseguier}, {Ricotti}, {Rosati}, {Welch}, {Windhorst}, {Xu}, {Zackrisson}, {Zanella}, \& {Zitrin}}]{messa2025}
{Messa}, M., {Vanzella}, E., {Loiacono}, F., {et~al.} 2025, arXiv e-prints, arXiv:2507.18705, \dodoi{10.48550/arXiv.2507.18705}

\bibitem[{{Mowla} {et~al.}(2022){Mowla}, {Iyer}, {Desprez}, {Estrada-Carpenter}, {Martis}, {Noirot}, {Sarrouh}, {Strait}, {Asada}, {Abraham}, {Brammer}, {Sawicki}, {Willott}, {Bradac}, {Doyon}, {Muzzin}, {Pacifici}, {Ravindranath}, \& {Zabl}}]{mowla2022}
{Mowla}, L., {Iyer}, K.~G., {Desprez}, G., {et~al.} 2022, \apjl, 937, L35, \dodoi{10.3847/2041-8213/ac90ca}

\bibitem[{{Mowla} {et~al.}(2024){Mowla}, {Iyer}, {Asada}, {Desprez}, {Tan}, {Martis}, {Sarrouh}, {Strait}, {Abraham}, {Brada{\v{c}}}, {Brammer}, {Muzzin}, {Pacifici}, {Ravindranath}, {Sawicki}, {Willott}, {Estrada-Carpenter}, {Jahan}, {Noirot}, {Matharu}, {Rihtar{\v{s}}i{\v{c}}}, \& {Zabl}}]{2024Natur.636..332M}
{Mowla}, L., {Iyer}, K., {Asada}, Y., {et~al.} 2024, \nat, 636, 332, \dodoi{10.1038/s41586-024-08293-0}

\bibitem[{{Nakajima} {et~al.}(2025){Nakajima}, {Ouchi}, {Harikane}, {Vanzella}, {Ono}, {Isobe}, {Nishigaki}, {Tsujimoto}, {Nakamura}, {Xu}, {Umeda}, \& {Zhang}}]{Nakajima.2025}
{Nakajima}, K., {Ouchi}, M., {Harikane}, Y., {et~al.} 2025, arXiv e-prints, arXiv:2506.11846, \dodoi{10.48550/arXiv.2506.11846}

\bibitem[{{Nanayakkara} {et~al.}(2025){Nanayakkara}, {Glazebrook}, {Schreiber}, {Chittenden}, {Brammer}, {Esdaile}, {Jacobs}, {Kacprzak}, {Kawinwanichakij}, {Kimmig}, {Labbe}, {Lagos}, {Marchesini}, {Mart{\`\i}nez-Mar{\`\i}n}, {Marsan}, {Oesch}, {Papovich}, {Remus}, \& {Tran}}]{nanayakkara2025}
{Nanayakkara}, T., {Glazebrook}, K., {Schreiber}, C., {et~al.} 2025, \apj, 981, 78, \dodoi{10.3847/1538-4357/ada6ac}

\bibitem[{{Nielsen} {et~al.}(2025){Nielsen}, {Steinhardt}, {Harper}, {McPartland}, \& {Sedgewick}}]{Nielsen2025}
{Nielsen}, E.~W., {Steinhardt}, C.~L., {Harper}, M., {McPartland}, C., \& {Sedgewick}, A. 2025, \aap, 700, A116, \dodoi{10.1051/0004-6361/202554507}

\bibitem[{{Osterbrock} \& {Ferland}(2006)}]{Osterbrock.2006}
{Osterbrock}, D.~E., \& {Ferland}, G.~J. 2006, {Astrophysics of gaseous nebulae and active galactic nuclei}

\bibitem[{{Pagel} {et~al.}(1979){Pagel}, {Edmunds}, {Blackwell}, {Chun}, \& {Smith}}]{Pagel.1979}
{Pagel}, B.~E.~J., {Edmunds}, M.~G., {Blackwell}, D.~E., {Chun}, M.~S., \& {Smith}, G. 1979, \mnras, 189, 95, \dodoi{10.1093/mnras/189.1.95}

\bibitem[{{Pallottini} \& {Ferrara}(2023)}]{pallottini_2023}
{Pallottini}, A., \& {Ferrara}, A. 2023, \aap, 677, L4, \dodoi{10.1051/0004-6361/202347384}

\bibitem[{{Papovich} {et~al.}(2022){Papovich}, {Simons}, {Estrada-Carpenter}, {Matharu}, {Momcheva}, {Trump}, {Backhaus}, {Brammer}, {Cleri}, {Finkelstein}, {Giavalisco}, {Ji}, {Jung}, {Kewley}, {Nicholls}, {Pirzkal}, {Rafelski}, \& {Weiner}}]{Papovich.2022}
{Papovich}, C., {Simons}, R.~C., {Estrada-Carpenter}, V., {et~al.} 2022, \apj, 937, 22, \dodoi{10.3847/1538-4357/ac8058}

\bibitem[{{Parlanti} {et~al.}(2025){Parlanti}, {Carniani}, {Venturi}, {Herrera-Camus}, {Arribas}, {Bunker}, {Charlot}, {D'Eugenio}, {Maiolino}, {Perna}, {{\"U}bler}, {B{\"o}ker}, {Cresci}, {Curti}, {Jones}, {Lamperti}, {P{\'e}rez-Gonz{\'a}lez}, {Del Pino}, \& {Zamora}}]{Parlanti.2025}
{Parlanti}, E., {Carniani}, S., {Venturi}, G., {et~al.} 2025, \aap, 695, A6, \dodoi{10.1051/0004-6361/202451692}

\bibitem[{{Pascale} {et~al.}(2025){Pascale}, {Dai}, {Frye}, \& {Beverage}}]{pascale2025}
{Pascale}, M., {Dai}, L., {Frye}, B.~L., \& {Beverage}, A.~G. 2025, \apjl, 988, L76, \dodoi{10.3847/2041-8213/aded93}

\bibitem[{{Perna} {et~al.}(2023){Perna}, {Arribas}, {Marshall}, {D'Eugenio}, {{\"U}bler}, {Bunker}, {Charlot}, {Carniani}, {Jakobsen}, {Maiolino}, {Rodr{\'\i}guez Del Pino}, {Willott}, {B{\"o}ker}, {Circosta}, {Cresci}, {Curti}, {Husemann}, {Kumari}, {Lamperti}, {P{\'e}rez-Gonz{\'a}lez}, \& {Scholtz}}]{Perna.2023}
{Perna}, M., {Arribas}, S., {Marshall}, M., {et~al.} 2023, \aap, 679, A89, \dodoi{10.1051/0004-6361/202346649}

\bibitem[{{Postnikova} \& {Bizyaev}(2023)}]{Postnikova.2023}
{Postnikova}, V.~K., \& {Bizyaev}, D. 2023, Astronomy Letters, 49, 151, \dodoi{10.1134/S1063773723040047}

\bibitem[{{Quintero} {et~al.}(2004){Quintero}, {Hogg}, {Blanton}, {Schlegel}, {Eisenstein}, {Gunn}, {Brinkmann}, {Fukugita}, {Glazebrook}, \& {Goto}}]{quintero2004}
{Quintero}, A.~D., {Hogg}, D.~W., {Blanton}, M.~R., {et~al.} 2004, \apj, 602, 190, \dodoi{10.1086/380601}

\bibitem[{{Rauscher}(2024)}]{Rauscher.2024}
{Rauscher}, B.~J. 2024, \pasp, 136, 015001, \dodoi{10.1088/1538-3873/ad1b36}

\bibitem[{{Rauscher} {et~al.}(2017){Rauscher}, {Arendt}, {Fixsen}, {Greenhouse}, {Lander}, {Lindler}, {Loose}, {Moseley}, {Mott}, {Wen}, {Wilson}, \& {Xenophontos}}]{Rauscher.2017}
{Rauscher}, B.~J., {Arendt}, R.~G., {Fixsen}, D.~J., {et~al.} 2017, \pasp, 129, 105003, \dodoi{10.1088/1538-3873/aa83fd}

\bibitem[{{Rieke} {et~al.}(2023){Rieke}, {Kelly}, {Misselt}, {Stansberry}, {Boyer}, {Beatty}, {Egami}, {Florian}, {Greene}, {Hainline}, {Leisenring}, {Roellig}, {Schlawin}, {Sun}, {Tinnin}, {Williams}, {Willmer}, {Wilson}, {Clark}, {Rohrbach}, {Brooks}, {Canipe}, {Correnti}, {DiFelice}, {Gennaro}, {Girard}, {Hartig}, {Hilbert}, {Koekemoer}, {Nikolov}, {Pirzkal}, {Rest}, {Robberto}, {Sunnquist}, {Telfer}, {Wu}, {Ferry}, {Lewis}, {Baum}, {Beichman}, {Doyon}, {Dressler}, {Eisenstein}, {Ferrarese}, {Hodapp}, {Horner}, {Jaffe}, {Johnstone}, {Krist}, {Martin}, {McCarthy}, {Meyer}, {Rieke}, {Trauger}, \& {Young}}]{Rieke.2023}
{Rieke}, M.~J., {Kelly}, D.~M., {Misselt}, K., {et~al.} 2023, \pasp, 135, 028001, \dodoi{10.1088/1538-3873/acac53}

\bibitem[{{Rigby} {et~al.}(2023){Rigby}, {Vieira}, {Phadke}, {Hutchison}, {Welch}, {Cathey}, {Spilker}, {Gonzalez}, {Adhikari}, {Aravena}, {Bayliss}, {Birkin}, {Bursk}, {Chapman}, {Dahle}, {Elicker}, {Fischer}, {Florian}, {Gladders}, {Hayward}, {Hewald}, {Kettler}, {Khullar}, {Kim}, {Law}, {Mahler}, {Malhotra}, {Murphy}, {Narayanan}, {Olivier}, {Rhoads}, {Sharon}, {Solimano}, {Thiruvengadam}, {Vizgan}, \& {Younker}}]{Rigby.2023}
{Rigby}, J.~R., {Vieira}, J.~D., {Phadke}, K.~A., {et~al.} 2023, arXiv e-prints, arXiv:2312.10465, \dodoi{10.48550/arXiv.2312.10465}

\bibitem[{{Rivera-Thorsen} {et~al.}(2025){Rivera-Thorsen}, {Welch}, {Hutchison}, {Hayes}, {Rigby}, {Kim}, {Choe}, {Florian}, {Bayliss}, {Khullar}, {Sharon}, {Dahle}, {Chisholm}, {Solhaug}, {Owens}, \& {Gladders}}]{Rivera-Thorsen.2025}
{Rivera-Thorsen}, T.~E., {Welch}, B., {Hutchison}, T., {et~al.} 2025, arXiv e-prints, arXiv:2510.11702, \dodoi{10.48550/arXiv.2510.11702}

\bibitem[{{Robertson}(2022)}]{Robertson.2022}
{Robertson}, B.~E. 2022, \araa, 60, 121, \dodoi{10.1146/annurev-astro-120221-044656}

\bibitem[{{Rogers} {et~al.}(2025){Rogers}, {Strom}, {Rudie}, {Trainor}, {von Raesfeld}, {Raptis}, {Korhonen Cuestas}, {Miller}, {Steidel}, {Maseda}, {Chen}, \& {Law}}]{Rogers.2025}
{Rogers}, N. S.~J., {Strom}, A.~L., {Rudie}, G.~C., {et~al.} 2025, arXiv e-prints, arXiv:2509.18257, \dodoi{10.48550/arXiv.2509.18257}

\bibitem[{{Rosa-Gonz{\'a}lez} {et~al.}(2002){Rosa-Gonz{\'a}lez}, {Terlevich}, \& {Terlevich}}]{Rosa-Gonzalez.2002}
{Rosa-Gonz{\'a}lez}, D., {Terlevich}, E., \& {Terlevich}, R. 2002, \mnras, 332, 283, \dodoi{10.1046/j.1365-8711.2002.05285.x}

\bibitem[{{Roy} {et~al.}(2025){Roy}, {Henry}, {Jones}, {Barisic}, {Sanders}, {Bundy}, {Malkan}, {Nanayakkara}, {Glazebrook}, {Heckman}, {Espejo Salcedo}, {Wang}, {Obreschkow}, \& {Treu}}]{Roy.2025}
{Roy}, N., {Henry}, A., {Jones}, T., {et~al.} 2025, arXiv e-prints, arXiv:2510.11326, \dodoi{10.48550/arXiv.2510.11326}

\bibitem[{{Rubin} {et~al.}(2010){Rubin}, {Weiner}, {Koo}, {Martin}, {Prochaska}, {Coil}, \& {Newman}}]{Rubin.2010}
{Rubin}, K. H.~R., {Weiner}, B.~J., {Koo}, D.~C., {et~al.} 2010, \apj, 719, 1503, \dodoi{10.1088/0004-637X/719/2/1503}

\bibitem[{{Sanders} {et~al.}(2023){Sanders}, {Shapley}, {Topping}, {Reddy}, \& {Brammer}}]{Sanders.2023}
{Sanders}, R.~L., {Shapley}, A.~E., {Topping}, M.~W., {Reddy}, N.~A., \& {Brammer}, G.~B. 2023, \apj, 955, 54, \dodoi{10.3847/1538-4357/acedad}

\bibitem[{{Sanders} {et~al.}(2015){Sanders}, {Shapley}, {Kriek}, {Reddy}, {Freeman}, {Coil}, {Siana}, {Mobasher}, {Shivaei}, {Price}, \& {de Groot}}]{Sanders.2015}
{Sanders}, R.~L., {Shapley}, A.~E., {Kriek}, M., {et~al.} 2015, \apj, 799, 138, \dodoi{10.1088/0004-637X/799/2/138}

\bibitem[{{Sanders} {et~al.}(2016){Sanders}, {Shapley}, {Kriek}, {Reddy}, {Freeman}, {Coil}, {Siana}, {Mobasher}, {Shivaei}, {Price}, \& {de Groot}}]{Sanders.2016}
---. 2016, \apj, 816, 23, \dodoi{10.3847/0004-637X/816/1/23}

\bibitem[{{Sanders} {et~al.}(2018){Sanders}, {Shapley}, {Kriek}, {Freeman}, {Reddy}, {Siana}, {Coil}, {Mobasher}, {Dav{\'e}}, {Shivaei}, {Azadi}, {Price}, {Leung}, {Fetherolf}, {de Groot}, {Zick}, {Fornasini}, \& {Barro}}]{Sanders.2018}
---. 2018, \apj, 858, 99, \dodoi{10.3847/1538-4357/aabcbd}

\bibitem[{{Sanders} {et~al.}(2020){Sanders}, {Shapley}, {Reddy}, {Kriek}, {Siana}, {Coil}, {Mobasher}, {Shivaei}, {Freeman}, {Azadi}, {Price}, {Leung}, {Fetherolf}, {de Groot}, {Zick}, {Fornasini}, \& {Barro}}]{Sanders.2020}
{Sanders}, R.~L., {Shapley}, A.~E., {Reddy}, N.~A., {et~al.} 2020, \mnras, 491, 1427, \dodoi{10.1093/mnras/stz3032}

\bibitem[{{Sanders} {et~al.}(2024){Sanders}, {Shapley}, {Topping}, {Reddy}, {Berg}, {Bouwens}, {Brammer}, {Carnall}, {Cullen}, {Dav{\'e}}, {Dunlop}, {Ellis}, {F{\"o}rster Schreiber}, {Furlanetto}, {Glazebrook}, {Illingworth}, {Jones}, {Kriek}, {McLeod}, {McLure}, {Narayanan}, {Oesch}, {Pahl}, {Pettini}, {Schaerer}, {Stark}, {Steidel}, {Tang}, {Clarke}, {Donnan}, \& {Kehoe}}]{Sanders.2024}
{Sanders}, R.~L., {Shapley}, A.~E., {Topping}, M.~W., {et~al.} 2024, arXiv e-prints, arXiv:2408.05273, \dodoi{10.48550/arXiv.2408.05273}

\bibitem[{{Sanders} {et~al.}(2025){Sanders}, {Shapley}, {Topping}, {Reddy}, {Berg}, {Khostovan}, {Bouwens}, {Brammer}, {Carnall}, {Cullen}, {Dav{\'e}}, {Dunlop}, {Ellis}, {F{\"o}rster Schreiber}, {Furlanetto}, {Glazebrook}, {Illingworth}, {Jones}, {Kriek}, {McLeod}, {McLure}, {Narayanan}, {Oesch}, {Pahl}, {Pettini}, {Schaerer}, {Stark}, {Steidel}, {Tang}, {Clarke}, {Donnan}, \& {Kehoe}}]{Sanders.2025}
---. 2025, arXiv e-prints, arXiv:2508.10099, \dodoi{10.48550/arXiv.2508.10099}

\bibitem[{{Scholtz} {et~al.}(2025){Scholtz}, {Curti}, {D'Eugenio}, {{\"U}bler}, {Maiolino}, {Marconcini}, {Smit}, {Perna}, {Witstok}, {Arribas}, {B{\"o}ker}, {Bunker}, {Carniani}, {Charlot}, {Cresci}, {Lamperti}, {Parlanti}, {P{\'e}rez-Gonz{\'a}lez}, {Rodr{\'\i}guez Del Pino}, \& {Venturi}}]{Scholtz.2025}
{Scholtz}, J., {Curti}, M., {D'Eugenio}, F., {et~al.} 2025, \mnras, 539, 2463, \dodoi{10.1093/mnras/staf518}

\bibitem[{{Setton} {et~al.}(2023){Setton}, {Dey}, {Khullar}, {Bezanson}, {Newman}, {Aguilar}, {Ahlen}, {Andrews}, {Brooks}, {de la Macorra}, {Dey}, {Eftekharzadeh}, {Font-Ribera}, {A Gontcho}, {Kremin}, {Juneau}, {Landriau}, {Meisner}, {Miquel}, {Moustakas}, {Pearl}, {Prada}, {Tarl{\'e}}, {Siudek}, {Weaver}, {Zhou}, \& {Zou}}]{Setton2023}
{Setton}, D.~J., {Dey}, B., {Khullar}, G., {et~al.} 2023, \apjl, 947, L31, \dodoi{10.3847/2041-8213/acc9b5}

\bibitem[{{Setton} {et~al.}(2024){Setton}, {Khullar}, {Miller}, {Bezanson}, {Greene}, {Suess}, {Whitaker}, {Antwi-Danso}, {Atek}, {Brammer}, {Cutler}, {Dayal}, {Feldmann}, {Fujimoto}, {Furtak}, {Glazebrook}, {Goulding}, {Kokorev}, {Labbe}, {Leja}, {Ma}, {Marchesini}, {Nanayakkara}, {Pan}, {Price}, {Siegel}, {Shipley}, {Weaver}, {van Dokkum}, {Wang}, \& {Williams}}]{Setton.2024}
{Setton}, D.~J., {Khullar}, G., {Miller}, T.~B., {et~al.} 2024, \apj, 974, 145, \dodoi{10.3847/1538-4357/ad6a18}

\bibitem[{{Shapley} {et~al.}(2015){Shapley}, {Reddy}, {Kriek}, {Freeman}, {Sanders}, {Siana}, {Coil}, {Mobasher}, {Shivaei}, {Price}, \& {de Groot}}]{Shapley.2015}
{Shapley}, A.~E., {Reddy}, N.~A., {Kriek}, M., {et~al.} 2015, \apj, 801, 88, \dodoi{10.1088/0004-637X/801/2/88}

\bibitem[{{Shapley} {et~al.}(2019){Shapley}, {Sanders}, {Shao}, {Reddy}, {Kriek}, {Coil}, {Mobasher}, {Siana}, {Shivaei}, {Freeman}, {Azadi}, {Price}, {Leung}, {Fetherolf}, {de Groot}, {Zick}, {Fornasini}, \& {Barro}}]{Shapley.2019}
{Shapley}, A.~E., {Sanders}, R.~L., {Shao}, P., {et~al.} 2019, \apjl, 881, L35, \dodoi{10.3847/2041-8213/ab385a}

\bibitem[{{Shapley} {et~al.}(2025){Shapley}, {Sanders}, {Topping}, {Reddy}, {Pahl}, {Oesch}, {Berg}, {Bouwens}, {Brammer}, {Carnall}, {Cullen}, {Dav{\'e}}, {Dunlop}, {Ellis}, {F{\"o}rster Schreiber}, {Furlanetto}, {Glazebrook}, {Illingworth}, {Jones}, {Kriek}, {McLeod}, {McLure}, {Narayanan}, {Pettini}, {Schaerer}, {Stark}, {Steidel}, {Tang}, {Clarke}, {Donnan}, \& {Kehoe}}]{Shapley.2025}
{Shapley}, A.~E., {Sanders}, R.~L., {Topping}, M.~W., {et~al.} 2025, \apj, 981, 167, \dodoi{10.3847/1538-4357/adaf98}

\bibitem[{{Shin} {et~al.}(2008){Shin}, {Strauss}, {Oguri}, {Inada}, {Falco}, {Broadhurst}, \& {Gunn}}]{Shin2000}
{Shin}, M.-S., {Strauss}, M.~A., {Oguri}, M., {et~al.} 2008, \aj, 136, 44, \dodoi{10.1088/0004-6256/136/1/44}

\bibitem[{{Shivaei} {et~al.}(2015){Shivaei}, {Reddy}, {Shapley}, {Kriek}, {Siana}, {Mobasher}, {Coil}, {Freeman}, {Sanders}, {Price}, {de Groot}, \& {Azadi}}]{Shivaei.2015}
{Shivaei}, I., {Reddy}, N.~A., {Shapley}, A.~E., {et~al.} 2015, \apj, 815, 98, \dodoi{10.1088/0004-637X/815/2/98}

\bibitem[{{Shivaei} {et~al.}(2018){Shivaei}, {Reddy}, {Siana}, {Shapley}, {Kriek}, {Mobasher}, {Freeman}, {Sanders}, {Coil}, {Price}, {Fetherolf}, {Azadi}, {Leung}, \& {Zick}}]{Shivaei.2018}
{Shivaei}, I., {Reddy}, N.~A., {Siana}, B., {et~al.} 2018, \apj, 855, 42, \dodoi{10.3847/1538-4357/aaad62}

\bibitem[{{Siana} {et~al.}(2010){Siana}, {Teplitz}, {Ferguson}, {Brown}, {Giavalisco}, {Dickinson}, {Chary}, {de Mello}, {Conselice}, {Bridge}, {Gardner}, {Colbert}, \& {Scarlata}}]{Siana.2010}
{Siana}, B., {Teplitz}, H.~I., {Ferguson}, H.~C., {et~al.} 2010, \apj, 723, 241, \dodoi{10.1088/0004-637X/723/1/241}

\bibitem[{{Simons} {et~al.}(2021){Simons}, {Papovich}, {Momcheva}, {Trump}, {Brammer}, {Estrada-Carpenter}, {Backhaus}, {Cleri}, {Finkelstein}, {Giavalisco}, {Ji}, {Jung}, {Matharu}, \& {Weiner}}]{Simons.2021}
{Simons}, R.~C., {Papovich}, C., {Momcheva}, I., {et~al.} 2021, \apj, 923, 203, \dodoi{10.3847/1538-4357/ac28f4}

\bibitem[{{Speagle}(2020)}]{Speagle2020}
{Speagle}, J.~S. 2020, \mnras, 493, 3132, \dodoi{10.1093/mnras/staa278}

\bibitem[{{Stark}(2016)}]{Stark.2016}
{Stark}, D.~P. 2016, \araa, 54, 761, \dodoi{10.1146/annurev-astro-081915-023417}

\bibitem[{{Steidel} {et~al.}(2016){Steidel}, {Strom}, {Pettini}, {Rudie}, {Reddy}, \& {Trainor}}]{Steidel.2016}
{Steidel}, C.~C., {Strom}, A.~L., {Pettini}, M., {et~al.} 2016, \apj, 826, 159, \dodoi{10.3847/0004-637X/826/2/159}

\bibitem[{{Steidel} {et~al.}(2014){Steidel}, {Rudie}, {Strom}, {Pettini}, {Reddy}, {Shapley}, {Trainor}, {Erb}, {Turner}, {Konidaris}, {Kulas}, {Mace}, {Matthews}, \& {McLean}}]{Steidel.2014}
{Steidel}, C.~C., {Rudie}, G.~C., {Strom}, A.~L., {et~al.} 2014, \apj, 795, 165, \dodoi{10.1088/0004-637X/795/2/165}

\bibitem[{{Strait} {et~al.}(2023){Strait}, {Brammer}, {Muzzin}, {Desprez}, {Asada}, {Abraham}, {Brada{\v{c}}}, {Iyer}, {Martis}, {Mowla}, {Noirot}, {Sarrouh}, {Sawicki}, {Willott}, {Gould}, {Grindlay}, {Matharu}, \& {Rihtar{\v{s}}i{\v{c}}}}]{Strait2023}
{Strait}, V., {Brammer}, G., {Muzzin}, A., {et~al.} 2023, \apjl, 949, L23, \dodoi{10.3847/2041-8213/acd457}

\bibitem[{{Strom} {et~al.}(2022){Strom}, {Rudie}, {Steidel}, \& {Trainor}}]{Strom.2022}
{Strom}, A.~L., {Rudie}, G.~C., {Steidel}, C.~C., \& {Trainor}, R.~F. 2022, \apj, 925, 116, \dodoi{10.3847/1538-4357/ac38a3}

\bibitem[{{Strom} {et~al.}(2018){Strom}, {Steidel}, {Rudie}, {Trainor}, \& {Pettini}}]{Strom.2018}
{Strom}, A.~L., {Steidel}, C.~C., {Rudie}, G.~C., {Trainor}, R.~F., \& {Pettini}, M. 2018, \apj, 868, 117, \dodoi{10.3847/1538-4357/aae1a5}

\bibitem[{{Strom} {et~al.}(2017){Strom}, {Steidel}, {Rudie}, {Trainor}, {Pettini}, \& {Reddy}}]{Strom.2017}
{Strom}, A.~L., {Steidel}, C.~C., {Rudie}, G.~C., {et~al.} 2017, \apj, 836, 164, \dodoi{10.3847/1538-4357/836/2/164}

\bibitem[{{Suess} {et~al.}(2022){Suess}, {Kriek}, {Bezanson}, {Greene}, {Setton}, {Spilker}, {Feldmann}, {Goulding}, {Johnson}, {Leja}, {Narayanan}, {Hall-Hooper}, {Hunt}, {Lower}, \& {Verrico}}]{Suess2022a}
{Suess}, K.~A., {Kriek}, M., {Bezanson}, R., {et~al.} 2022, \apj, 926, 89, \dodoi{10.3847/1538-4357/ac404a}

\bibitem[{{Suzuki} {et~al.}(2017){Suzuki}, {Kodama}, {Onodera}, {Shimakawa}, {Hayashi}, {Tadaki}, {Koyama}, {Tanaka}, {Sobral}, {Smail}, {Best}, {Khostovan}, {Minowa}, \& {Yamamoto}}]{Suzuki.2017}
{Suzuki}, T.~L., {Kodama}, T., {Onodera}, M., {et~al.} 2017, \apj, 849, 39, \dodoi{10.3847/1538-4357/aa8df3}

\bibitem[{{Tapia-Contreras} {et~al.}(2025){Tapia-Contreras}, {Tissera}, {Sillero}, {Gonzalez-Jara}, {Casanueva-Villarreal}, {Pedrosa}, {Bignone}, {Padilla}, \& {Dom{\'\i}nguez-Tenreiro}}]{Tapia-Contreras.2025}
{Tapia-Contreras}, B., {Tissera}, P.~B., {Sillero}, E., {et~al.} 2025, \aap, 700, A69, \dodoi{10.1051/0004-6361/202554013}

\bibitem[{{Tissera} {et~al.}(2019){Tissera}, {Rosas-Guevara}, {Bower}, {Crain}, {Lagos}, {Schaller}, {Schaye}, \& {Theuns}}]{Tissera.2019}
{Tissera}, P.~B., {Rosas-Guevara}, Y., {Bower}, R.~G., {et~al.} 2019, \mnras, 482, 2208, \dodoi{10.1093/mnras/sty2817}

\bibitem[{{Topping} {et~al.}(2025){Topping}, {Sanders}, {Shapley}, {Pahl}, {Reddy}, {Stark}, {Berg}, {Clarke}, {Cullen}, {Dunlop}, {Ellis}, {Schreiber}, {Illingworth}, {Jones}, {Narayanan}, {Pettini}, \& {Schaerer}}]{Topping.2025}
{Topping}, M.~W., {Sanders}, R.~L., {Shapley}, A.~E., {et~al.} 2025, \mnras, 541, 1707, \dodoi{10.1093/mnras/staf903}

\bibitem[{{Tremonti} {et~al.}(2004){Tremonti}, {Heckman}, {Kauffmann}, {Brinchmann}, {Charlot}, {White}, {Seibert}, {Peng}, {Schlegel}, {Uomoto}, {Fukugita}, \& {Brinkmann}}]{Tremonti.2004}
{Tremonti}, C.~A., {Heckman}, T.~M., {Kauffmann}, G., {et~al.} 2004, \apj, 613, 898, \dodoi{10.1086/423264}

\bibitem[{{Trussler} {et~al.}(2024){Trussler}, {Conselice}, {Adams}, {Austin}, {Caruana}, {Harvey}, {Li}, {Lovell}, {Seeyave}, {Vijayan}, \& {Wilkins}}]{trussler2024}
{Trussler}, J., {Conselice}, C., {Adams}, N., {et~al.} 2024, arXiv e-prints, arXiv:2404.07163, \dodoi{10.48550/arXiv.2404.07163}

\bibitem[{{Venturi} {et~al.}(2024){Venturi}, {Carniani}, {Parlanti}, {Kohandel}, {Curti}, {Pallottini}, {Vallini}, {Arribas}, {Bunker}, {Cameron}, {Castellano}, {Ferrara}, {Fontana}, {Gallerani}, {Gelli}, {Maiolino}, {Ntormousi}, {Pacifici}, {Pentericci}, {Salvadori}, \& {Vanzella}}]{Venturi.2024}
{Venturi}, G., {Carniani}, S., {Parlanti}, E., {et~al.} 2024, \aap, 691, A19, \dodoi{10.1051/0004-6361/202449855}

\bibitem[{{Vincenzo} {et~al.}(2016){Vincenzo}, {Belfiore}, {Maiolino}, {Matteucci}, \& {Ventura}}]{Vincenzo.2016}
{Vincenzo}, F., {Belfiore}, F., {Maiolino}, R., {Matteucci}, F., \& {Ventura}, P. 2016, \mnras, 458, 3466, \dodoi{10.1093/mnras/stw532}

\bibitem[{Virtanen {et~al.}(2020)Virtanen, Gommers, Oliphant, Haberland, Reddy, Cournapeau, Burovski, Peterson, Weckesser, Bright, {van der Walt}, Brett, Wilson, Millman, Mayorov, Nelson, Jones, Kern, Larson, Carey, Polat, Feng, Moore, {VanderPlas}, Laxalde, Perktold, Cimrman, Henriksen, Quintero, Harris, Archibald, Ribeiro, Pedregosa, {van Mulbregt}, \& {SciPy 1.0 Contributors}}]{Virtanen.2020.SciPy}
Virtanen, P., Gommers, R., Oliphant, T.~E., {et~al.} 2020, Nature Methods, 17, 261, \dodoi{10.1038/s41592-019-0686-2}

\bibitem[{{Wang} {et~al.}(2022){Wang}, {Jones}, {Vulcani}, {Treu}, {Morishita}, {Roberts-Borsani}, {Malkan}, {Henry}, {Brammer}, {Strait}, {Brada{\v{c}}}, {Boyett}, {Calabr{\`o}}, {Castellano}, {Fontana}, {Glazebrook}, {Kelly}, {Leethochawalit}, {Marchesini}, {Santini}, {Trenti}, \& {Yang}}]{2022ApJ...938L..16W}
{Wang}, X., {Jones}, T., {Vulcani}, B., {et~al.} 2022, \apjl, 938, L16, \dodoi{10.3847/2041-8213/ac959e}

\bibitem[{{Wang} {et~al.}(2025){Wang}, {Shen}, {Vogelsberger}, {Li}, {Kannan}, {Puchwein}, {Smith}, {Borrow}, {Garaldi}, {Keating}, {Zier}, {McClymont}, {Tacchella}, {Ni}, \& {Hernquist}}]{Wang.2025}
{Wang}, Z., {Shen}, X., {Vogelsberger}, M., {et~al.} 2025, \mnras, 544, 2675, \dodoi{10.1093/mnras/staf1677}

\bibitem[{{Weibel} {et~al.}(2024){Weibel}, {de Graaff}, {Setton}, {Miller}, {Oesch}, {Brammer}, {Lagos}, {Whitaker}, {Williams}, {Baggen}, {Bezanson}, {Boogaard}, {Cleri}, {Greene}, {Hirschmann}, {Hviding}, {Kuruvanthodi}, {Labb{\'e}}, {Leja}, {Maseda}, {Matthee}, {McConachie}, {Naidu}, {Roberts-Borsani}, {Schaerer}, {Suess}, {Valentino}, {van Dokkum}, \& {Wang}}]{Weibel.2024}
{Weibel}, A., {de Graaff}, A., {Setton}, D.~J., {et~al.} 2024, arXiv e-prints, arXiv:2409.03829, \dodoi{10.48550/arXiv.2409.03829}

\bibitem[{{Zhang} {et~al.}(2017){Zhang}, {Yan}, {Bundy}, {Bershady}, {Haffner}, {Walterbos}, {Maiolino}, {Tremonti}, {Thomas}, {Drory}, {Jones}, {Belfiore}, {S{\'a}nchez}, {Diamond-Stanic}, {Bizyaev}, {Nitschelm}, {Andrews}, {Brinkmann}, {Brownstein}, {Cheung}, {Li}, {Law}, {Roman Lopes}, {Oravetz}, {Pan}, {Storchi Bergmann}, \& {Simmons}}]{Zhang.2017}
{Zhang}, K., {Yan}, R., {Bundy}, K., {et~al.} 2017, \mnras, 466, 3217, \dodoi{10.1093/mnras/stw3308}

\bibitem[{{Zhu} {et~al.}(2025){Zhu}, {Suess}, {Kriek}, {Setton}, {Bezanson}, {Donofrio}, {Feldmann}, {Goulding}, {Greene}, {Narayanan}, \& {Spilker}}]{zhu2025}
{Zhu}, P., {Suess}, K.~A., {Kriek}, M., {et~al.} 2025, \apj, 981, 60, \dodoi{10.3847/1538-4357/adaf1c}

\end{thebibliography}
\bibliographystyle{aasjournal}



\end{document}